%% file: LHCb-PAPER-2019-027.tex
\pdfoutput=1
\documentclass[12pt,a4paper]{article}
\usepackage{ifthen} 
\newboolean{pdflatex}
\setboolean{pdflatex}{true} 

\newboolean{articletitles}
\setboolean{articletitles}{true} 

\newboolean{uprightparticles}
\setboolean{uprightparticles}{false} 

\def\paperauthors{LHCb collaboration} 
\def\paperasciititle{Determination of quantum numbers for several excited charmed mesons observed in B-->D*+pi-pi- decays} 
\def\papertitle{Determination of quantum numbers for several excited charmed mesons observed in $\Bm \to \Dstarp \pim \pim$ decays}
\def\paperkeywords{{High Energy Physics}, {LHCb}} 
\def\papercopyright{\the\year\ CERN for the benefit of the LHCb collaboration} 
\def\paperlicence{CC-BY-4.0 licence}
\def\paperlicenceurl{https://creativecommons.org/licenses/by/4.0/}
\def\epem     {\ensuremath{e^+e^-}\xspace}
\def\Dstarpi {\ensuremath{\Dstarp \pim}\xspace}
\def\Dzpi {\ensuremath{\Dz \pip}\xspace}
\def\Dppim {\ensuremath{\Dp \pim}\xspace}

\def\Done {\ensuremath{D_1(2420)}\xspace}

\def\Donew {\ensuremath{D_1(2430)}\xspace}
\def\Dstartwo {\ensuremath{D^*_2(2460)}\xspace}

\def\Donem {\ensuremath{D^*_{1}(2600)}\xspace}
\def\Donemj {\ensuremath{D^*_J(2600)}\xspace}

\def\Dtwom {\ensuremath{D_{2}(2740)}\xspace}

\def\Dthree {\ensuremath{D^*_{3}(2750)}\xspace}

\def\Dzero {\ensuremath{D_0(2550)}\xspace}

\def\Dtwo {\ensuremath{D(2740)}\xspace}
\def\invfb   {\ensuremath{\mbox{\,fb}^{-1}}\xspace}

\def\photos     {\mbox{\textsc{Photos}}\xspace}

\def\calR         {{\ensuremath{\cal R}\xspace}}

\input{preamble}
\usepackage{longtable} 

\begin{document}

\renewcommand{\thefootnote}{\fnsymbol{footnote}}
\setcounter{footnote}{1}
\input{title-LHCb-PAPER}


\renewcommand{\thefootnote}{\arabic{footnote}}
\setcounter{footnote}{0}

\pagestyle{plain} 
\setcounter{page}{1}
\pagenumbering{arabic}

\section{Introduction}
\label{sec:Introduction}
Charmed-meson spectroscopy provides a powerful test of quark-model predictions in the Standard Model.
Many charmed-meson states predicted in the 1980ies (see e.g. Ref.~\cite{Godfrey:1985xj} and references within, and Ref.~\cite{Godfrey:2015dva} for a recent update) have not yet been observed experimentally.
The study of $D \pi$ final states enables a search for
natural spin-parity resonances, ($P=(-1)^J$, labeled as $D^*$) while the study of $D^* \pi$ final states provides the possibility of studying both natural and unnatural spin-parity states, except for the $J^P=0^+$ case, which is forbidden due to angular momentum and parity conservation.
Above the ground $1S$ states ($D$, $D^*$), two of the orbital $1P$ excitations, \Done and \Dstartwo, are experimentally well established~\cite{Tanabashi:2018oca} since they have relatively narrow widths ($\sim$30\mev).\footnote{The system of units where $c=1$ is adopted throughout.} One of the broad $1P$ states, $D^*_0(2400)$, has been studied by the \belle, \babar and \lhcb  \ collaborations in exclusive \B decays~\cite{Abe:2003zm,Aubert:2009wg,LHCb-PAPER-2015-007,LHCb-PAPER-2014-070,LHCb-PAPER-2016-026}.
Another broad $1P$ state, \Donew, has been observed by the \belle \ collaboration in the amplitude analysis of 560 $\Bm \to \Dstarp \pim \pim$ decays~\cite{Abe:2003zm}. The study of the $B \to D^{**}l\nu$ decay by the \babar~\cite{Aubert:2008ea} and \belle~\cite{Liventsev:2007rb} collaborations gives contradicting results on the production of \Donew in semileptonic $B$ decays.

The search for excited charmed mesons, labeled $D_J$, can be performed using two different approaches: using inclusive reactions, or through amplitude analyses
of exclusive \B decays.
In inclusive $D^{(*)}\pi$ production, where production of any $J^P$ state is permitted,  large data samples are obtained, however in addition to a large combinatorial background.
In three-body $D_J$ decays it is also possible to perform an angular analysis and therefore distinguish between natural and unnatural spin-parity assignments.
The amplitude analysis of \B decays, on the other hand, although often with limited data sample size, allows a full spin-parity analysis of the charmed mesons present in the decay. In addition, backgrounds are usually rather low and comparatively well understood.

Using the first approach, the \babar~\cite{delAmoSanchez:2010vq} (in \epem annihilations) and \lhcb~\cite{Aaij:2013sza} (in $pp$ interactions) collaborations, have
analyzed 
the inclusive production of the \Dppim, \Dzpi and \Dstarpi final states.
Both collaborations observe four resonances, labeled by the Particle Data Group (PDG) as \Dzero, \Donemj, \Dtwo and \Dthree~\cite{Tanabashi:2018oca}. The \Dzero and \Dtwo decay angular distributions
are consistent with an unnatural spin-parity, while the \Donemj and \Dthree states are assigned natural parities.
The \Dthree resonance was also observed in \B-decay amplitude analyses of $\Bz \to \Dzb\pip\pim$~\cite{Aaij:2015sqa} and $\Bm \to \Dp \pim \pip$~\cite{Aaij:2016fma} by the \lhcb collaboration, where quantum numbers were determined to be $J^P=3^-$.
For the \Dzero meson, angular distributions are consistent with a $J^P=0^-$ assignment, however for the other states no definite assignment exists.

This paper reports on the study of $D_J$ spectroscopy in the $\Dstarp\pim$ system through an  amplitude analysis of the $\Bm \to \Dstarp \pim \pim$ decay.\footnote{The inclusion of charge-conjugate processes is implied, unless stated.} The data sample corresponds to a total integrated luminosity of 4.7\invfb of $pp$ collisions collected at center-of-mass energies of
7, 8 and 13\tev with the \lhcb detector.
The 7 and 8\tev dataset is labeled in the following as ``Run 1'' data, and the 13\tev dataset as ``Run 2'' data.

The article is organized as follows. Section~\ref{sec:detector} gives details on the LHCb detector, while Sec.~\ref{sec:selection} is devoted to the description of the data selection procedure. Section~\ref{sec:data} describes the data features and Sec.~\ref{sec:back}
is devoted to the handling of the background and the efficiency model. In Sec.~\ref{sec:daly} the amplitude analysis model
is described,
while Sec.~\ref{sec:qmi} and Sec.~\ref{sec:fits} give details on the fits to the data. The measurements of the partial branching fractions are given in Sec.~\ref{sec:br} and results are summarized in Sec.~\ref{sec:sum}.

\section{LHCb detector}
\label{sec:detector}

The \lhcb detector~\cite{Alves:2008zz,Aaij:2014jba} is a single-arm forward
spectrometer covering the \mbox{pseudorapidity} range $2<\eta <5$,
designed for the study of particles containing \bquark or \cquark
quarks. The detector includes a high-precision tracking system
consisting of a silicon-strip vertex detector 
surrounding the $pp$ interaction region, a large-area silicon-strip detector
located upstream of a dipole magnet with a bending power of about
$4{\rm\,Tm}$, and three stations of silicon-strip detectors and straw
drift tubes placed downstream of the magnet.
The polarity of the dipole magnet is reversed periodically throughout data-taking.
The tracking system provides a measurement of the momentum of charged particles with relative uncertainty that varies from $0.5\,\%$ at low momentum ($3\gev$) to $1.0\,\%$ at $200\gev$.
The minimum distance of a track to a primary vertex (PV) (defined as the location of a reconstructed $pp$ collision) the impact parameter (IP), is measured with a resolution of $(15+29/\pt)\mum$,
where \pt is the component of the momentum transverse to the beam, in~\gev.
Different types of charged hadrons are distinguished using information
from two ring-imaging Cherenkov detectors. 
Photon, electron and hadron candidates are identified by a calorimeter system consisting of scintillating-pad and preshower detectors, an electromagnetic and a hadronic calorimeter. 
Muons are identified by a system composed of alternating layers of iron and multiwire proportional chambers. 

The trigger consists of a hardware stage, based on information from the calorimeter and muon systems, followed by a software stage, in which all tracks with $\pt>500~(300)\mev$ are reconstructed for data collected at $\sqrt{s}=7\tev$ (8--13\tev).
The software trigger used in this analysis requires a two-, three- or four-track secondary vertex with significant displacement from the primary $pp$ interaction vertices. 
At least one charged particle must have $\pt > 1.7\gev$ and be inconsistent with originating from any PV.
A multivariate algorithm~\cite{Gligorov:2012qt} is used for the identification of secondary vertices consistent with the decay of a \bquark hadron.

In the offline selection, the objects that fired the trigger are associated with reconstructed particles.  
Selection requirements can therefore be made not only on the software trigger that fired, but also on whether the decision was due to the signal candidate, other particles produced in the $pp$ collision, or a combination of both. Both cases are retained for further analysis.

Simulated samples are used to characterize the detector response to signal and
certain types of backgrounds.
In the simulation, $pp$ collisions are generated using
\pythia~\cite{Sjostrand:2006za,*Sjostrand:2007gs} with a specific \lhcb
configuration~\cite{LHCb-PROC-2010-056}.  Decays of hadronic particles
are described by \evtgen~\cite{Lange:2001uf}, in which final state
radiation is generated using \photos~\cite{Golonka:2005pn}. The
interaction of the generated particles with the detector and its
response are implemented using the \geant
toolkit~\cite{Allison:2006ve, *Agostinelli:2002hh} as described in
Ref.~\cite{LHCb-PROC-2011-006}.

\section{Selection requirements}
\label{sec:selection}

The selection of the \Bm meson is performed using the decay chain
\begin{equation}
pp \to \Bm X, \ \Bm \to \Dstarp \pim \pim, \ \Dstarp \to \Dz \pip, \ \Dz \to \Km \pip,
\label{eq:signal}
\end{equation}
where $X$ represents a system composed of any collection of charged or neutral particles.
After applying selections on the quality of the \Bm candidate tracks,
further requirements are applied on their momenta, $p$, and transverse momenta, $\pt$.
The \Dz meson is reconstructed through its $\Km \pip$ decay, applying loose particle identification criteria on both particles and good vertex quality requirements.
The remaining tracks associated to the \Bm final state form a $\pip \pim \pim$ system which defines the \Bm decay vertex.
Very loose particle-identification criteria are applied to the three pions together with good vertex-quality and impact-parameter constraints.
The invariant mass of the above $\pip \pim \pim$ system is required to be below the physical boundary $m(\pip \pim \pim)<3.6 \gev$. 
In the data collected at $\sqrt{s}=7$ and 8\tev (48.5\% of the total dataset), 
the requirement is $m(\pip \pim \pim)<3.0 \gev$, which also removes 1.2\% of the signal.
Although the loss in the Run 1 data is rather small, it produces a nonnegligible distortion in the \Bm Dalitz plot and in the
$\pim \pim$ invariant-mass distribution.

The momentum scale is calibrated using samples of $\decay{\jpsi}{\mumu}$ 
and $\decay{\Bu}{\jpsi\Kp}$~decays collected concurrently
with the~data sample used for this analysis~\cite{LHCb-PAPER-2012-048,LHCb-PAPER-2013-011}.
The~relative accuracy of this
procedure is estimated to be $3 \times 10^{-4}$ using samples of other
fully reconstructed $\bquark$~hadrons, $\PUpsilon$~and
$\KS$~mesons.

\begin{figure}[b]
\centering
\small
\includegraphics[width=0.49\textwidth]{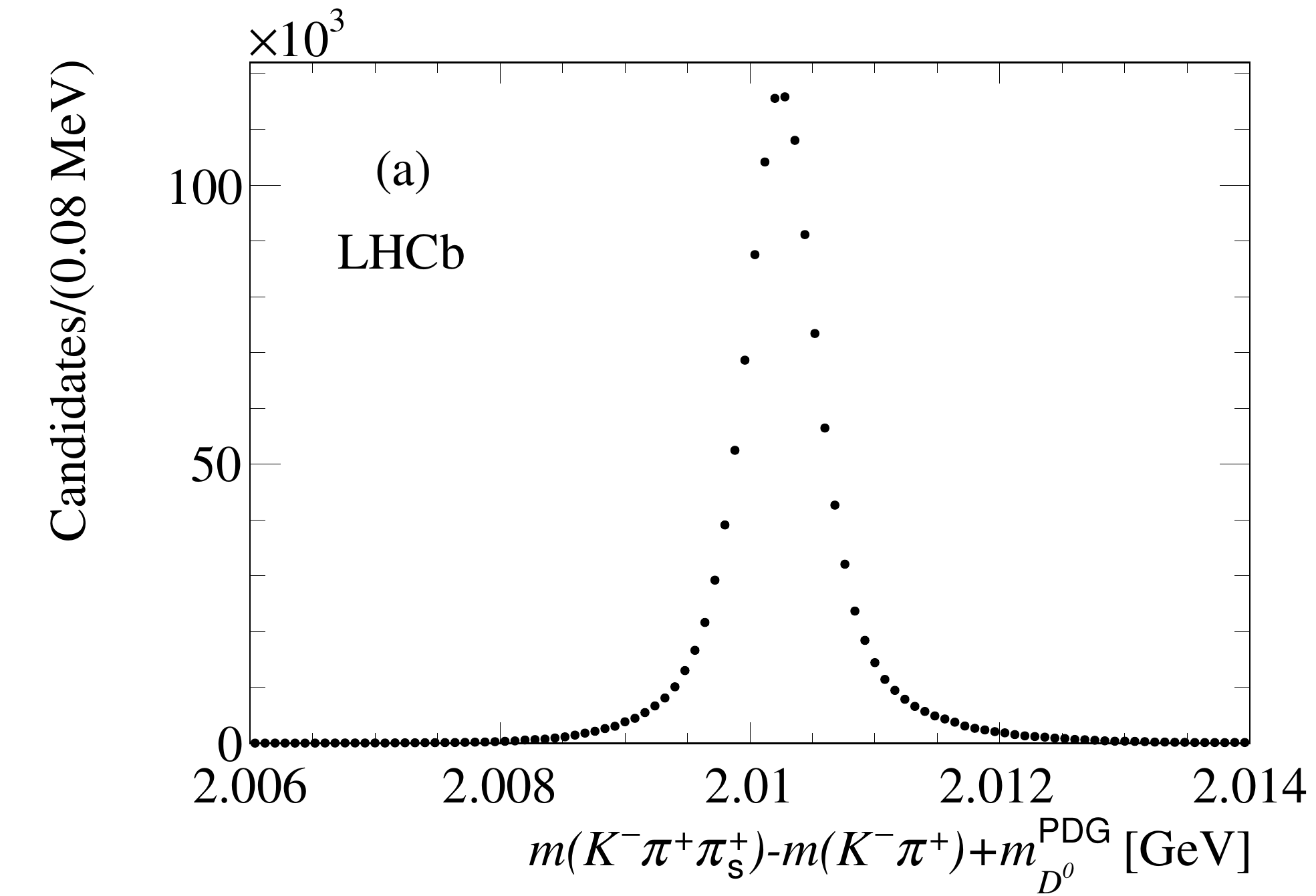}
\includegraphics[width=0.49\textwidth]{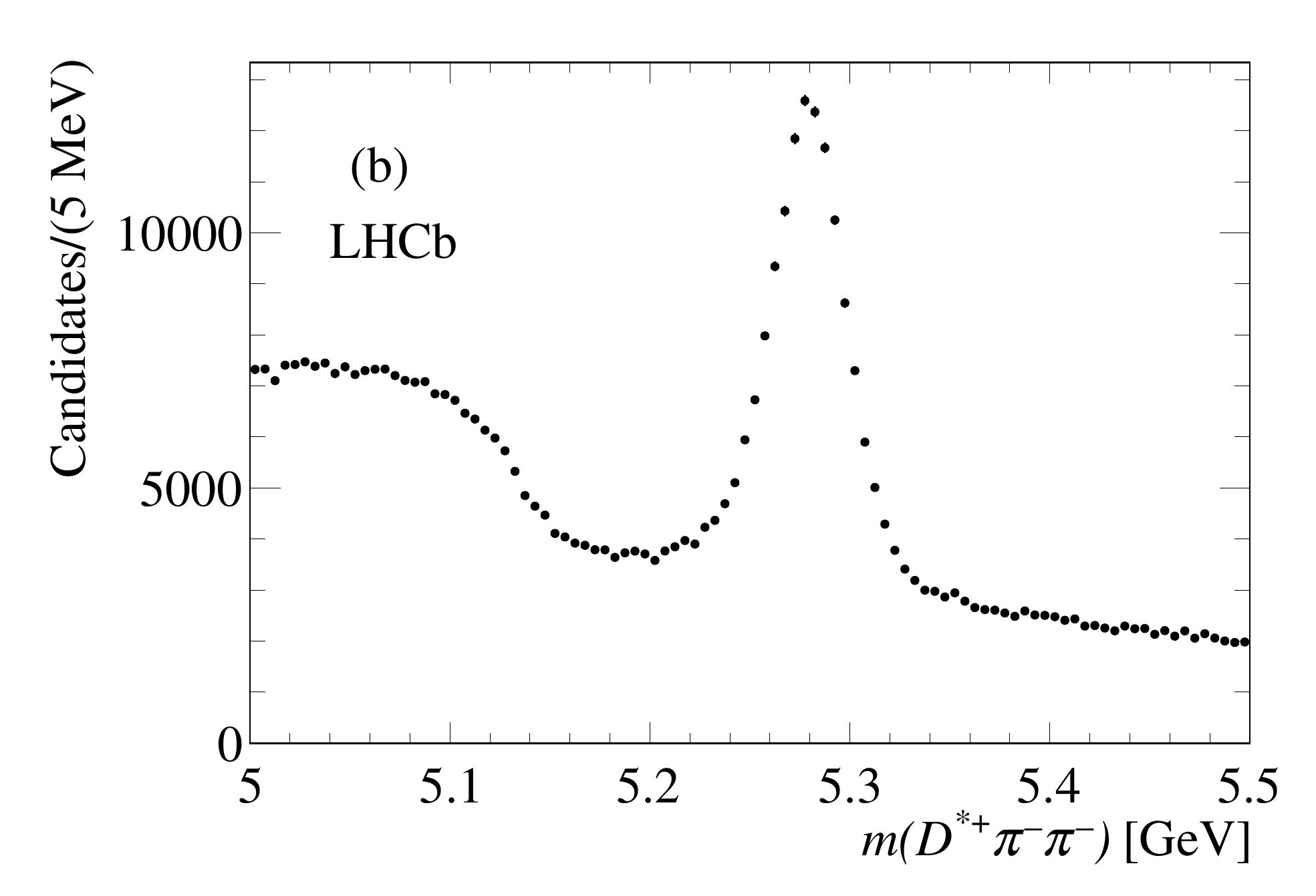}
\caption{\small\label{fig:fig1} Distribution of (a)~$m(\Km \pip \pi^+_{\rm s})-m(\Km \pip)+m_{D^0}^{\rm PDG}$ and (b)~$\Dstarp \pim \pim$ invariant masses for candidates after the selection on the $\chisqndf$ from the fit to the \Bm decay tree.}
\end{figure}

Figure~\ref{fig:fig1}(a) shows the $\Dz \pip$ mass spectrum, computed as $m(\Km \pip \pi^+_{\rm s})-m(\Km \pip)+m_{D^0}^{\rm PDG}$ (here $\pi^+_{\rm s}$ indicates the ``slow pion'' from the \Dstarp decay and $m_{D^0}^{\rm PDG}$ indicates the known \Dz mass value), where a clean
\Dstarp signal can be observed.

The \Dstarp candidate is selected within $3.5\sigma$ of the fitted $D^{*+}$ mass value, where $\sigma=0.45 \mev$ is the effective mass resolution obtained from a fit to the $\Dz \pip$ mass spectrum with a sum of two Gaussian functions. 
The $\Bm \to \Dstarp \pim \pim$ decay is affected by background from $\Bzb \to \Dstarp \pim$ decays combined with a random \pim candidate in the event. This contribution populates the high-mass sideband of the \Bm signal and is removed if either $\Dstarp \pim$ have a mass within $2\sigma$ the \Bzb known value~\cite{Tanabashi:2018oca}, where $\sigma=18.5\mev$ is obtained from a Gaussian fit to the \Bzb mass distribution.
The resulting $\Dstarp \pim \pim$ mass distribution is shown in Fig~\ref{fig:fig1}(b), where the \Bm signal can be observed over a significant background.

A significant source of background is due to $B^- \to D^{*+} \pim \pim \piz$ or $\Bzb \to D^{*+} \pim \pim \pip$ decays, where a pion is not reconstructed.
However the $B^- \to D^{*+} \pim \pim$ mass combination from these final states populate the low-mass sideband of the \Bm signal and do not extend into
the signal
region. A coherent source of background which could affect the signal region is due to $\Bzb \to D^{*+} \pim \piz$ decays, where the \piz meson is not reconstructed but is replaced by a random \pim candidate from the event. In this case the $\Dstarp \pim$ system could have
a definite spin-parity configuration, however this contribution is found to be negligible. A possible source of background comes from the $\Bm \to \Dstarp \Km \pim$ decay,
where the \Km candidate is misidentified as a pion. 
This background is kinematically
confined in the lower sideband of the \Bm signal. Its contribution 
relative to $B^- \to D^{*+} \pim \pim$ has been measured in Ref.~\cite{Aaij:2017dtg} and found to be negligible. Therefore the background under the \Bm
signal is dominated by combinatorial background.

To reduce the combinatorial background while keeping enough signal for an amplitude analysis, a multivariate selection is employed,
in the form of a likelihood ratio defined, for each event, as
\begin{equation}
\calR = \sum_{i=1}^{6} {\rm log}(P_{\rm s}(i)/P_{\rm b}(i)),
\label{eq:like}
\end{equation}
where $i$ runs over a set of 6 variables and $P_{\rm s}(i)$ and $P_{\rm b}(i)$ are probability density functions (PDFs) of the signal and background contributions, respectively.
The signal PDFs are obtained from simulated signal samples, while background PDFs are obtained from the \Bm sideband regions, defined within $4.5-6.6\sigma$ on either side of the \Bm  mass peak, where
$\sigma$ is obtained from the fit to the $\Dstarp \pim \pim$ mass spectrum defined below.
The variables used are: the \Bm decay length significance, defined as the ratio between the decay length and its uncertainty;
the \Bm transverse momentum; the $\chi^2$ of the primary vertex associated to the \Bm meson; the \Bm and \Dstarp impact parameters with respect to the primary vertex; and the $\chisqndf$ from the fit to the $B^- \to (\Dz(\to K^- \pi^+) \pi^+_{\rm s}) \pi^- \pi^-$ decay tree.

\begin{figure}[tb]
\centering
\small
\includegraphics[width=0.75\textwidth]{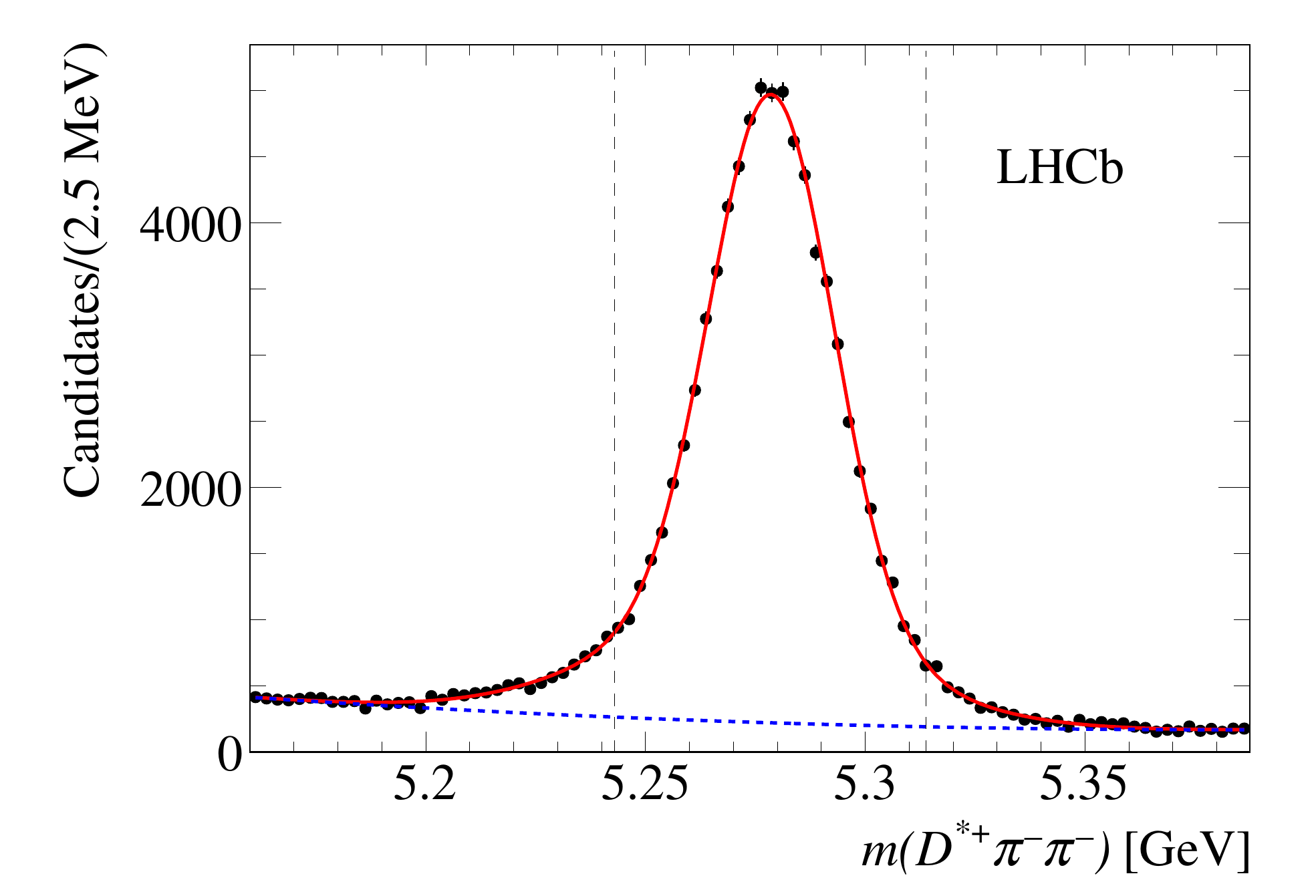}
\caption{\small\label{fig:fig2} Mass distribution for $\Dstarp \pim \pim$ candidates after the selection $\calR>0.5$. The full (red) line  is the result from the fit while the dotted (blue) line describes the background. The vertical dashed lines indicate the signal region.}
\end{figure}

The choice of the selection value on the variable \calR\ is performed using an optimization procedure where the $\Dstarp \pim \pim$ mass spectrum of candidates selected with increasing cut on \calR\ is fitted. The fits are performed using two Gaussian functions with a common mean to describe the \Bm signal and a quadratic
function for the background. Defining $\sigma$ as the weighted mean of the widths of the two Gaussian functions, the signal region is defined within $2\sigma$, where $\sigma=17.7\mev$.
For each selection the fit estimates the signal and background yields, $N_{\rm sig}$ and $N_{\rm bkg}$. From these quantities the purity $p=\frac{N_{\rm sig}}{N_{\rm sig}+N_{\rm bkg}}$ and the significance $s=\frac{N_{\rm sig}}{\sqrt{N_{\rm sig}+N_{\rm bkg}}}$ are evaluated. To obtain both the largest purity and significance, the figure of merit $s \cdot p$ is evaluated. It has its maximum at $\calR>0.5$ which is taken as the default selection. For this selection, Fig.~\ref{fig:fig2} shows the resulting $\Dstarp \pim \pim$ mass spectrum where the \Bm signal is observed over small background.

For the above selection the signal purity is $p=0.9$, while the efficiency is 81.9\%. The yield within the signal region is 79\,120, of which 48.5\% and 51.5\% are from Run 1 and Run 2, respectively.
The purities of the two data sets are found to be the same.
The number of events with multiple \Bm candidate combinations is negligible. 

\section{\boldmath The $\Bm \to \Dstarp \pim \pim$ Dalitz plot}
\label{sec:data}

\begin{figure}[bt]
\centering
\small
\includegraphics[width=0.75\textwidth]{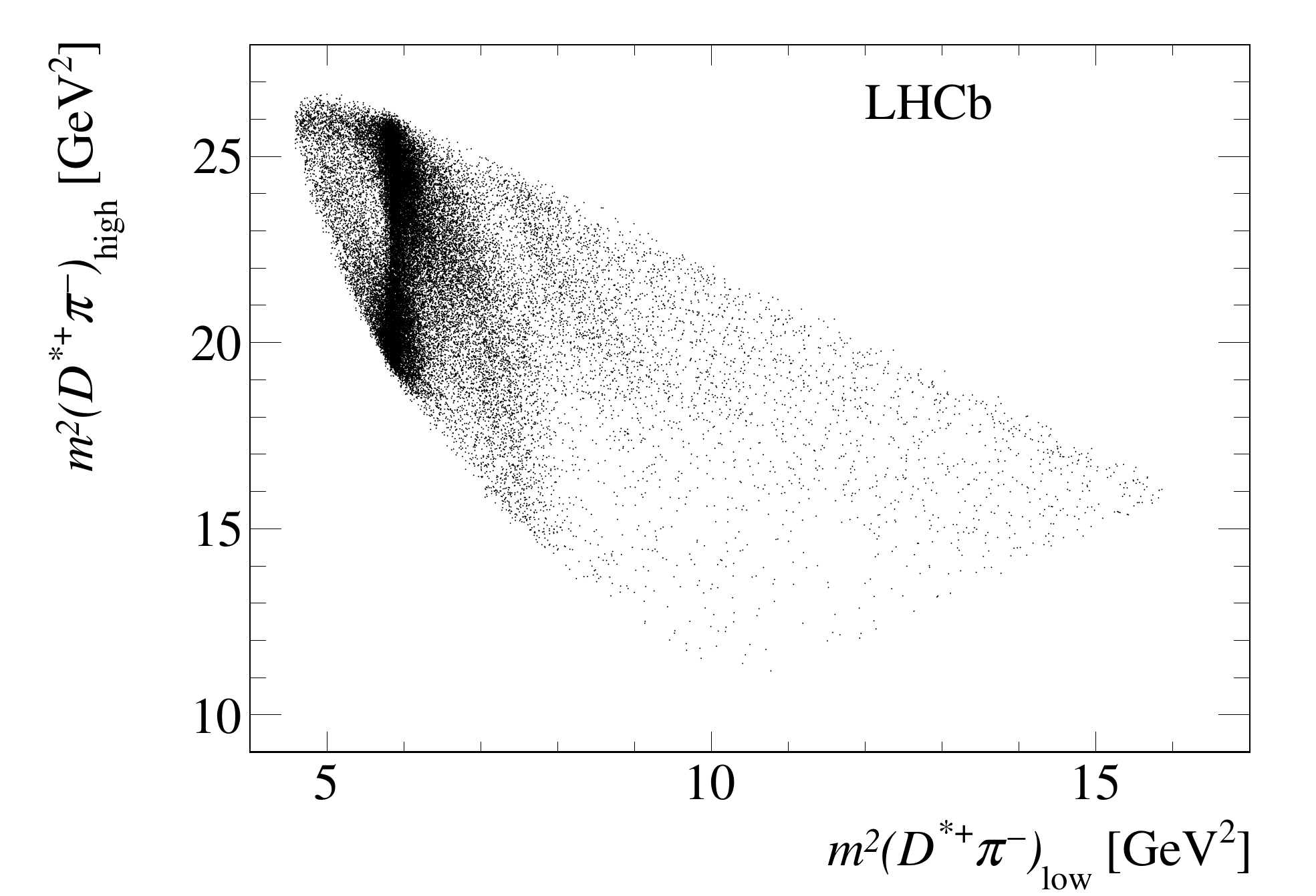}
\caption{\small\label{fig:fig3} Dalitz plot distribution for $\Bm \to \Dstarp \pim \pim$ candidates in Run 2 data.}
\end{figure}

The $\Bm \to \Dstarp \pim \pim$ decay mode contains two indistinguishable \pim mesons, giving two $\Dstarp \pim$ mass
combinations. In the following, $m(\Dstarp \pim)_{\rm low}$ and $m(\Dstarp \pim)_{\rm high}$ indicate the lower and higher values of the two mass combinations, respectively. The \Bm Dalitz plot, described as a function of $m^2(\Dstarp \pim)_{\rm high}$ and $m^2(\Dstarp \pim)_{\rm low}$, is shown in Fig.~\ref{fig:fig3} for Run 2 data.  

\begin{figure}[bt]
\centering
\small
\includegraphics[width=0.75\textwidth]{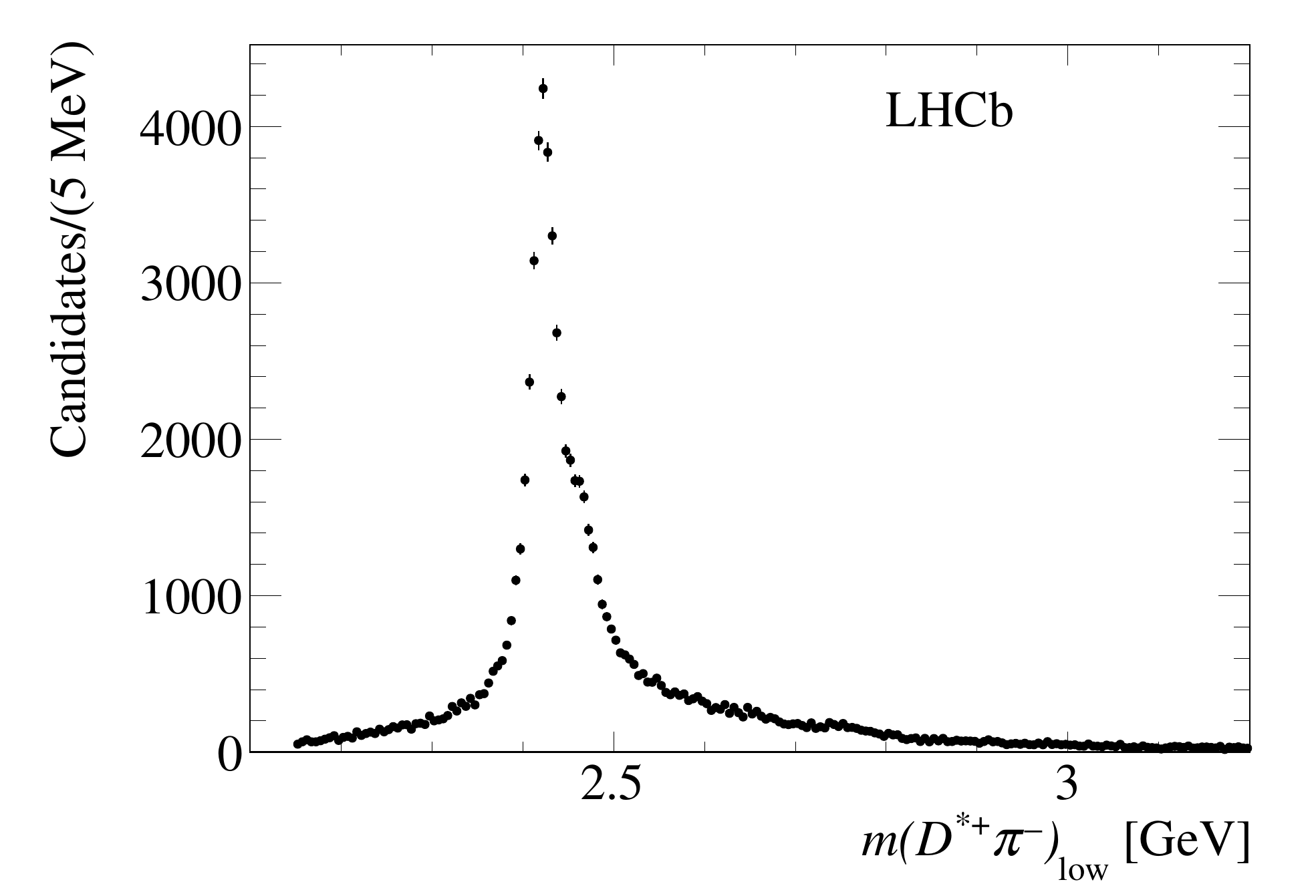}
\caption{\small\label{fig:fig4} Distribution of 
$m(\Dstarp \pim)_{\rm low}$ for the total dataset.}
\end{figure}

The Dalitz plot contains clear vertical bands in the $6\gev^2$ mass region, due to the presence
of the well-known \Done and \Dstartwo resonances. The presence of further weaker bands can be observed
in the higher mass region. The prominent presence of the above two resonances can be observed in the $m(\Dstarp \pim)_{\rm low}$ projection, shown in Fig.~\ref{fig:fig4} for the total dataset.
On the other hand, the presence of additional states is rather weak in the mass projection and therefore an angular analysis is needed to separate the different contributions.

\begin{figure}[tb]
\centering
\small
\includegraphics[width=0.32\textwidth]{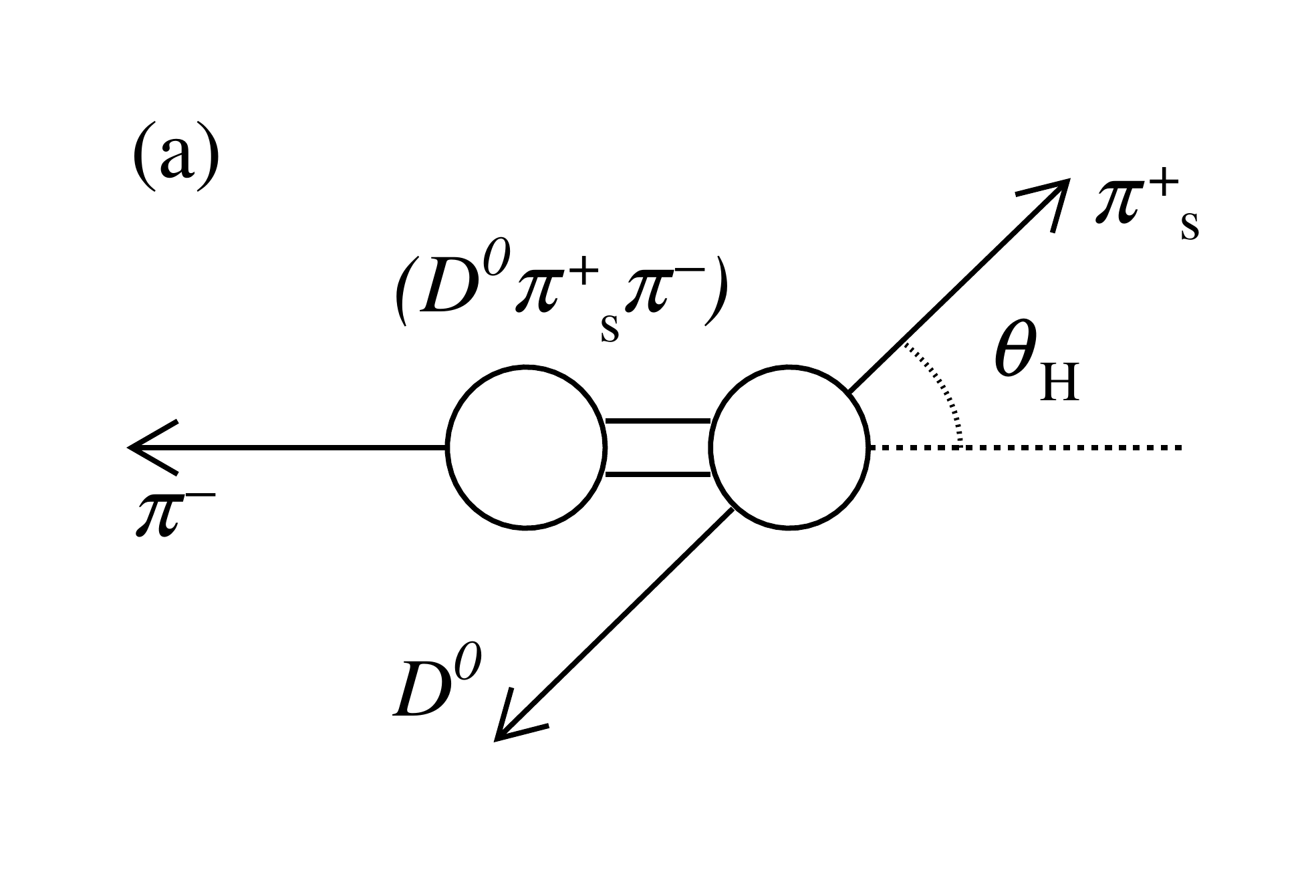}
\includegraphics[width=0.32\textwidth]{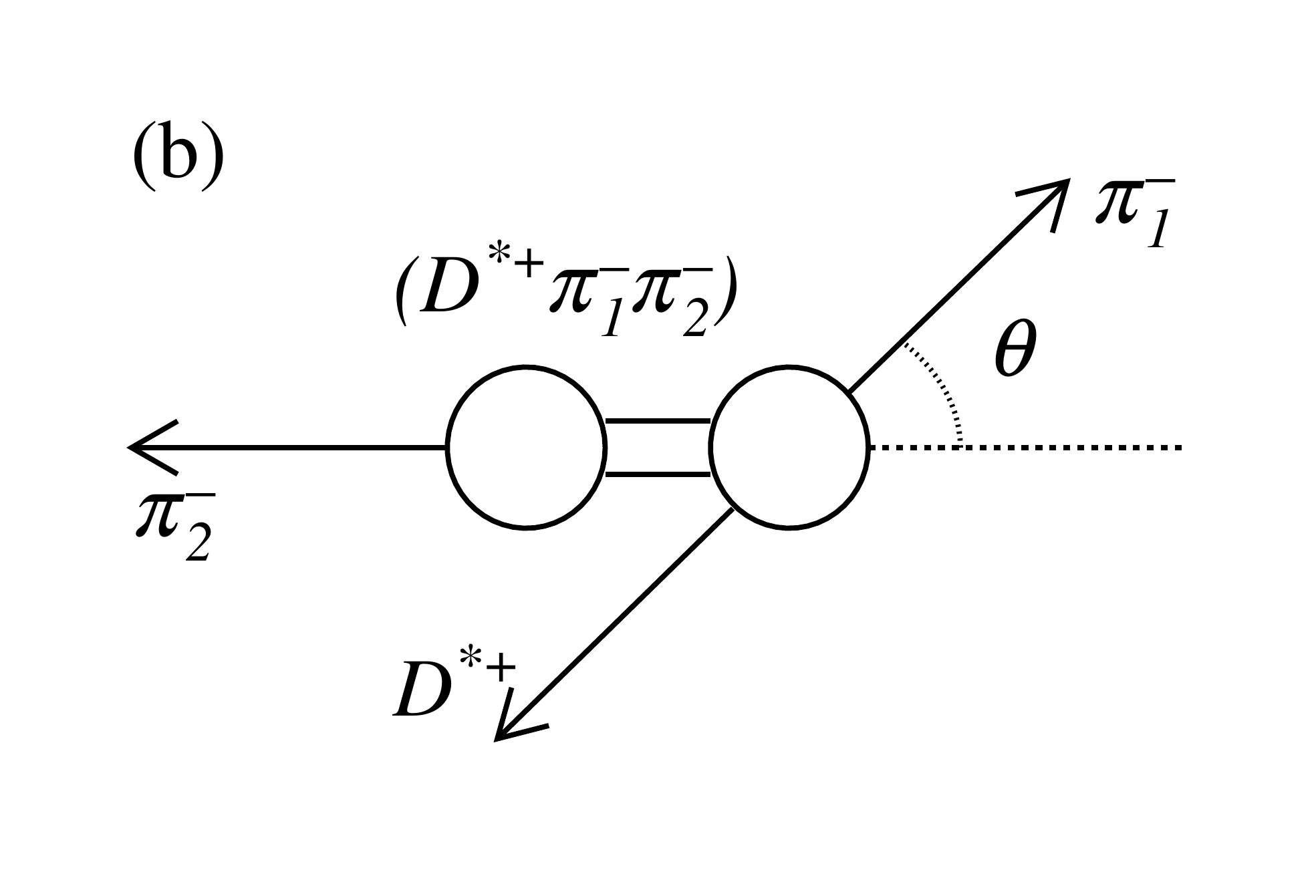}
\includegraphics[width=0.32\textwidth]{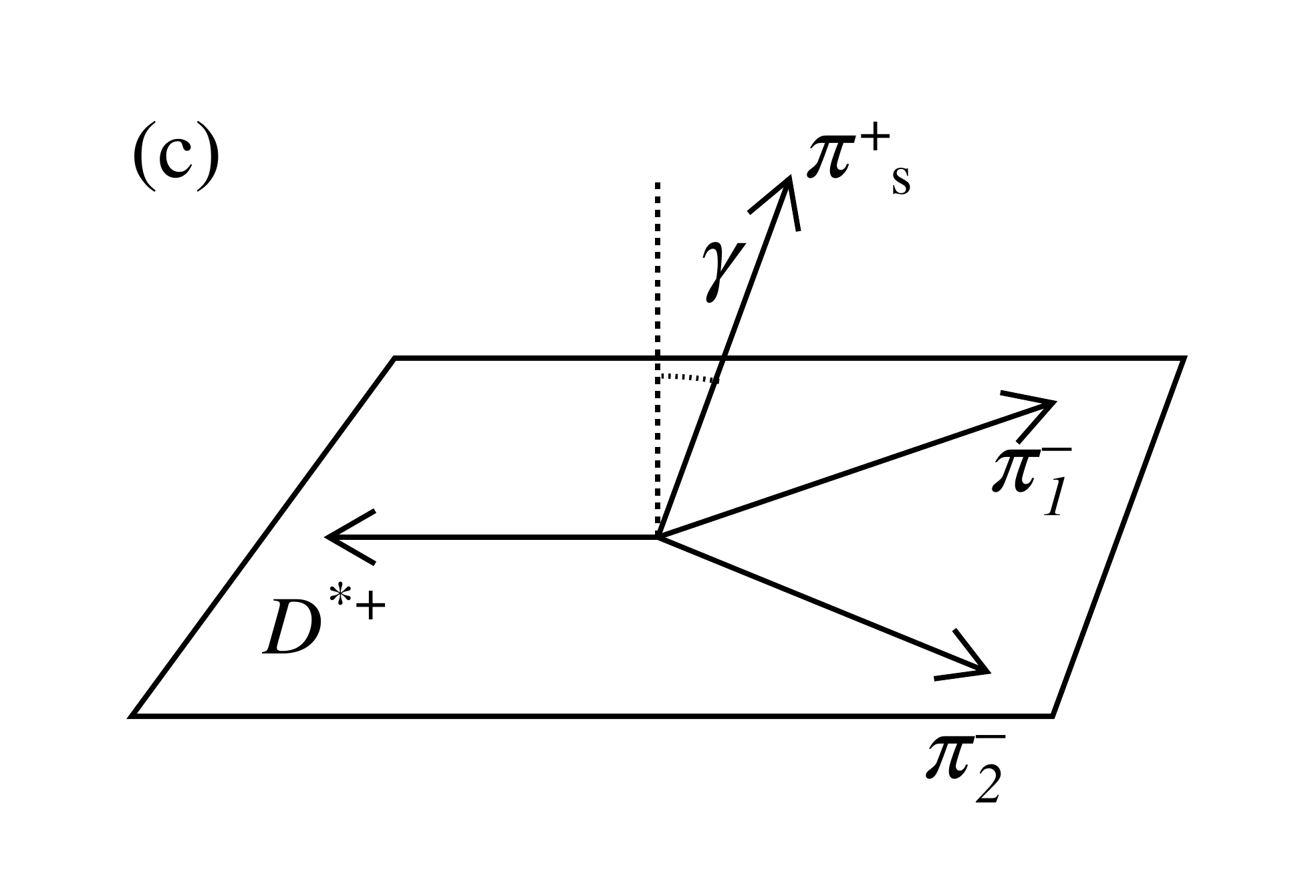}
\caption{\small\label{fig:fig5} Definition of the angles (a) $\theta_H$, (b) $\theta$ and (c) $\gamma$ .}
\end{figure}

The following angles are useful in discriminating between different $J^P$ contributions:
\begin{description}
\item[$\theta_H$,] the helicity angle defined as the angle between the $\pi^+_{\rm s}$ direction
in the $\Dz \pi^+_{\rm s}$ rest frame and the $\Dz \pi^+_{\rm s}$ direction in the $\Dz \pi^+_{\rm s} \pim$ rest frame (see Fig.~\ref{fig:fig5}(a));

\item[$\theta$,] the helicity angle defined as the angle formed by the $\pi^-_1$ direction in the $\Dstarp \pi^-_1$ rest frame
and the $\Dstarp \pi^-_1$ direction in the $\Dstarp \pi^-_1 \pi^-_2$ rest frame (see Fig.~\ref{fig:fig5}(b));

\item[$\gamma$,] the angle in the $\Dstarp \pim \pim$ rest frame 
formed by the $\pi^+_{\rm s}$ direction in the $\Dz \pi^+_{\rm s}$ rest frame and the normal to the
$\Dstarp \pim \pim$ plane (see Fig.~\ref{fig:fig5}(c)).
\end{description}

\begin{figure}[tb]
\centering
\small
\includegraphics[width=0.85\textwidth]{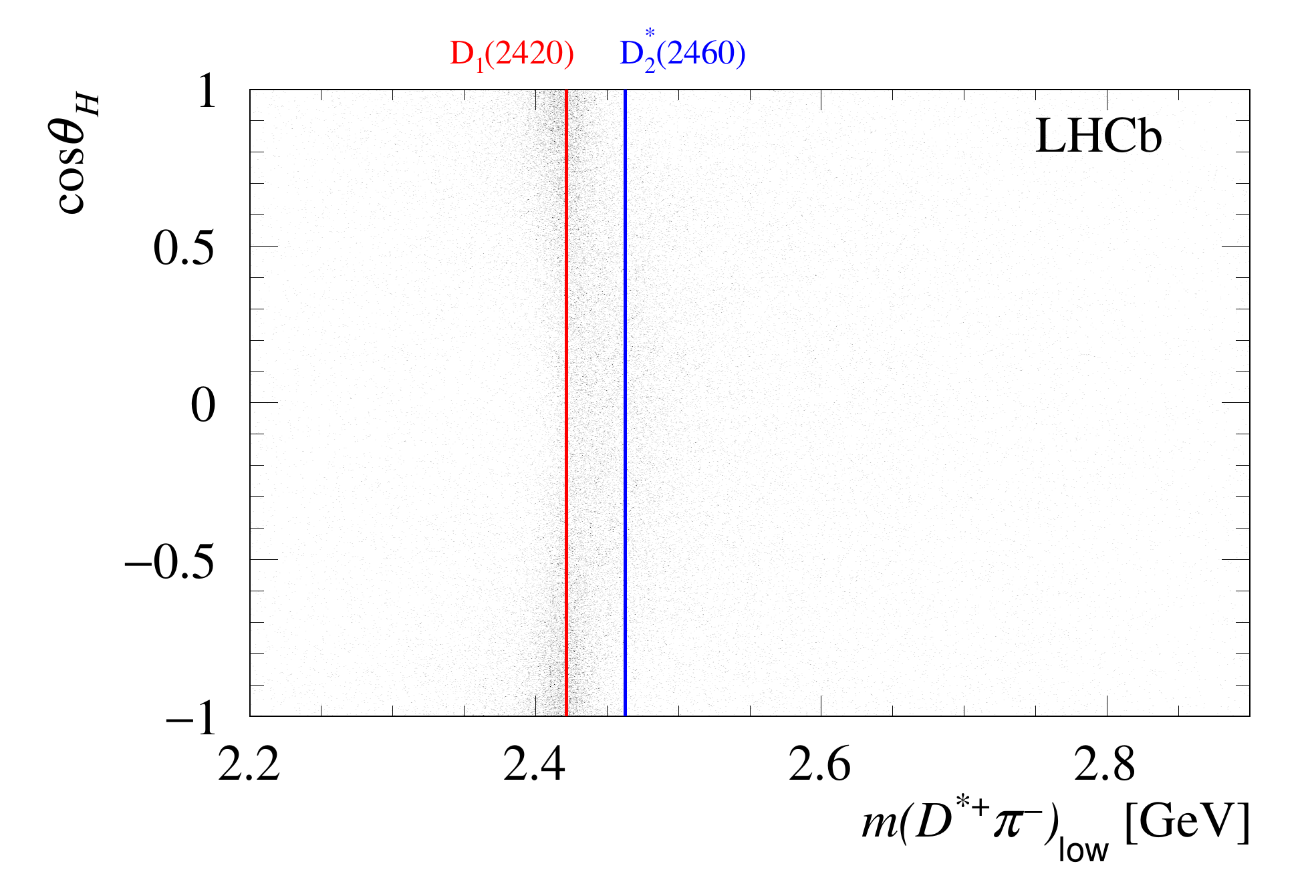}
\caption{\small\label{fig:fig6} Two-dimensional distribution of $\cos \theta_H$ vs. $m(\Dstarp \pim)_{\rm low}$. The vertical lines indicate the positions of the \Done and \Dstartwo resonances.}
\end{figure}

The angle $\theta_H$ is useful to discriminate between natural and
unnatural spin-parity resonances for which the expected angular distributions are $\sin^2 \theta_H$ and
$1 - h \cos^2 \theta_H$ (where $h<1$ depends on the properties of the decay), respectively, except for $J^P=0^-$ where a $\cos^2\theta_H$ distribution is expected.
Figure~\ref{fig:fig6} shows the two-dimensional distribution of $\cos \theta_H$ vs. $m(\Dstarp \pim)_{\rm low}$. The two vertical bands are due to the \Done and \Dstartwo states which exhibit the expected $1 - h \cos^2 \theta_H$ and $\sin^2 \theta_H$ distributions, respectively. 

\section{Background and efficiency}
\label{sec:back}

\subsection{Background model}

\begin{figure}[tb]
\centering
\small
\includegraphics[width=0.49\textwidth]{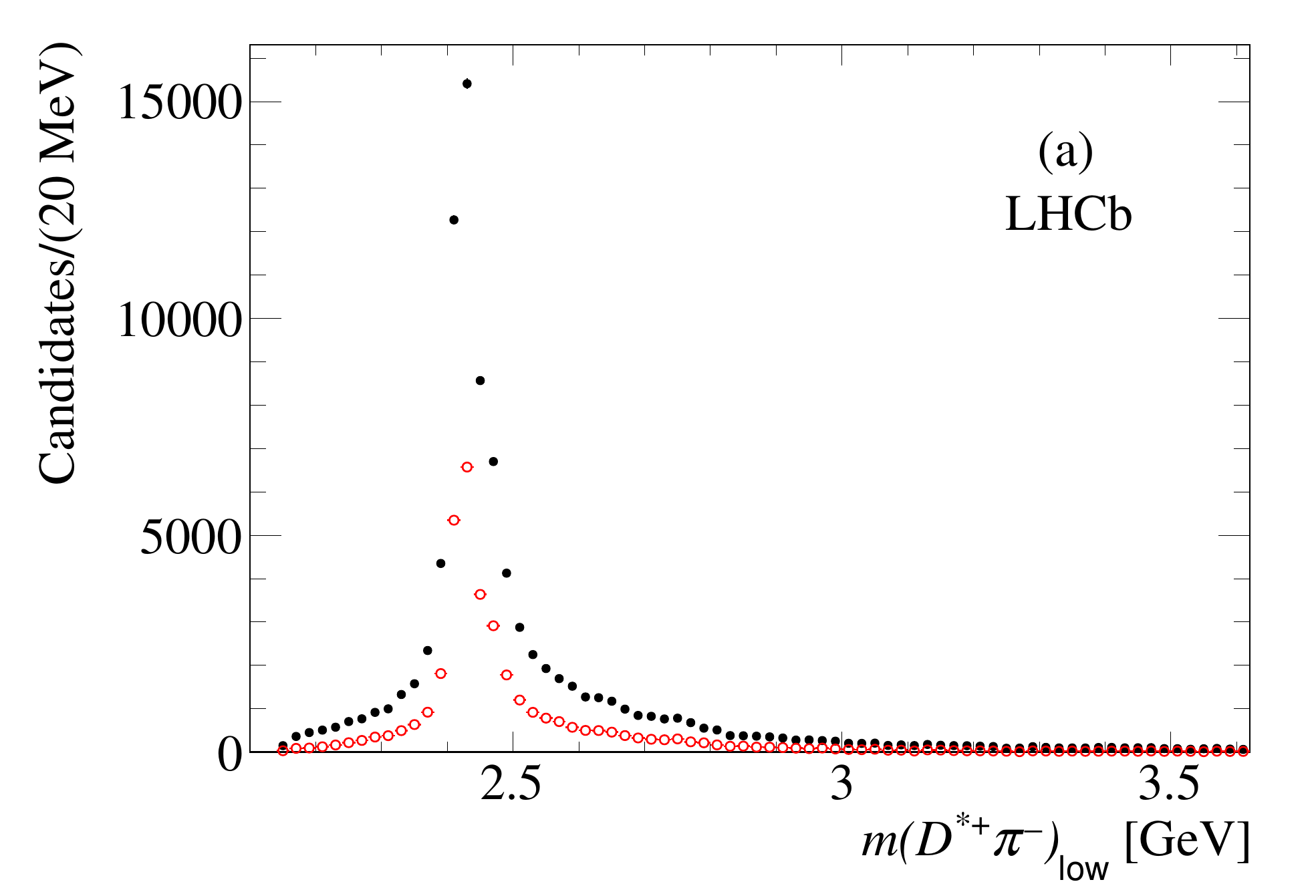}
\includegraphics[width=0.49\textwidth]{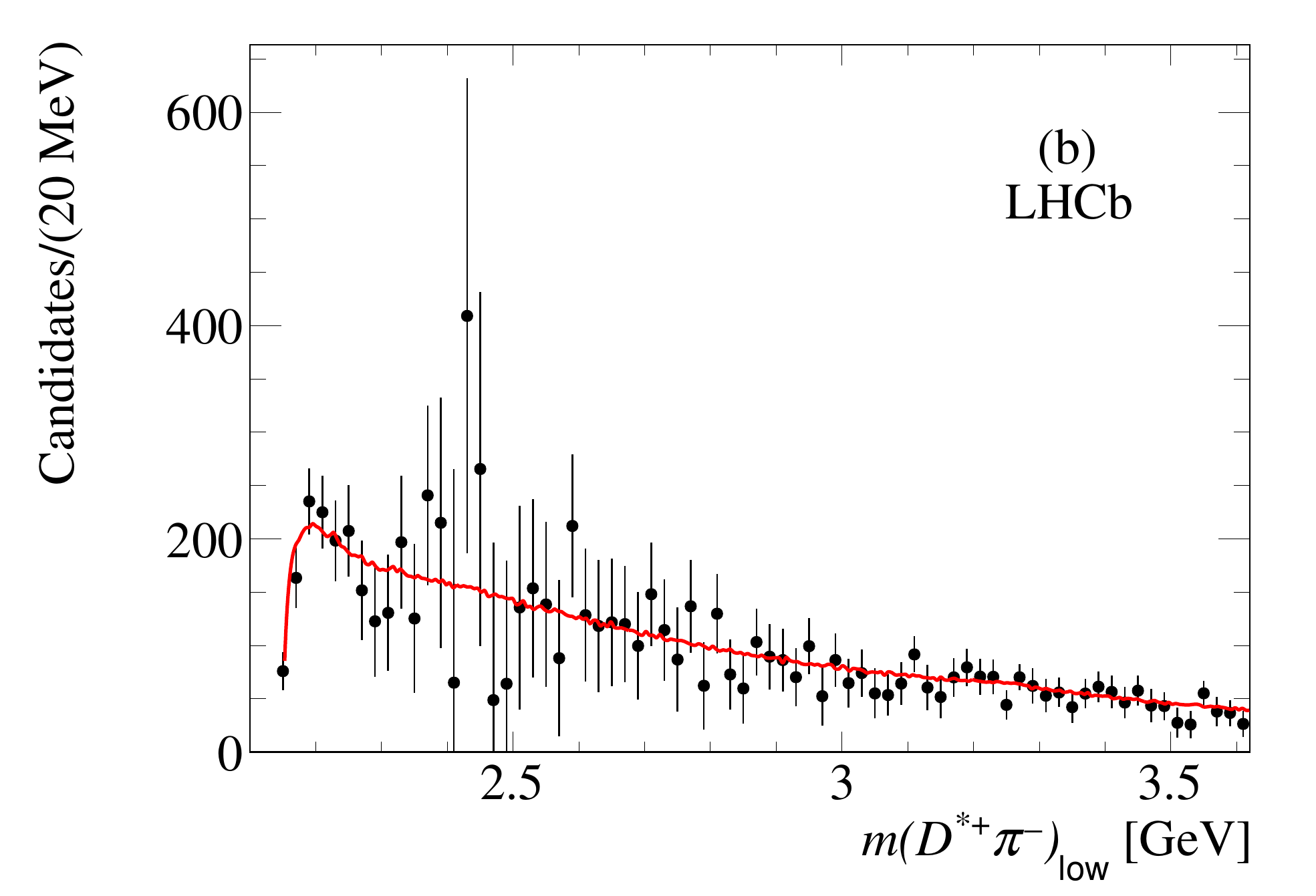}
\includegraphics[width=0.49\textwidth]{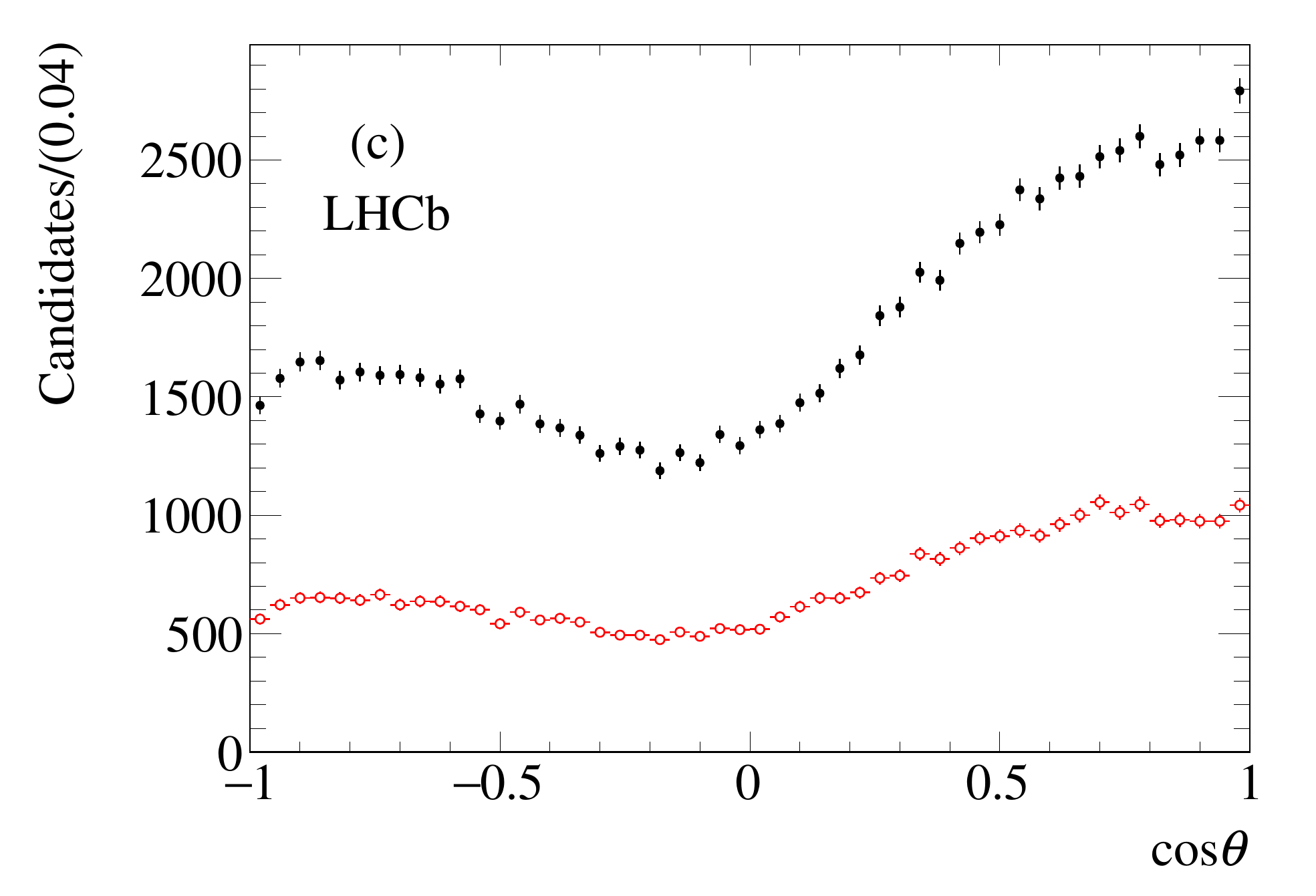}
\includegraphics[width=0.49\textwidth]{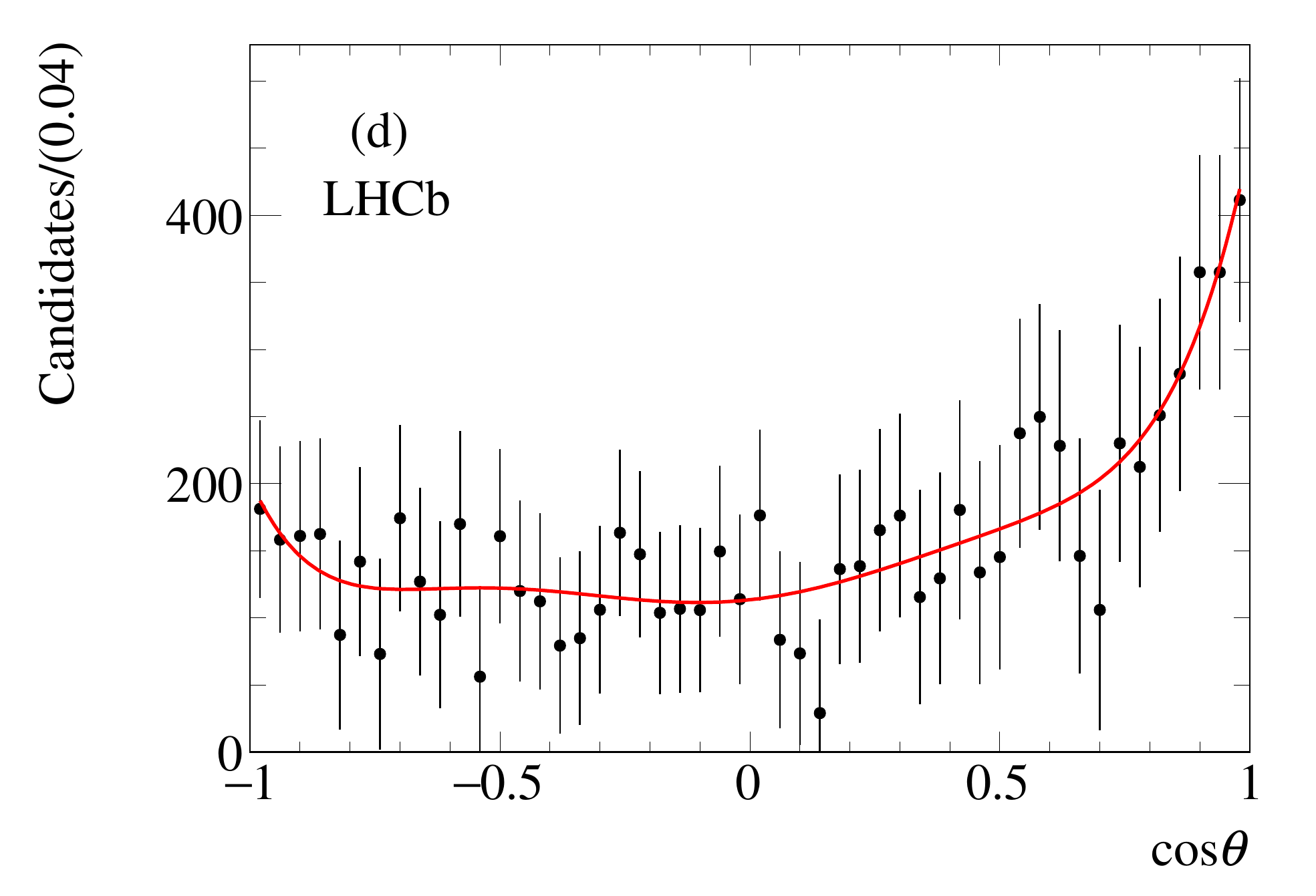}
\caption{\small\label{fig:fig7}
Distribution for low/high-purity selection (filled black/open red points) for (a)~$m(\Dstarp \pim)_{\rm low}$ and (c)~$\cos \theta$.
Signal-subtracted distributions for (b)~$m(\Dstarp \pim)_{\rm low}$ and (d)~$\cos \theta$ with superimposed fit curves described in the text.}
\end{figure}

The background model is obtained from the data in the signal region using the method of signal subtraction.
Using the \calR\ variable defined in Eq.~\eqref{eq:like}, two datasets are extracted, (a) with low purity ($\calR > 0.0$, $p_a=0.865$ and signal yield $N_a = 77\,644$), (b) with high purity ($\calR > 2.5$, $p_b= 0.949$ and signal yield $N_b = 34\,019$). The background distribution for a given variable is then obtained by subtracting the high-purity distributions, scaled by the factor $N_a/N_b$, from the low-purity distributions.
The variables $m(\Dstarp \pim)_{\rm low}$ and $\cos \theta$ are found to be independent and different for signal and background, therefore the resulting background model is obtained by the product of the PDFs of these distributions.
Figure~\ref{fig:fig7}(a) shows the $m(\Dstarp \pim)_{\rm low}$ distribution for the low-purity (filled points) and high-purity (open points) selections. Figure~\ref{fig:fig7}(b) shows the signal-subtracted distribution, where no significant resonant structure is seen. The superimposed
curve is the result from a fit performed using the sum of two exponential functions multiplied by the \Bm phase-space factor.
Similarly, Fig.~\ref{fig:fig7}(c) shows the $\cos \theta$ distribution for the low-purity and high-purity samples, and
Fig.~\ref{fig:fig7}(d) shows the signal-subtracted distribution. The curve is the result of a fit using a $6^{\rm th}$ order polynomial function.

\subsection{Efficiency}
\label{sec:effy}

The efficiency is computed using simulated samples of signal decays analyzed using the same procedure as for data. Due to the different trigger and reconstruction methods, the efficiency is computed separately for Run 1  and Run 2 data.
It is found that the efficiency mainly depends on the variables, $m(\Dstarp \pim)$ and $\cos \theta$. Weak or no dependence is found on other variables.
The efficiency model is obtained by dividing the simulated sample into 22 slices in ${\rm log}(m(\Dstarp \pim)/\mev)$ and fitting the $\cos \theta$ distribution for each slice
using $5^{\rm th}$ order polynomial functions. The efficiency for a given value of $\log(m(\Dstarp \pim)/\mev)$ is then obtained by linear interpolation between two adjacent bins.

\begin{figure}[tb]
\centering
\small
\includegraphics[width=0.49\textwidth]{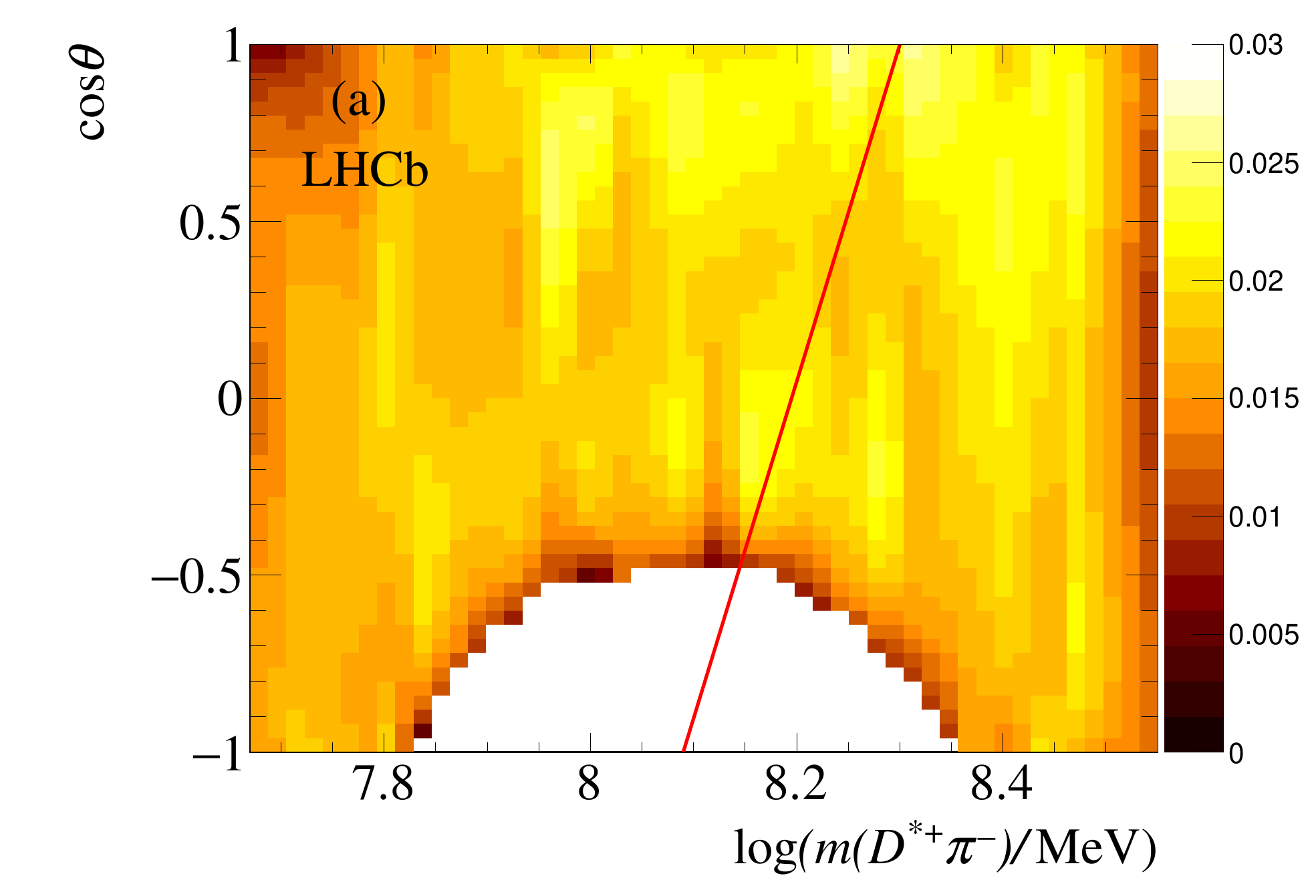}
\includegraphics[width=0.49\textwidth]{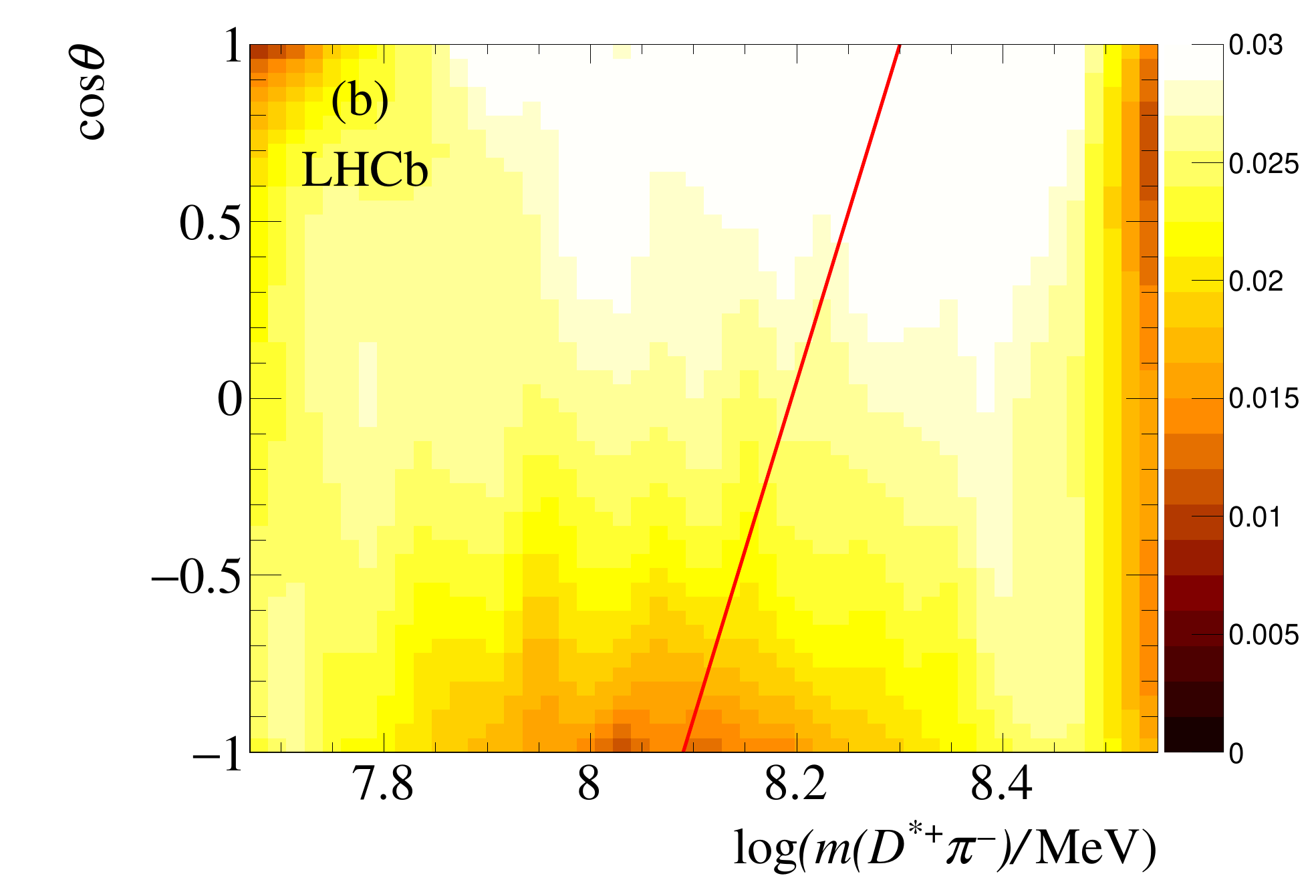}
\caption{\small\label{fig:fig8}
Interpolated efficiency as a function of $\log(m(\Dstarp \pim)/\mev)$ and $\cos \theta$ for (a)~Run~1 and (b)~Run~2 data.
The line indicates the regions where $\log(m(\Dstarp \pim)/\mev)$ is taken as
$\log(m(\Dstarp \pim)_{\rm low}/\mev)$ (left side of the line) and as $\log(m(\Dstarp \pim)_{\rm high}/\mev)$ (right side of the
line).  
}
\end{figure}

Figure~\ref{fig:fig8} shows the interpolated efficiency maps in the $(\log(m(\Dstarp \pim)/\mev), \cos \theta)$ plane, separately for Run 1 and Run 2. The empty region in Run 1 data is caused by the requirement $m(\pip \pim \pim)<3.0$ GeV. 
Although this region is populated by a small fraction of signal, estimated using Run 2 data,
this introduces some uncertainty in the description of the Run 1 data.

The mass resolution is studied as function of $m(\Dstarp \pim)$ using simulation.
For each slice in $m(\Dstarp \pim)$ the difference between the generated and reconstructed mass is computed and the resulting distributions are fitted using the sum of two Gaussian functions. The effective resolution $\sigma$, increases almost linearly from $\sigma=4\mev$ at $2.4\gev$ to $\sigma=7\mev$ at $3.0\gev$. This value of the mass resolution is much smaller than the minimum width of the known resonances present in the $\Dstarp \pim$ mass spectrum (for \Done, $\Gamma=31\mev$), therefore, in the following, resolution effects are ignored.

\section{Amplitude analysis}
\label{sec:daly}

An amplitude analysis of the four-body decay
\begin{equation}
\Bm \to (\Dz \pip)_{D^{*+}} \pi^-_1 \pi^-_2,
\label{eq:bm}
\end{equation}
\noindent
where $\pi^-_1$ and $\pi_2$ indicate the two indistinguishable pions, is performed to extract the fractions and the phases of the charmed resonances contributing to the decay and to measure
their parameters and quantum numbers. 
All the amplitudes are symmetrized according to the exchange of the $\pi^-_1$ with $\pi^-_2$ mesons.

\subsection{Description of the amplitudes}

The amplitudes contributing to the decay are parameterized using the nonrelativistic Zemach tensors formalism~\cite{Zemach:1963bc,Dionisi:1980hi,Filippini:1995yc}.
It is assumed that reaction~(\ref{eq:bm}) proceeds as

\begin{equation}
B^- \to R^0 \pi^-_2
\label{eq:brdec}
\end{equation}
\noindent
where $R^0$ is an intermediate charmed meson resonance which decays as

\begin{equation}
R^0 \to \Dstarp (\to \Dz \pi^+_{\rm s}) \pi^-_1. 
\label{eq:rdec}
\end{equation}

Reaction (\ref{eq:brdec}) is a weak decay and does not conserve parity while
reaction (\ref{eq:rdec}) is a strong decay and conserves both angular
momentum and parity.
The four particles in the final state are labeled as
\begin{equation}
 \Dz(1), \pi^+_{\rm s}(2), \pi^-_1(3), \pi^-_2(4).
\end{equation}
In the description of the decay $R^0 \to \Dz \pi^+_{\rm s} \pi^-_1$, the 3-vectors $\boldsymbol{p_i}$ ($i$=1,2,3) indicate the momenta of the particles  in the $\Dz \pi^+_{\rm s} \pi^-_1$ rest frame and $L$ indicates the angular momentum between the  $\Dz \pi^+_{\rm s}$ system and the $\pi^-_1$ meson.
For the resonance $\Dstarp$ decaying as $\Dstarp  \to \Dz \pi^+_{\rm s}$, the decay products $\Dz$ and $\pi^+_{\rm s}$ having 3-momenta $\boldsymbol{p_1}$, $\boldsymbol{p_2}$ and masses $m_1$ and $m_2$, the 3-vector $\boldsymbol{t_3}$ is defined as
\begin{equation}
\boldsymbol{t_3} = \boldsymbol{p_1} - \boldsymbol{p_2} - (\boldsymbol{p_1} + \boldsymbol{p_2})\frac{m_1^2-m_2^2}{m_{12}^2},
\end{equation}
with $m_{12}$ indicating the $\Dz \pi^+_{\rm s}$ invariant mass.
To describe the decay $B^- \to R^0 \pi^-_2$, the 3-vector
$\boldsymbol{q_4}$ indicates the momentum of $\pi^-_2$ in the $B^-$ rest frame.

The amplitudes are obtained as follows:
\begin{itemize}
\item{}
a symmetric and traceless tensor of rank $L$, $\boldsymbol{P}$, constructed with $\boldsymbol{p_3}$ is used to describe the orbital angular momentum $L$;
\item{}
a symmetric and traceless tensor of rank $S$, $\boldsymbol{T}$, constructed with  $\boldsymbol{t_3}$ is used to describe the spin of intermediate resonances;
\item{}
  the tensors $\boldsymbol{P}$ and $\boldsymbol{T}$ are then combined into a tensor $\boldsymbol{J}$ of rank $J$ to obtain the total spin $J$ of the $\Dz \pip \pi^-_1$ system;
\item{}
  a symmetric and traceless tensor, $\boldsymbol{Q}$, of rank $J$ constructed with ${\boldsymbol{q_4}}$ is used to describe the orbital angular momentum between $R^0$ and $\pi^-_2$;
\item{}
  the scalar product of the two tensors $\boldsymbol{J}$ and $\boldsymbol{Q}$ gives the scalar which represents the 0 spin of the \Bm meson.
\end{itemize}

The resonance $R^0$, having a given $J^P$,
decays into a $1^-$ resonance ($D^{*+}$) and a $0^-$ particle with a given orbital angular momentum $L$.
In a first approach, the resonance lineshape is described by a complex relativistic Breit--Wigner function, ${\rm BW}(m_{123})$, with appropriate Blatt--Weisskopf centrifugal barrier factors~\cite{BW,Aaij:2016fma} which are computed assuming a radius $4.5\, {\rm GeV}^{-1}$. In a second approach the resonance lineshapes are described by the quasi-model-independent method (QMI)~\cite{Aitala:2005yh,Lees:2015zzr}, described later.

The list of the amplitudes used in the present analysis is given in Table~\ref{tab:tab1}.
The $J^P=0^-$ nonresonant contribution term is omitted because it is found to be negligible.
\begin{table}
  \centering
  \caption{List of the amplitudes used in the present analysis. }
{\small
\begin{tabular}{lcl}
  \hline
   \noalign{\vskip4pt}
   $J^P$  & $L$ & Amplitude \cr
      \noalign{\vskip4pt}
      \hline
         \noalign{\vskip4pt}
         $0^-$ & 1 & ${\rm BW}(m_{123})[\boldsymbol{t_3 \cdot p_3}]$ \cr
         \noalign{\vskip4pt}   
         \hline
           \noalign{\vskip4pt} 
           $1^+S$ & 0 & ${\rm BW}(m_{123})[\boldsymbol{t_3} \cdot \boldsymbol{q_4}]$ \cr
              \noalign{\vskip4pt}
              \hline
            \noalign{\vskip4pt}     
            $1^+D$ & 2 & ${\rm BW}(m_{123})[\boldsymbol{p_3}(\boldsymbol{t_3} \cdot \boldsymbol{p_3}) - \frac{1}{3}(\boldsymbol{p_3} \cdot \boldsymbol{p_3})\boldsymbol{t_3}]\cdot \boldsymbol{q_4}$ \cr
           \noalign{\vskip4pt}    
           \hline
       \noalign{\vskip4pt}       
       $1^-$ & 1 & ${\rm BW}(m_{123})[(\boldsymbol{t_3} \times \boldsymbol{p_3}) \cdot \boldsymbol{q_4}]$ \cr
        \noalign{\vskip4pt}  
        \hline
          \noalign{\vskip4pt} 
$2^-P$ & 1 &  ${\rm BW}(m_{123})[\frac{1}{2}(t_3^i p_3^j + t_3^jp_3^i)-\frac{1}{3}(\boldsymbol{t_3}\cdot
            \boldsymbol{p_3}) \delta^{ij}]\cdot[q^i_4q^j_4 - \frac{1}{3}|q_4|^2\delta^{ij}]$\cr
             \noalign{\vskip4pt}
             \hline
              \noalign{\vskip4pt}  
$2^-F$ & 3 &  ${\rm BW}(m_{123})[(\boldsymbol{t_3} \cdot \boldsymbol{p_3}) (p^i_3p^j_3 
            - \frac{1}{3}|p_3|^2\delta^{ij})]
                \cdot[q^i_4q^j_4 - \frac{1}{3}|q_4|^2\delta^{ij}]$\cr
          \noalign{\vskip4pt}         
          \hline
         \noalign{\vskip4pt}    
$2^+$ & 2 &  ${\rm BW}(m_{123})[\frac{1}{2}[(\boldsymbol{t_3}\times \boldsymbol{p_3})^ip^j_3+
            p^i_3(\boldsymbol{t_3}\times \boldsymbol{p_3})^j] - \frac{1}{3}[{\bf
               (t_3} \times \boldsymbol{p_3}) \cdot \boldsymbol{p_3}]\delta^{ij}\cdot[q^i_4q^j_4 - \frac{1}{3}|q_4|^2\delta^{ij}]$\cr
              \noalign{\vskip4pt}
\hline
$3^-$ & 3 & ${\rm BW}(m_{123})[(\boldsymbol{t_3} \times \boldsymbol{p_3})^ip^j_3p^k_3 + (\boldsymbol{t_3} \times
    \boldsymbol{p_3})^kp^i_3p^j_3 + (\boldsymbol{t_3} \times \boldsymbol{p_3})^jp^i_3p^k_3)$ - \cr
    & & $\frac{1}{5}[\delta^{ij}(\boldsymbol{t_3} \times
            \boldsymbol{p_3})^k + \delta^{ik}(\boldsymbol{t_3} \times \boldsymbol{p_3})^j + \delta^{jk}(\boldsymbol{t_3}
            \times \boldsymbol{p_3})^i]|p_3|^2]  \cdot $ \cr
   & &  $\frac{1}{5} [(p_4^ip_4^jp_4^k + p_4^kp_4^ip_4^j  + p_4^jp_4^ip_4^k) -
            p_4^2(\delta^{ij}p_4^k + \delta^{ik}p_4^j +
            \delta^{jk}p_4^i)]$ \cr
    \noalign{\vskip4pt} 
\hline
\end{tabular}
}
\label{tab:tab1}
\end{table}
 
\subsection{Amplitude analysis fit}

The amplitude analysis of the $\Dstarp \pim \pim$ candidates in the \Bm mass region is performed using unbinned maximum-likelihood fits.
The likelihood function is written as
\begin{equation}
\mathcal{L} = \prod_{n=1}^N \bigg[p \cdot \epsilon(\boldsymbol{x_n})\frac{\sum_{i,j} c_i c_j^* A_i(\boldsymbol{x_n}) A_j^*(\boldsymbol{x_n})}{\sum_{i,j} c_i c_j^* I_{A_i A_j^*}}+(1-p)B(\boldsymbol{z_n})\bigg]
\end{equation}
\noindent where
\begin{itemize}
\item $N$ is the number of candidates in the signal region;
\item for the $n^{\rm th}$ event, $\boldsymbol{x_n}$ is the set of variables describing the 4-body $B$ meson decay;
\item $\epsilon(\boldsymbol{x_n})$ is the efficiency function;
\item $A_i(\boldsymbol{x_n})$ represents the $i^{\rm th}$ complex signal-amplitude contribution;
\item $c_i$ is the complex intensity of the $i^{\rm th}$ signal component; the $c_i$ parameters are allowed to vary during the fit process;
\item $I_{A_i A_j^*}=\int A_i (\boldsymbol{x})A_j^*(\boldsymbol{x}) \epsilon(\boldsymbol{x}) {\rm d}\boldsymbol{x}$ are normalization
 integrals; numerical integration is performed on phase-space-generated decays with the \Bm signal lineshape generated according to the experimental distribution;
 \item{} $p$ is the signal purity obtained from the fit to the $\Dstarp \pim \pim$ mass spectrum;
 \item{} $B(\boldsymbol{z_n})$ is the normalized background contribution, parameterized as a function of the two variables described in Sec.~\ref{sec:effy}.
\end{itemize}

The efficiency-corrected fraction $f_i$ due to a resonant or nonresonant contribution $i$ is defined as follows
\begin{equation}
f_i \equiv \frac {|c_i|^2 \int |A_i(\boldsymbol{x})|^2 {\rm d}\boldsymbol{x}}{\sum_{i,j} c_i c_j^* I_{A_i A_j^*}}.
    \label{eq:frac}
\end{equation}
 The $f_i$ fractions do not necessarily sum to 1 because of interference effects. The uncertainty for each $f_i$ fraction is evaluated by propagating the full covariance matrix obtained from the fit.
Similarly, the efficiency-corrected interference fractional contribution $f_{ij}$, for $i<j$ is defined as
\begin{equation}
f_{ij} \equiv \frac{\int 2 \Real[c_ic_j^* A_i(\boldsymbol{x}) A^*_j(\boldsymbol{x})] {\rm d}\boldsymbol{x}}{\sum_{i,j} c_i c_j^* I_{A_i A_j^*}}.
    \label{eq:frac_int}
\end{equation}

The amplitude analysis is started by including, one by one, all the possible charmed resonance contributions with masses and widths listed in Ref.~\cite{Tanabashi:2018oca}.
Resonances are kept if a significant likelihood increase ($\Delta(2\log\mathcal{L})>3$) is observed. The list of the states giving significant contributions at the end of the process 
is given in the upper section of Table~\ref{tab:tab2}.
The fit procedure is tested on pseudo-experiments using different combinations of amplitudes, input fractions and phases, obtaining a good
agreement between generated and fitted values.

The quality of the description of the data is tested by the \chisqndf, 
defined as the sum of two \chisq values, calculated from the two two-dimensional distributions $(m', \cos \theta_H)$ and $(\cos \theta, \cos \gamma)$ as
\begin{equation}
\chi^2=\sum_i(N^{\rm model}_i-N^{\rm data}_i)^2/N^{\rm model}_i.
\label{eq:chi}
\end{equation}
Here $N^{\rm model}_i$ and $N^{\rm data}_i$ are the fit predictions and observed yields in each bin of the two-dimensional distributions. The variable ${\rm ndf}$ is defined as ${\rm ndf}=N_{\rm cells}-N_{\rm par}$, where $N_{\rm cells}$ is the number of bins having at least 6 entries and $N_{\rm par}$ is the number of free parameters in the fit.
The variable $m'$, defined in the range 0--1, is computed as 
\begin{equation}
  m' = \frac{1}{\pi}{\rm arccos}\bigg(2\frac{m(\pim \pim) - m^{\min}_{\pim \pim}}{m^{\max}_{\pim \pim}-m^{\min}_{\pim \pim}}-1\bigg)
    \end{equation}
where $m^{\max}_{\pim \pim} = m_{B^-} - m_{\Dstarp}$ and $m^{\min}_{\pim \pim}=2m_{\pim}$. 

\section{Fits to the data using quasi-model-independent amplitudes}
\label{sec:qmi}

It is found in several analyses that the mass terms of some amplitudes may not be well described by Breit--Wigner functions, because they are broad or because additional
contributions may be present at higher mass. Therefore, for a given value of $J^P$, a quasi-model-independent method is tested to describe the amplitude, while leaving all the other resonances described by Breit--Wigner functions.
The method is also used to perform a scan of the mass spectrum to search for additional resonances.

The $\Dstarp \pim$ mass spectrum is divided into 31 slices with nonuniform bin widths and, for a given contributing resonance, the complex Breit--Wigner term is replaced by a set of 31 complex coefficients (magnitude anf phase) which are free to float. These values are 
fixed to arbitrary values in one bin, at a mass value in the $2.42-2.60\gev$ range, depending on the amplitude and therefore the set of additional free parameters is reduced to 60.

The largest amplitude, usually the $1^+D$ amplitude, is taken as the reference wave. 
Due to the large number of fit parameters, QMI amplitudes can only be introduced one by one. 
The fit is performed using as free parameters the real and imaginary parts of the amplitude in each bin of the $\Dstarp \pim$ mass spectrum.
The search for the QMI parameters is performed using a random search, starting from zero in each mass bin.
The fitted solution is then given as input to a second iteration modifying the value for the fixed bin to the average value obtained from the two adjacent bins.
Obvious spikes are smoothed in the input of the second iteration.
Normally the second iteration converges and is able to compute the full covariance matrix.
The fitted QMI amplitude is then modeled through a cubic-spline interpolation function.

The method is tested using different initial values for the first iteration. In all cases the fit converges to the same solution.
It is also tested with simulation obtaining good agreement between input and fitted values of the amplitudes.

The process starts with a QMI fit to the $J^P=1^+S$ amplitude, including all the amplitudes listed in the upper part of Table~\ref{tab:tab2} and described by Breit--Wigner functions with
initial parameter values fixed to those reported in Ref.~\cite{Tanabashi:2018oca}.
In this fit, due to significant interference effects
between the $1^+D$, the narrow $1^+S$ and the broad $1^+S$ amplitudes, the narrow $1^+D$/$1^+S$ \Done parameters, described by a Breit--Wigner function,  are left free as well. The resulting parameters for \Done resonance are given in Table~\ref{tab:tab2}.
Statistical significances are computed as the fitted fraction
divided by its statistical uncertainty. 

The QMI $J^P=1^+S$ amplitude is then fixed and a QMI analysis of the $J^P=0^-$ is performed.
The process continues by fixing the $J^P=0^-$ QMI amplitude and leaving, one by one, free Breit--Wigner parameters for all the resonances listed in the upper part of Table~\ref{tab:tab2}.
The parameters of the \Dstartwo resonances are fixed to the world averages because they are well determined.
The process is iterative, with QMI analyses of the $J^P=1^+S$ and $J^P=0^-$ amplitudes and free parameters for the resonances described by
Breit--Wigner functions repeated several times, until the process converges and no significant variation
of the free parameters is observed.
The resulting fitted parameters of \Donem, \Dtwom, and \Dthree amplitudes are listed in Table~\ref{tab:tab2}.
To obtain the parameters of the broad \Donew resonance, a fit is performed with the QMI model for the  $J^P=1^+S$ amplitude replaced by the  Breit--Wigner function model.
Similarly, to obtain the \Dzero parameters, the QMI model for the  $J^P=0^-$ amplitude is replaced by the Breit--Wigner function.
The presence of a broad $J^P=1^+D$ \Donew contribution has been tested but excluded from the final fit. Its effect, due to the presence the broad $J^P=1^+S$ resonance, is to produce large interference effects so that the total fraction increases to large and rather unphysical values without significantly improving the fit quality. 

\begin{table}[t]
 \caption{\small Resonance parameters from the amplitude analysis. The first uncertainty is statistical, the second systematic.
 The upper part reports the resonance parameters obtained from the amplitude analysis described in Sec.~\ref{sec:qmi}, the lower part those obtained from
 the mixing analysis described in Sec.~\ref{sec:mix}. The labels indicating the spin-parity of \Dzero, \Donem, and \Dtwom resonances are updated, with respect to those reported in Ref.~\cite{Tanabashi:2018oca}, according to the results from the amplitude analysis reported in this work. The \Dstartwo parameters are fixed to the world averages.}
 \label{tab:tab2}
 \begin{center}
\begin{tabular}{c | c | r@{}c@{}l | r@{}c@{}l | c}
Resonance & $J^P$ & \multicolumn{3}{c|}{Mass [MeV]} & \multicolumn{3}{c|}{Width [MeV]} & Significance ($\sigma$)  \cr
\hline
\Done   & $1^+$ & 2424.8 $\pm$ & \, 0.1 \, & $\pm$ 0.7 & 33.6 $\pm$ & \, 0.3 \, & $\pm$ 2.7  &  \cr
\Donew   & $1^+$ & 2411 $\pm$ & \, 3 \, & $\pm$ 9  & 309  $\pm$ & \, 9   \, & $\pm$  28  &  \cr
\Dstartwo & $2^+$ & 2460.56 $\pm$ & \, 0.35 \, &  & 47.5 $\pm$ & \, 1.1 \, & & \cr
\Dzero   & $0^-$ & 2518   $\pm$ & \, 2   \, & $\pm$ 7  & 199  $\pm$ & \, 5   \, & $\pm$  17  & 53 \cr
\Donem & $1^-$ & 2641.9 $\pm$ & \, 1.8 \, & $\pm$ 4.5 & 149  $\pm$ & \, 4   \, & $\pm$  20  & 60 \cr
\Dtwom   & $2^-$ & 2751   $\pm$ & \, 3   \, & $\pm$ 7  & 102  $\pm$ & \, 6   \, & $\pm$  26  & 16 \cr
\Dthree & $3^-$ & 2753   $\pm$ & \, 4   \, & $\pm$ 6   & 66   $\pm$ & \, 10  \, & $\pm$  14  & 8.7 \cr
\hline
$D_1$   & $1^+$      & 2423.7 $\pm$ & \, 0.1 \, & $\pm$ 0.8 & 31.5 $\pm$ & \, 0.1 \, & $\pm$ 2.1  &  \cr
$D_1'$   & $1^+$     &  2452  $\pm$ & \, 4   \, & $\pm$ 15  & 444  $\pm$ & \, 11  \, & $\pm$ 36   & \cr
\hline
\end{tabular}
\end{center}
\end{table}

\begin{figure}[tb]
\centering
\small
\includegraphics[width=0.45\textwidth]{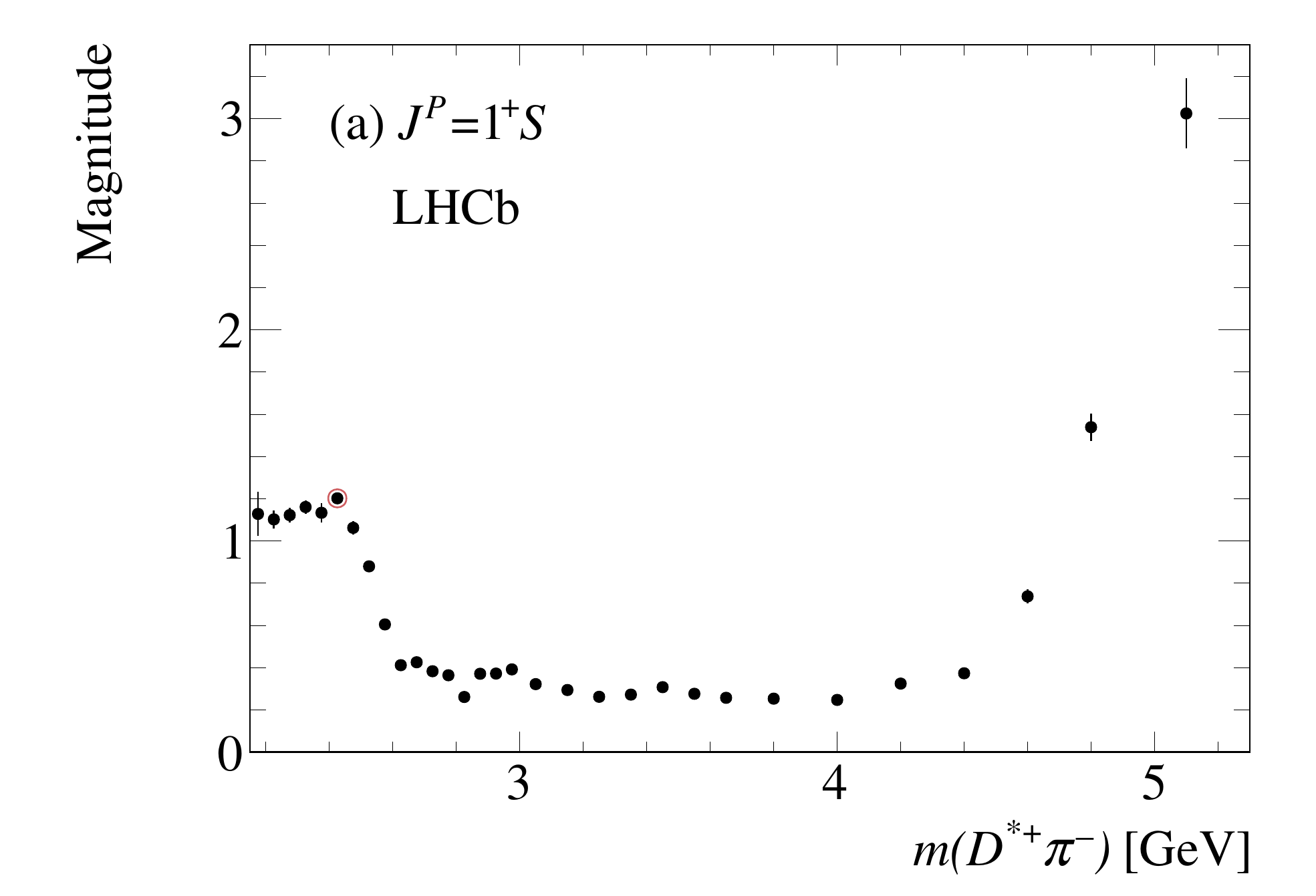}
\includegraphics[width=0.45\textwidth]{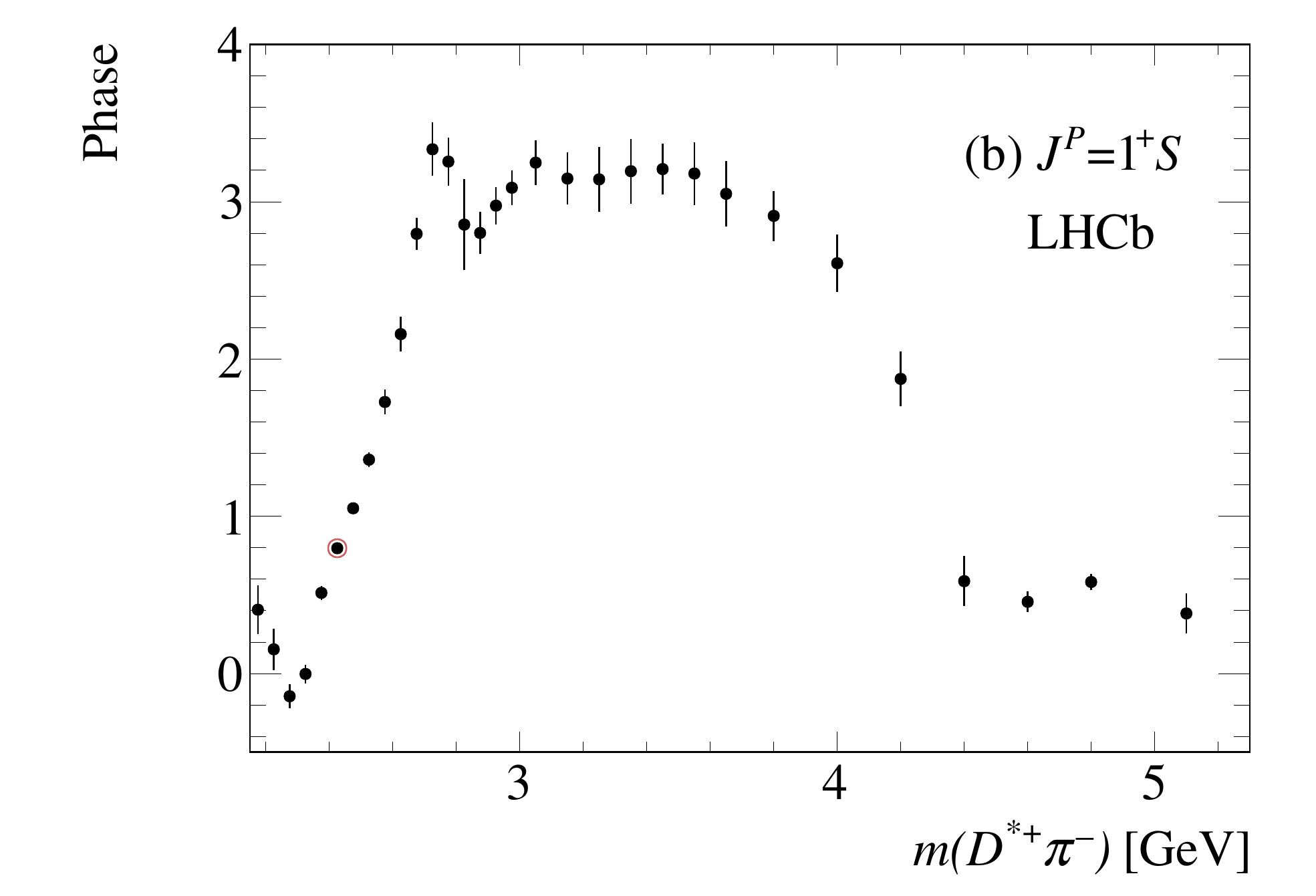}
\caption{\small\label{fig:fig9} (a)~Magnitude and (b)~phase of the $1^+S$ amplitude from the QMI method. The red circles indicate the mass bin where the
complex amplitude has been fixed.}
\end{figure}

Figure~\ref{fig:fig9} shows the fitted magnitude and phase of the $1^+S$ amplitude.
The presence of a broad structure can be noted close to threshold with a corresponding phase motion as expected for a resonance. The magnitude and phase
show further activity in the $2.8\gev$ mass region, suggesting the presence of an additional $1^+S$ resonance.
However, the introduction of a new $1^+S$ Breit-Wigner resonance with floating parameters in that mass region does not produce a significant contribution. 
The high-mass enhancement in the amplitude, on the
other hand, is due to symmetrization effects due to the presence of two identical pions. 

\begin{figure}[t]
\centering
\small
\includegraphics[width=0.49\textwidth]{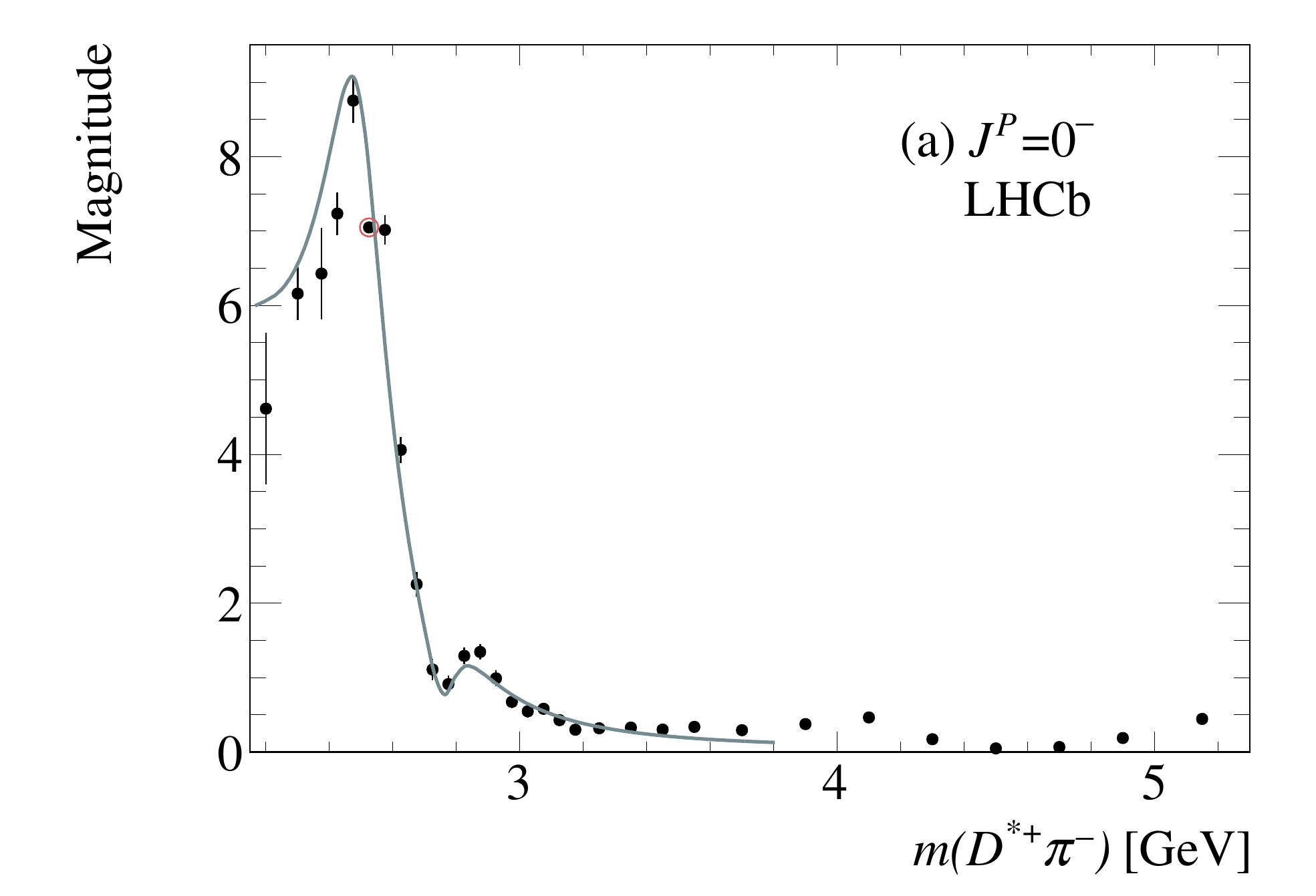}
\includegraphics[width=0.49\textwidth]{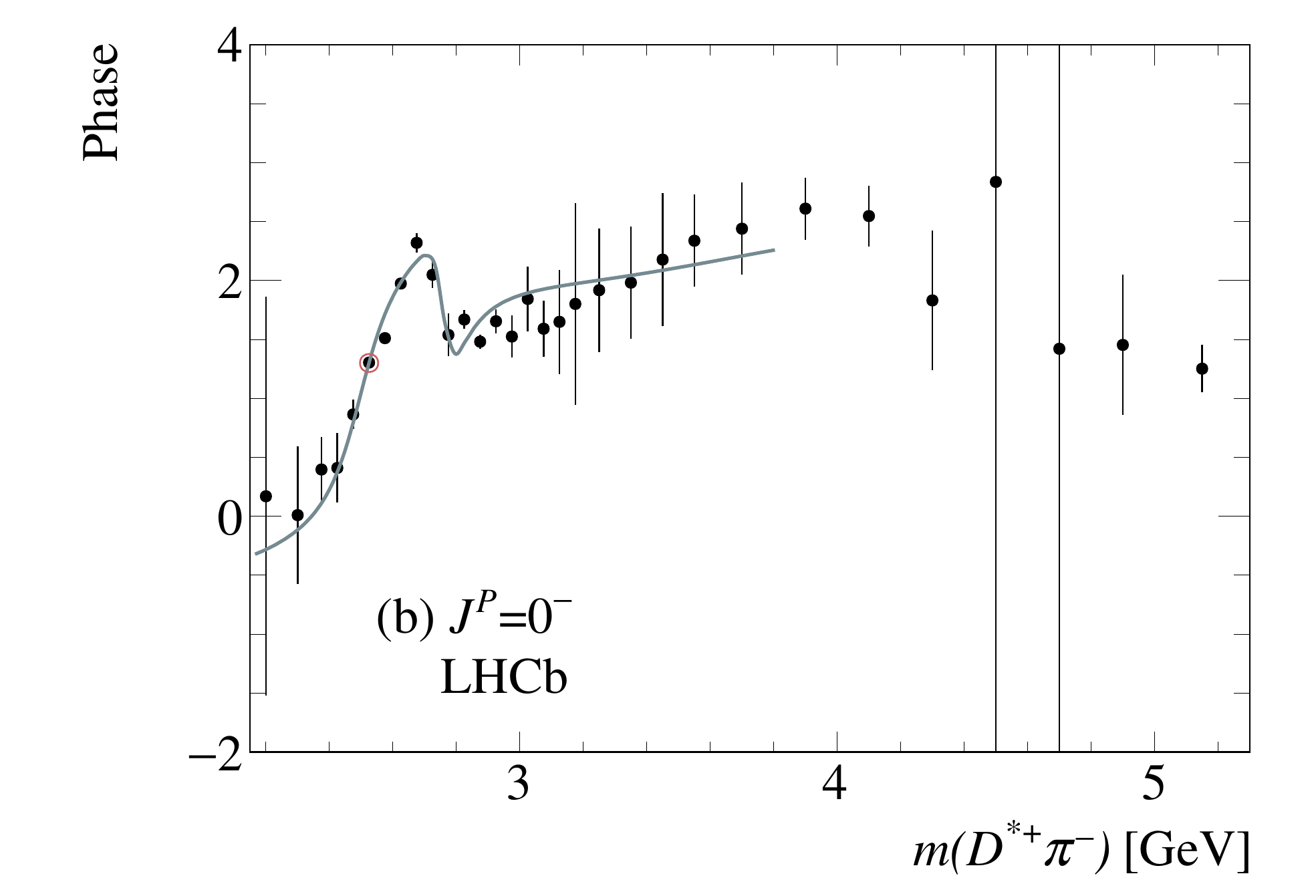}
\caption{\small\label{fig:fig10} (a)~Magnitude and (b)~phase of the $J^P=0^-$ amplitude from the QMI method. The red circle indicates the mass bin where the
complex amplitude has been fixed. The curves are the result from the fit described in the text.}
\end{figure}

The $J^P=0^-$ QMI magnitude and phase are shown in Fig.~\ref{fig:fig10}.
In addition to the \Dzero resonance, further activity can be seen in the 2.8\gev mass region both in amplitude and
phase, suggesting the presence of a possible new excited $D_0'$ resonance.
The $J^P=0^-$ amplitude and phase distributions are fitted using the model
\begin{equation}
  C(m) = ps(m)e^{-am} + c_1 {\rm BW}_{D_0}(m,m_0,\Gamma_0)e^{i \alpha} + c_2 {\rm BW}_{D_0'}(m,m_1,\Gamma_1)e^{i \beta},
\end{equation}
where $ps(m)$ is the $D^{*+}$ momentum in the $D^{*+} \pim$ center-of-mass frame and $a$, $c_1$, $c_2$, $\alpha$ and $\beta$ are free parameters.
The parameters of the \Dzero resonance ($m_0,\Gamma_0$) are fixed to the values extracted from the amplitude analysis (see Table~\ref{tab:tab2}), while the parameters of the
$D_0'$ resonance, ($m_1,\Gamma_1$) are free.
The first term in the above equation represents a threshold $J^P=0^-$ nonresonant term.
The fit is performed in terms of real and imaginary parts of the amplitude and then converted into amplitude and phase
when projected on the data in Fig.~\ref{fig:fig10}.
The $D'_0$ fitted parameters are
\begin{equation}
   m(D'_0) = 2782 \pm 13 \mev, \ \Gamma(D'_0)=146 \pm 23 \mev
\end{equation}
and the significance, computed as the ratio between the fitted fraction divided by the statistical uncertainty, is 3.2$\sigma$.
However, an attempt to include this new possible resonance in the amplitude analysis gives a fraction consistent with zero. 

\begin{figure}[t]
\centering
\small
\includegraphics[width=0.49\textwidth]{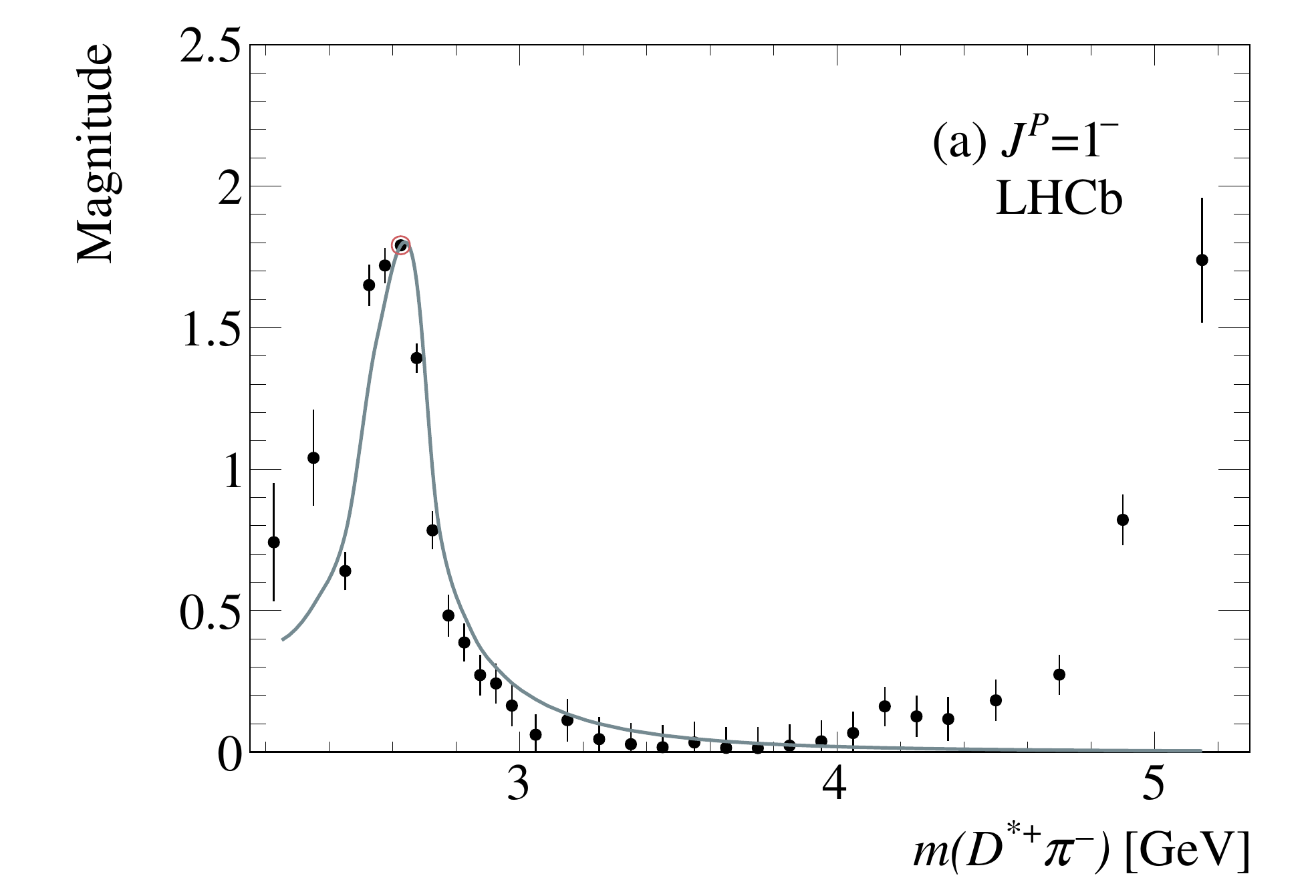}
\includegraphics[width=0.49\textwidth]{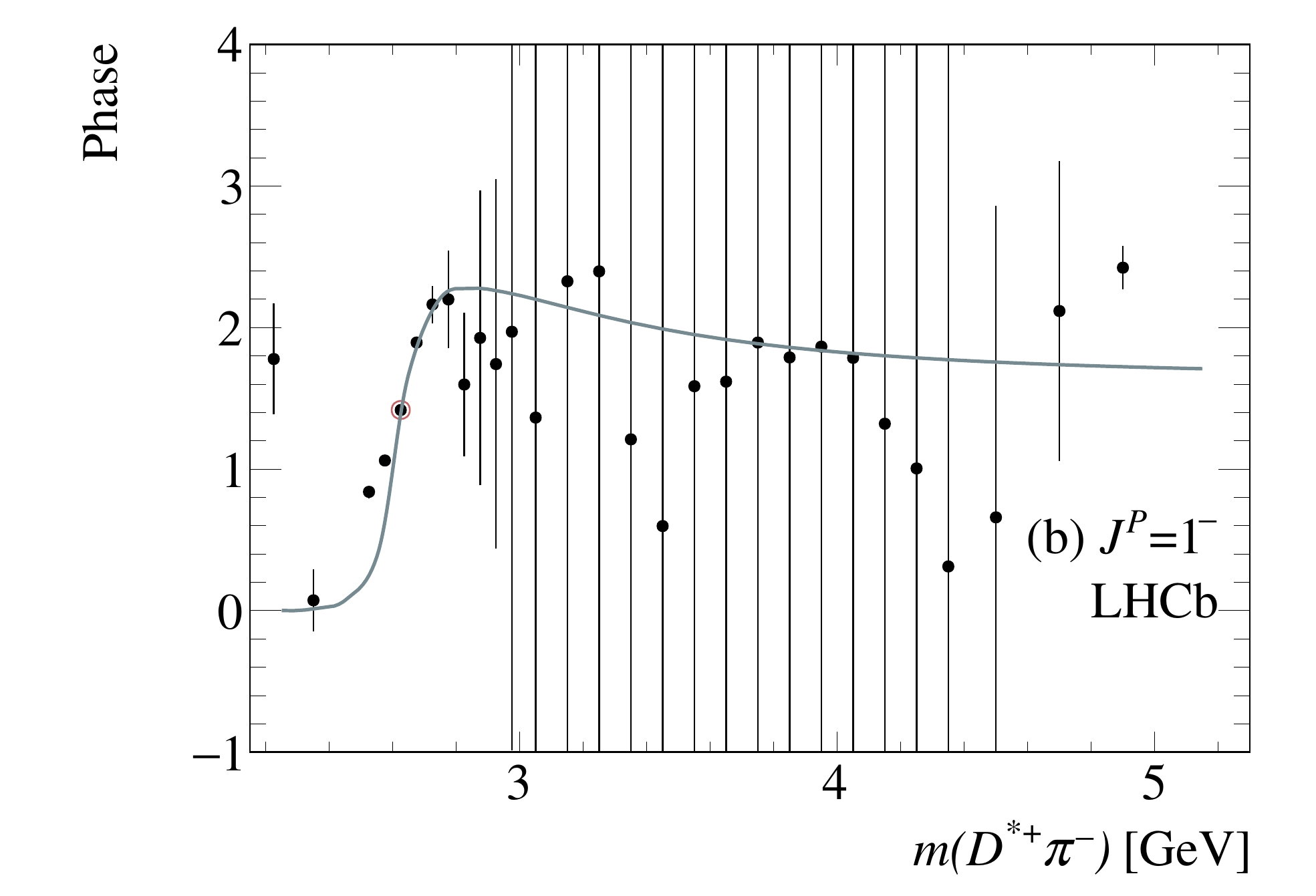}
\caption{\small\label{fig:fig11} (a)~Magnitude and (b)~phase of the $J^P=1^-$ amplitude from the QMI method. The red circle indicates the mass bin where the
complex amplitude has been fixed. The curves represent the Breit--Wigner function describing the \Donem resonance.}
\end{figure}

\begin{figure}[tb]
\centering
\small
\includegraphics[width=0.49\textwidth]{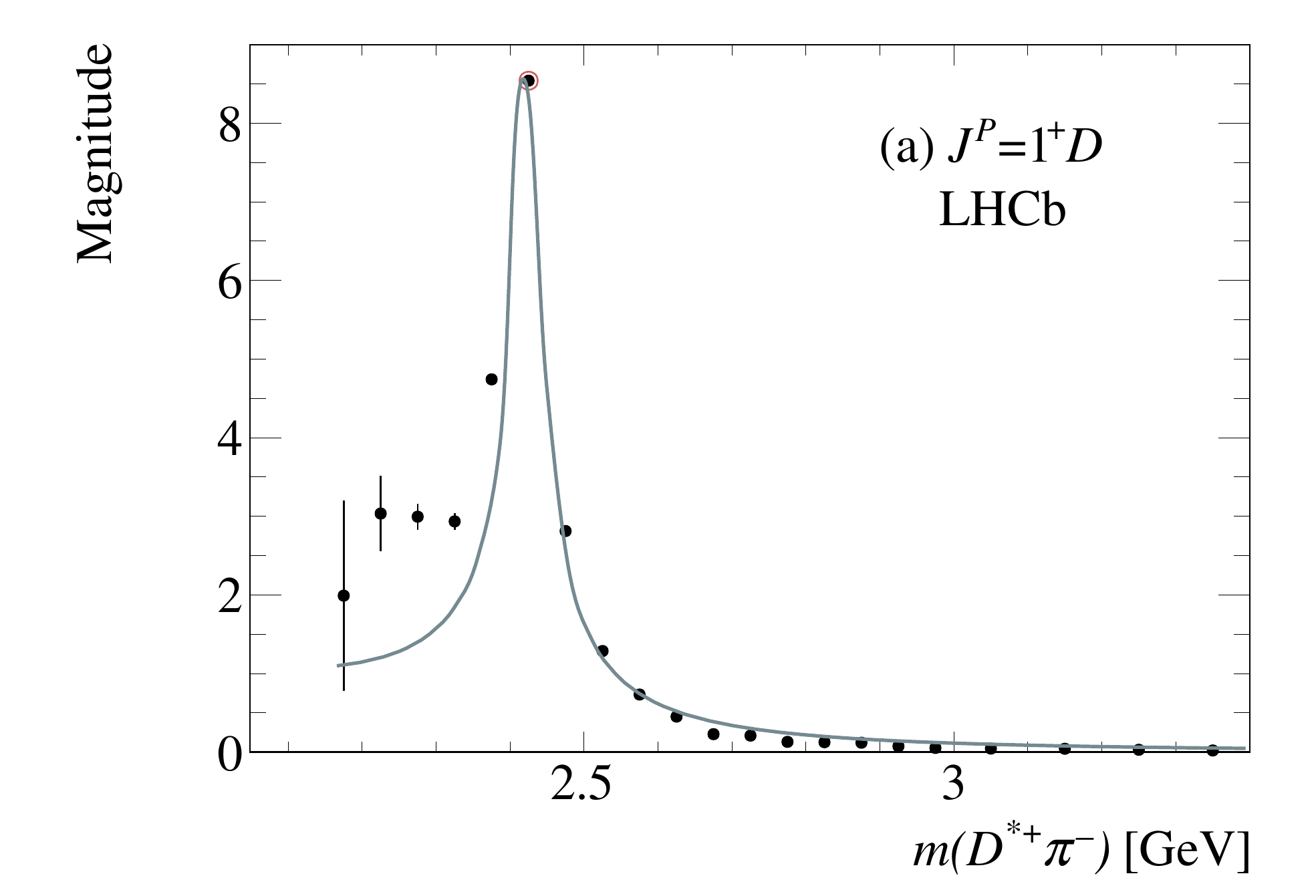}
\includegraphics[width=0.49\textwidth]{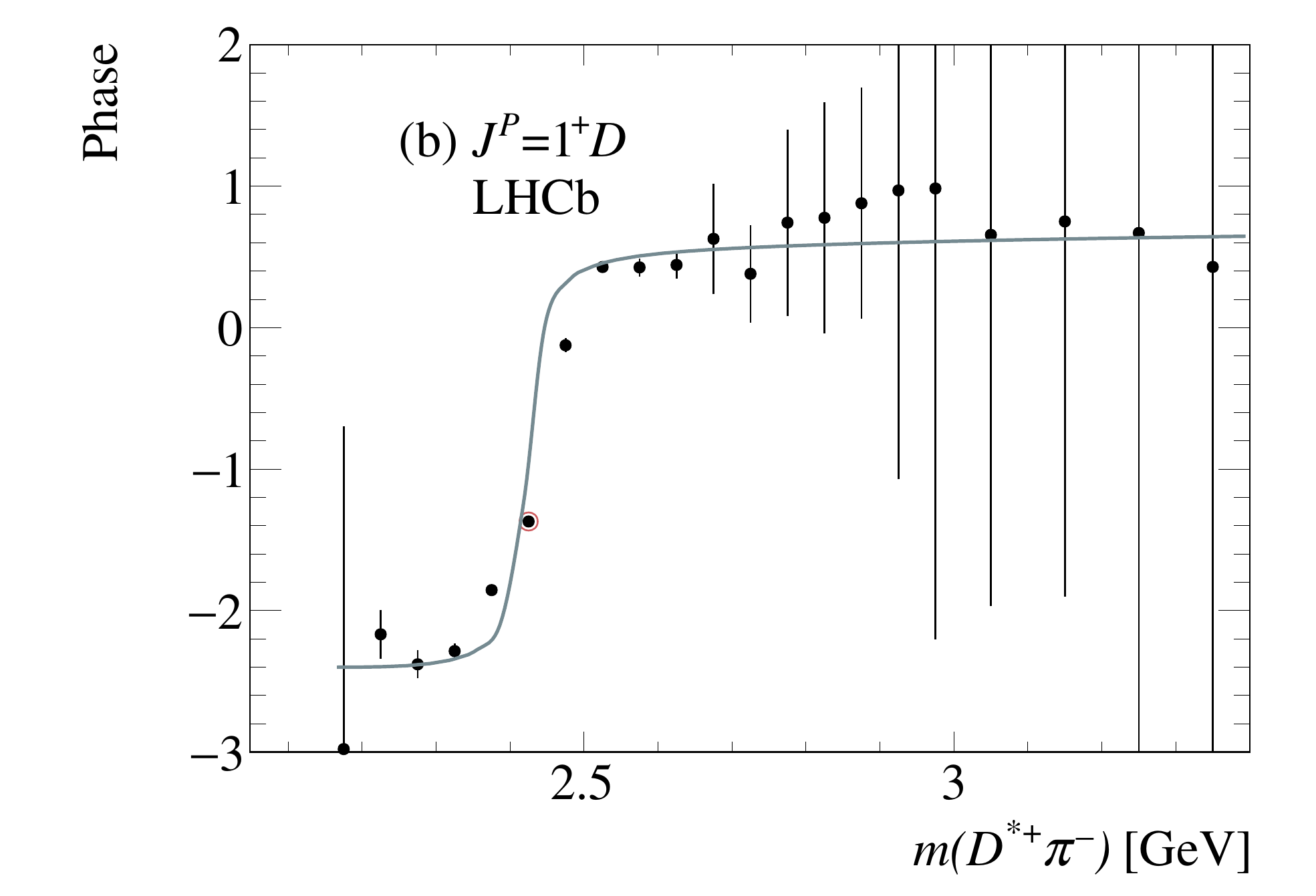}
\caption{\small\label{fig:fig12} (a)~Magnitude and (b)~phase of the $J^P=1^+D$ amplitude from the QMI method. The red circle indicates the mass bin where the
complex amplitude has been fixed. The curves represent the Breit--Wigner function describing the \Done resonance.}
\end{figure}

\begin{figure}[tb]
\centering
\small
\includegraphics[width=0.49\textwidth]{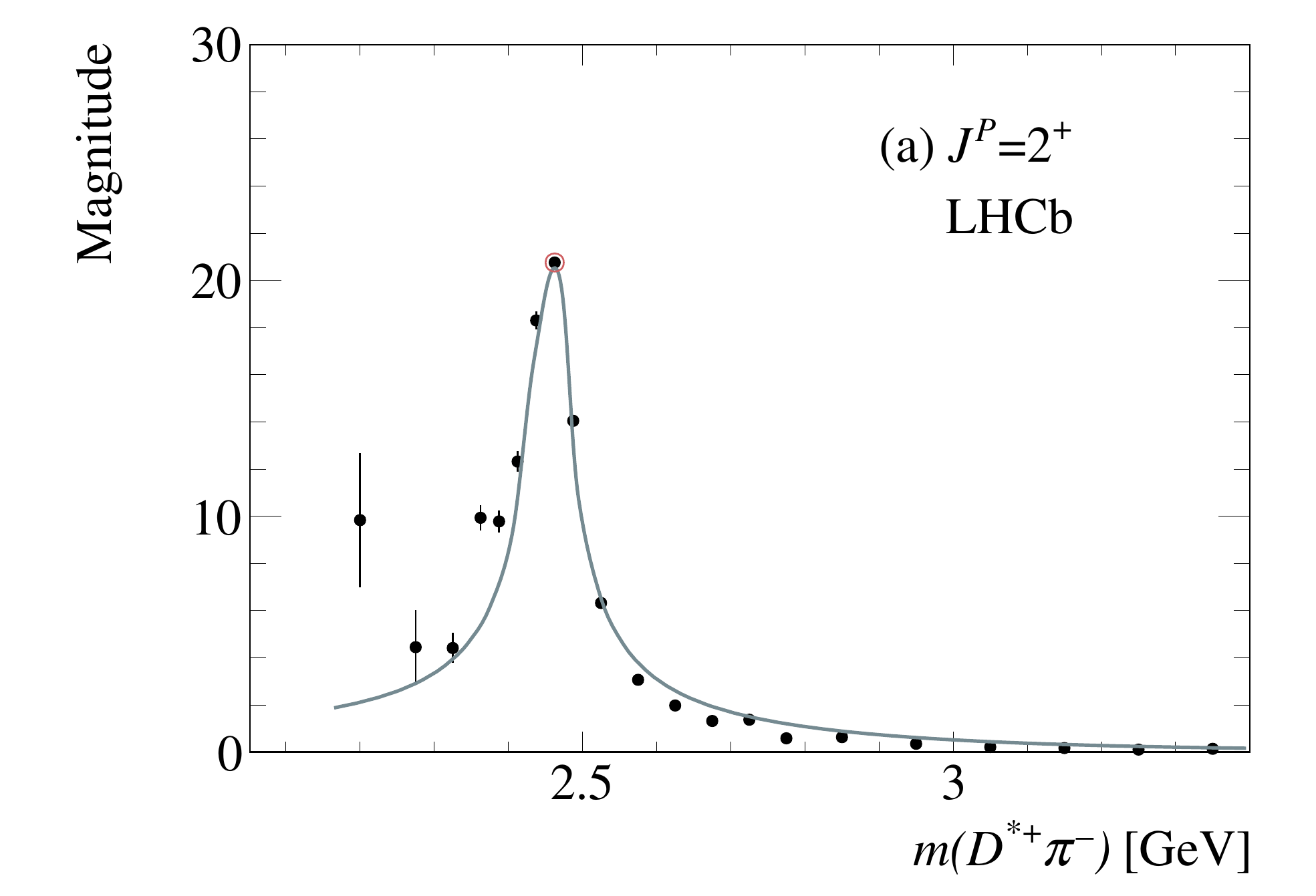}
\includegraphics[width=0.49\textwidth]{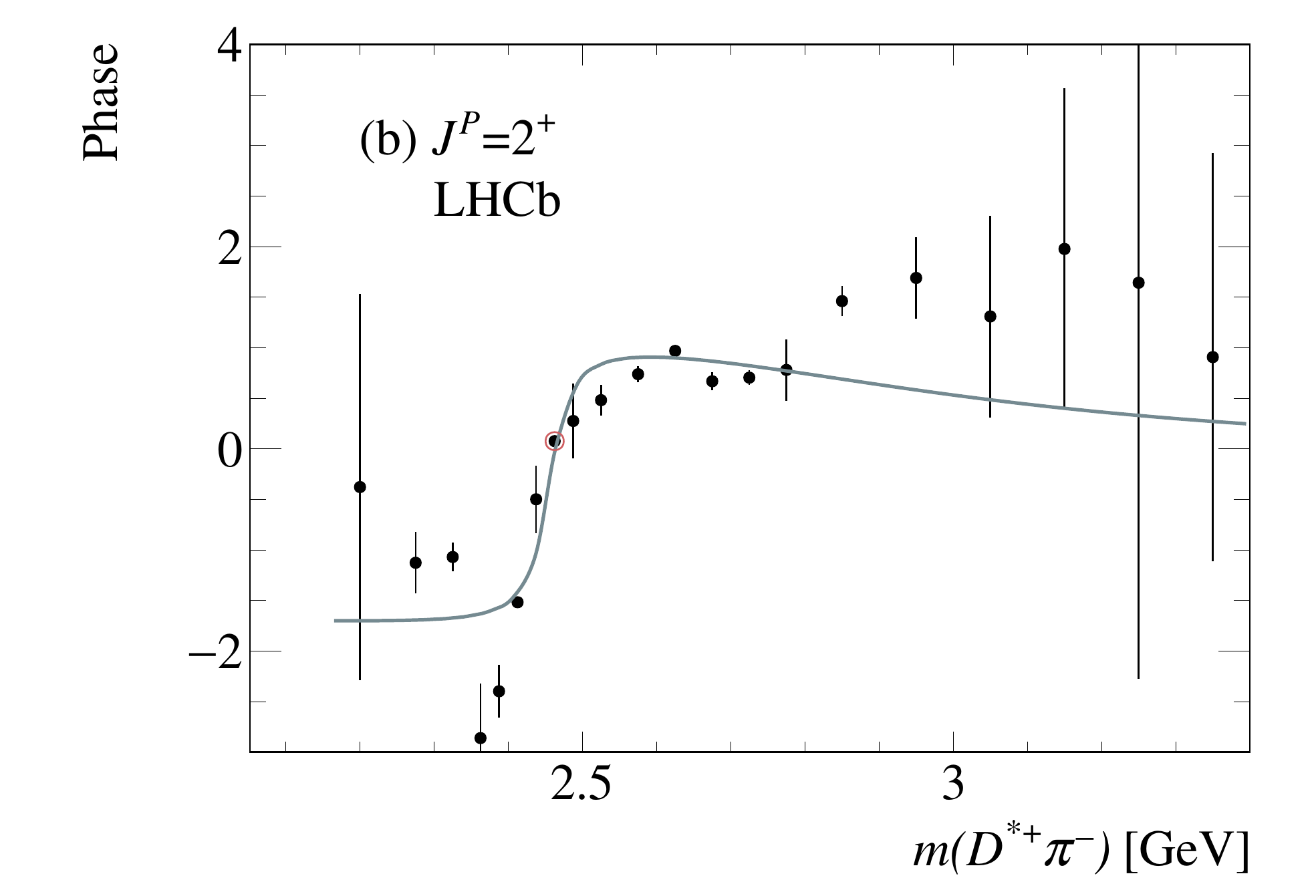}
\caption{\small\label{fig:fig13} (a)~Magnitude and (b)~phase of the $J^P=2^+$ amplitude from the QMI method. The red circle indicates the mass bin where the
complex amplitude has been fixed. The curves represent the Breit--Wigner function describing the \Dstartwo resonance.}
\end{figure}

To search for additional states, the QMI method is used for the most significant amplitudes, \ie those with  $J^P=1^-$ (Fig.~\ref{fig:fig11}), $J^P=1^+D$ (Fig.~\ref{fig:fig12}), and $J^P=2^+$ (Fig.~\ref{fig:fig13}).
In mass regions where the amplitude is consistent with zero the phase is not well measured and therefore statistical uncertainties are large.
Superimposed on the QMI amplitudes are the Breit--Wigner functions, with arbitrary normalizations, describing the \Donem (Fig.~\ref{fig:fig11}) and \Done resonances (Fig.~\ref{fig:fig12}), respectively, using the fitted parameters given in
Table~\ref{tab:tab2}. Similarly, the $J^P=2^+$ amplitude is shown in Fig.~\ref{fig:fig13} with \Dstartwo resonance parameters fixed to the values reported in Ref.~\cite{Tanabashi:2018oca}.

A good agreement between the results from the QMI method and the expected lineshape of the Breit--Wigner description of the resonances is found.
In the case of the $J^P=1^-$ amplitude no additional structure can be seen, and the enhancement at high mass can be associated to the reflection due to
the presence of two identical \pim mesons.
Some amplitudes as $J^P=1^+D$, $J^P=2^+$ and $J^P=1^-$ evidence some points off from the Breit--Wigner behavior in the threshold region. Since in these regions phase space is limited, these effects can be due to cross-feeds from other partial waves.

\section{Fit results}
\label{sec:fits}

The data are fitted using three different models described below.
\begin{itemize}
\item{} The $J^P=1^+S$ and $J^P=0^-$ are described by QMI amplitudes. This model gives the best description of the data and is used to search for new states and obtain the Breit--Wigner parameters for several resonances.
\item{} All the amplitudes are described by relativistic Breit--Wigner functions. This model is used to obtain Breit--Wigner parameters for the \Donew and \Dzero resonances and measure the partial branching fractions for ${\cal{B}}(B^- \to R^0 \pi^-)$, where $R^0$ indicates the charmed meson intermediate state. 
\item{} Mixing is allowed between the $1^+$ amplitudes. This model allows to measure the $D_1$ and $D_1'$ Breit--Wigner parameters and their mixing angle and phase.
\end{itemize}

\subsection{Results from the QMI model}
\label{sec:subqmi}

In this fitting model the $J^P=1^+S$ and $J^P=0^-$ are described by QMI, while all the other 
amplitudes are described by relativistic Breit--Wigner functions with
parameters given in Table~\ref{tab:tab2}.
The results from the fit are given in Table~\ref{tab:tab3}. The dominance of the \Done $(1^+D)$ resonance can be noted, with important contributions from $1^+S$ QMI and \Dstartwo amplitudes. The sum of fractions is larger than 100\%, indicating important interference effects.

\begin{table}[pt]
\centering
\caption{\small Fit results from the amplitude analysis for the model where the $J^P=1^+S$ and $J^P=0^-$ amplitudes are described by QMI. The first uncertainty is statistical, the second systematic.
}
\begin{tabular}{r | r | r@{}c@{}l | r@{}c@{}l}
Resonance & $J^P$ & \multicolumn{3}{c|}{fraction (\%)} & \multicolumn{3}{c}{phase (rad)} \cr
\hline
\Done & $1^+D$  & 59.8 $\pm$  & \, 0.3 \, & $\pm$ 2.9  & 0 \cr
$ 1^+S$ QMI  & $ 1^+S$ & 28.3 $\pm$ & \, 0.3 \, & $\pm$ 1.9  & $-$1.19 $\pm$ & \, 0.01 \, & $\pm$ 0.15 \cr
\Dstartwo&  $2^+$  & 15.3 $\pm$ & \, 0.2 \, & $\pm$ 0.3   & $-$0.71 $\pm$ & \, 0.01 \, & $\pm$ 0.48 \cr
\Done & $1^+S$  & 2.8  $\pm$ & \, 0.2 \, & $\pm$ 0.5   & 1.43  $\pm$ & \, 0.02 \, & $\pm$ 0.31 \cr
$ 0^-$ QMI   & $ 0^-$  & 10.6 $\pm$ & \,0.2 \,  & $\pm$ 0.7   & 1.94  $\pm$ & \, 0.01 \, & $\pm$ 0.19 \cr
\Donem& $1^-$   & 6.0  $\pm$ & \, 0.1 \, & $\pm$ 0.6  & 1.20   $\pm$ & \, 0.02 \, & $\pm$ 0.05 \cr 
\Dtwom  & $2^-P$ &  1.9 $\pm$  & \, 0.1 \, & $\pm$ 0.4  & $-1$.57  $\pm$ & \,0.04 \, & $\pm$ 0.15 \cr
\Dtwom  & $2^-F$ &  3.2 $\pm$  & \, 0.2 \, & $\pm$ 1.1  &  1.11 $\pm$ & \,0.04 \, & $\pm$ 0.29 \cr
\Dthree& $3^-$  & 0.35 $\pm$  & \, 0.04 \, & $\pm$ 0.05  & $-$1.17 $\pm$ & \,0.07 \, & $\pm$ 0.31 \cr
\hline
 Sum &  &   128.2 $\pm$ & \, 0.6 \, & $\pm$ 3.8 &  \cr
\hline
\end{tabular}
\label{tab:tab3}
\end{table}

\begin{figure}[ptb]
\centering
\small
\includegraphics[width=0.49\textwidth]{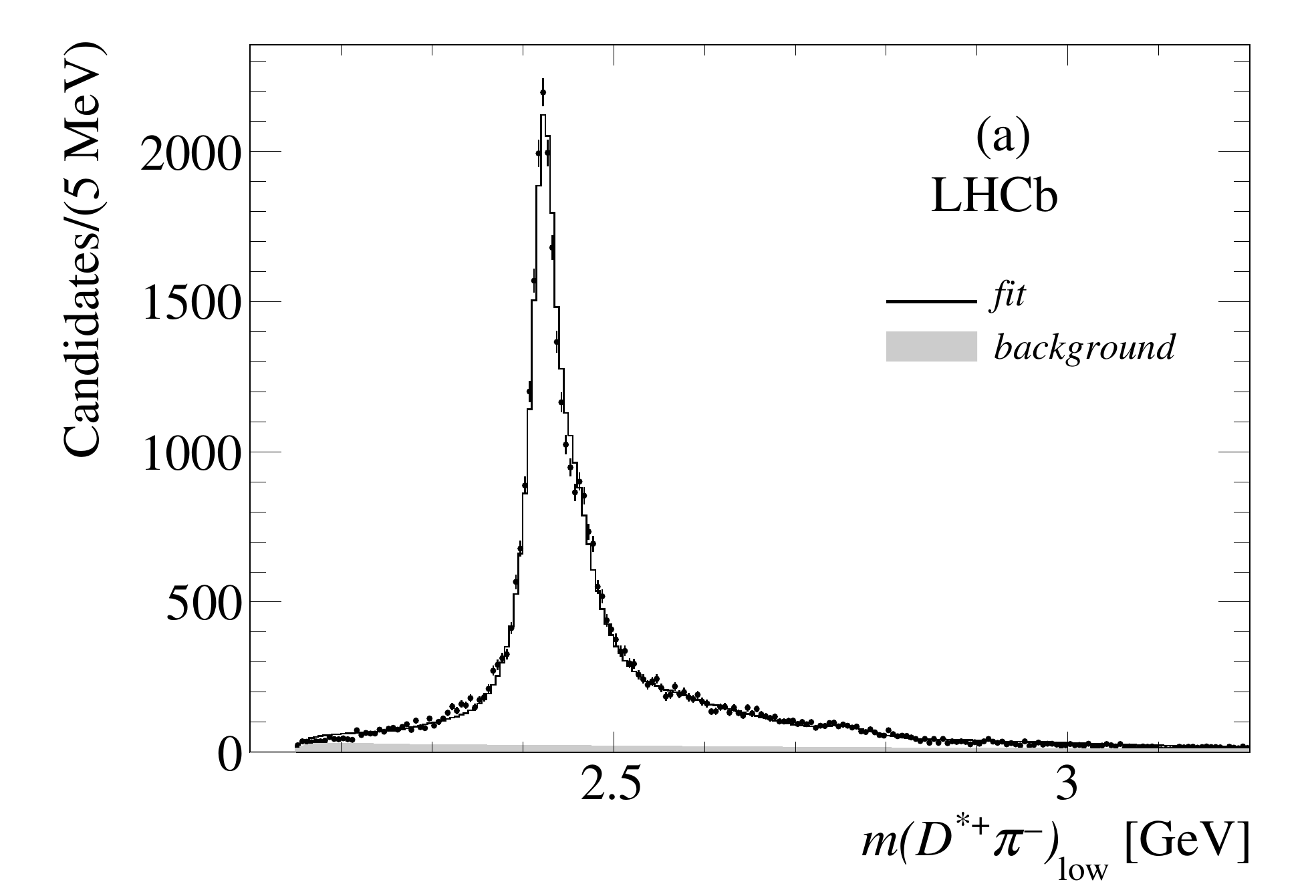}
\includegraphics[width=0.49\textwidth]{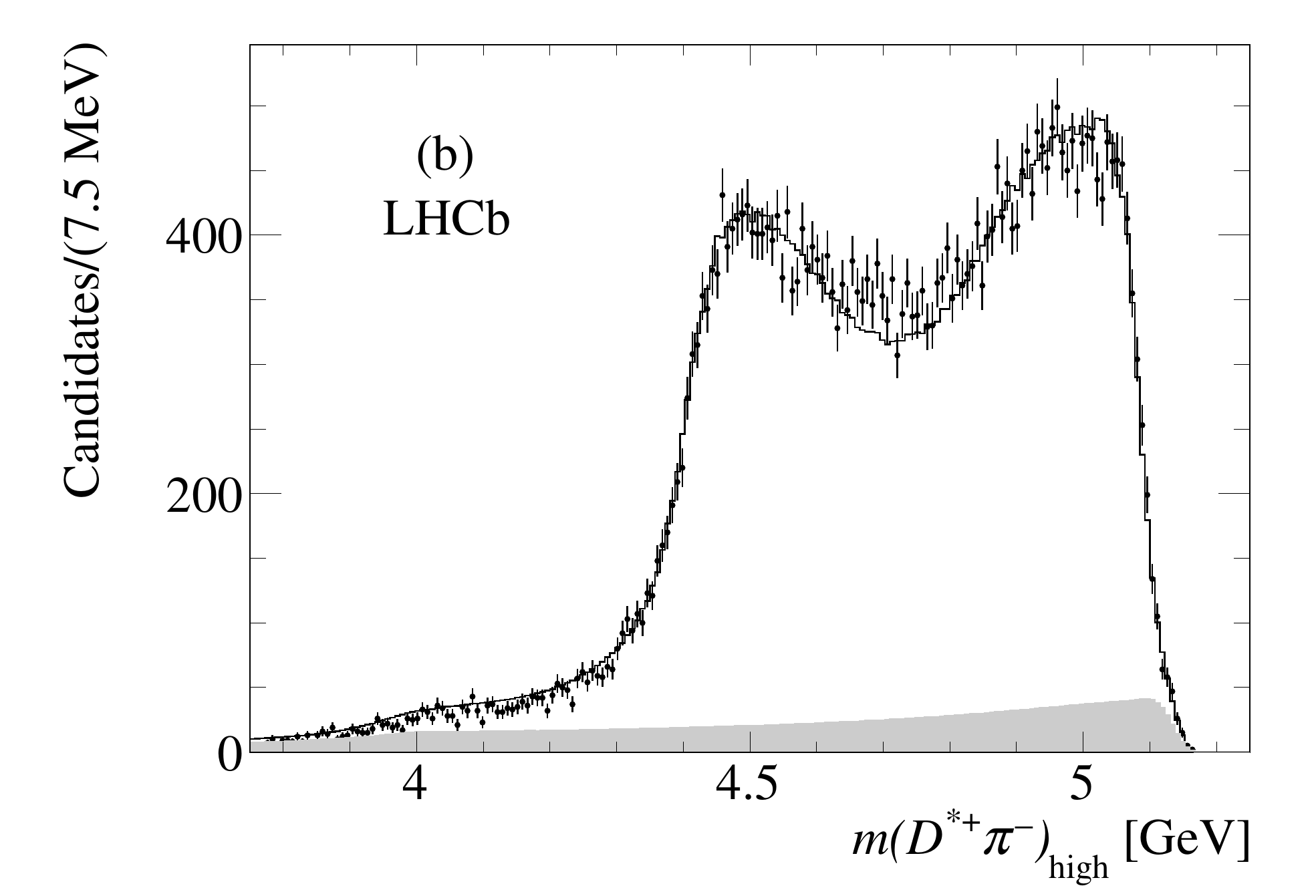}
\includegraphics[width=0.49\textwidth]{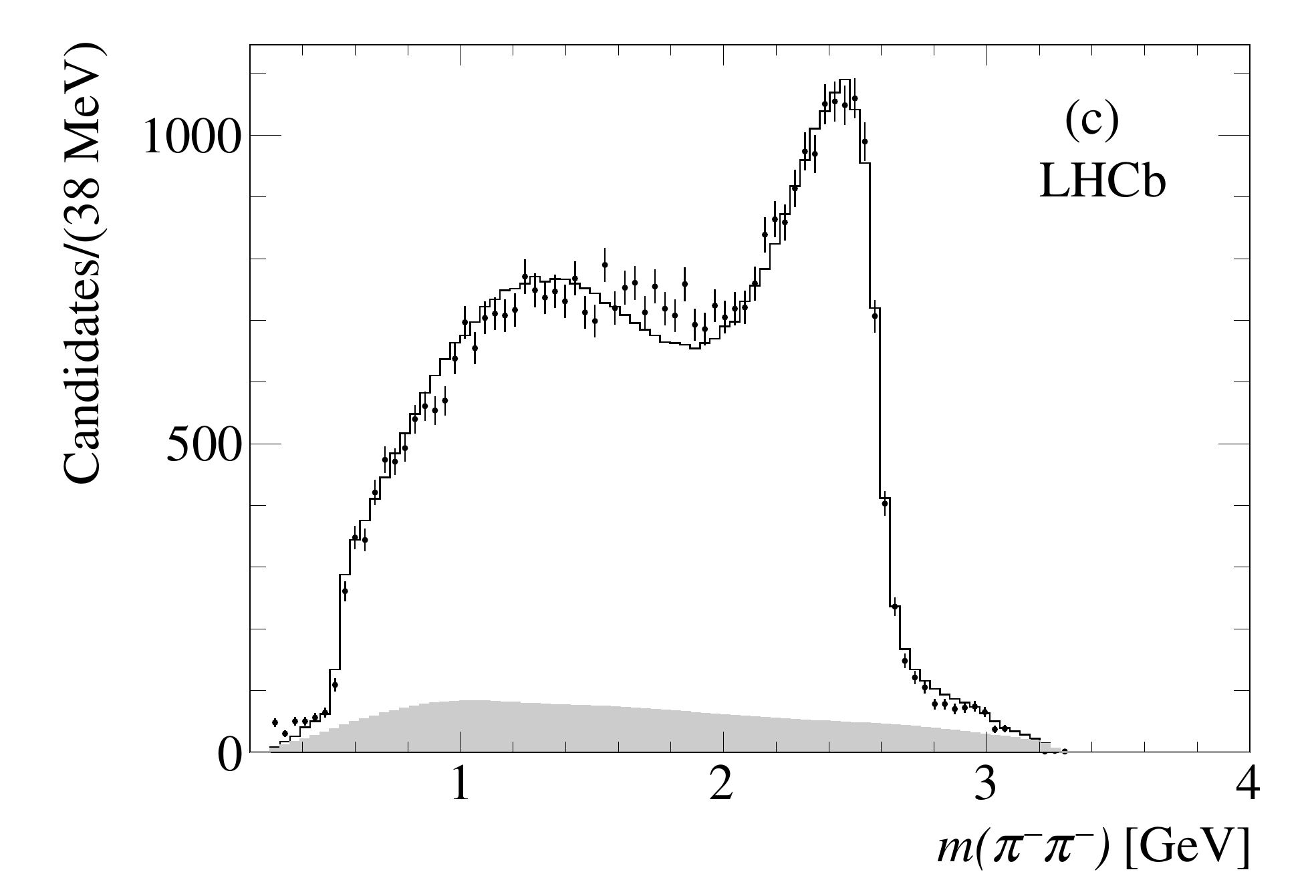}
\caption{\small\label{fig:fig14}  Projections of the fit to Run 2 data from the QMI fitting model. The background contribution is shown in gray. Data are represented with filled dots and the line is the results from the fit.
  }
\end{figure}

\begin{figure}[tb]
\centering
\small
\includegraphics[width=0.48\textwidth]{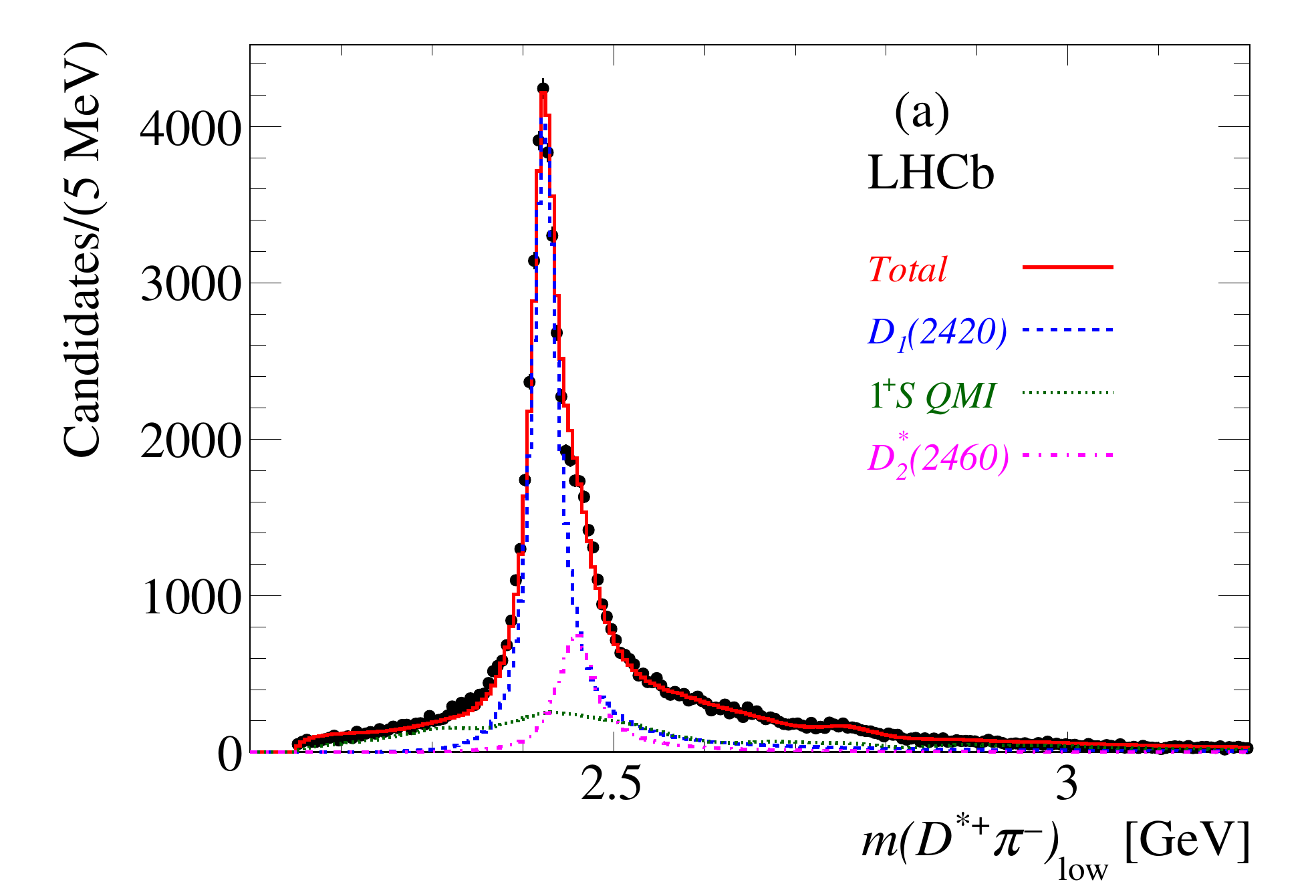}
\includegraphics[width=0.48\textwidth]{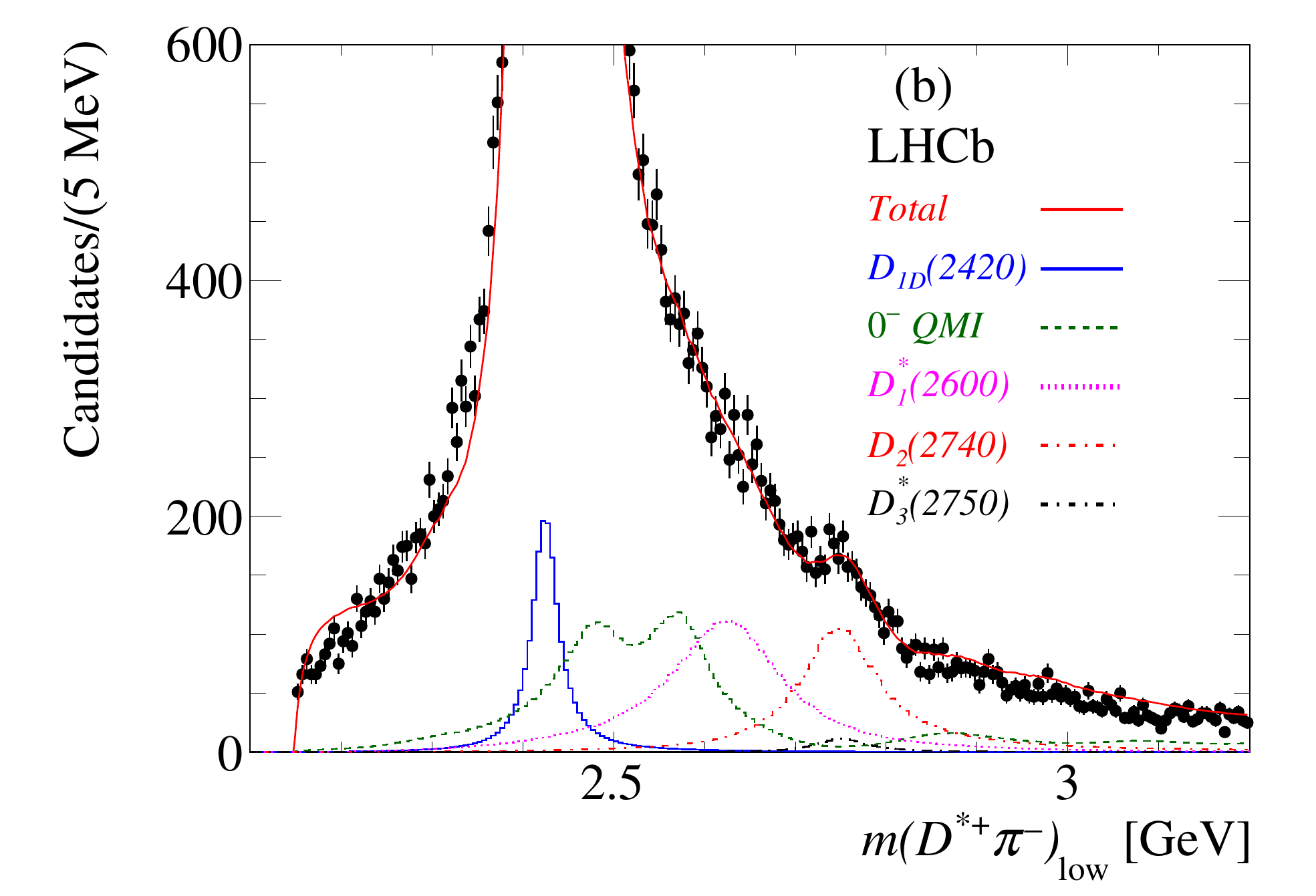}\\
\includegraphics[width=0.48\textwidth]{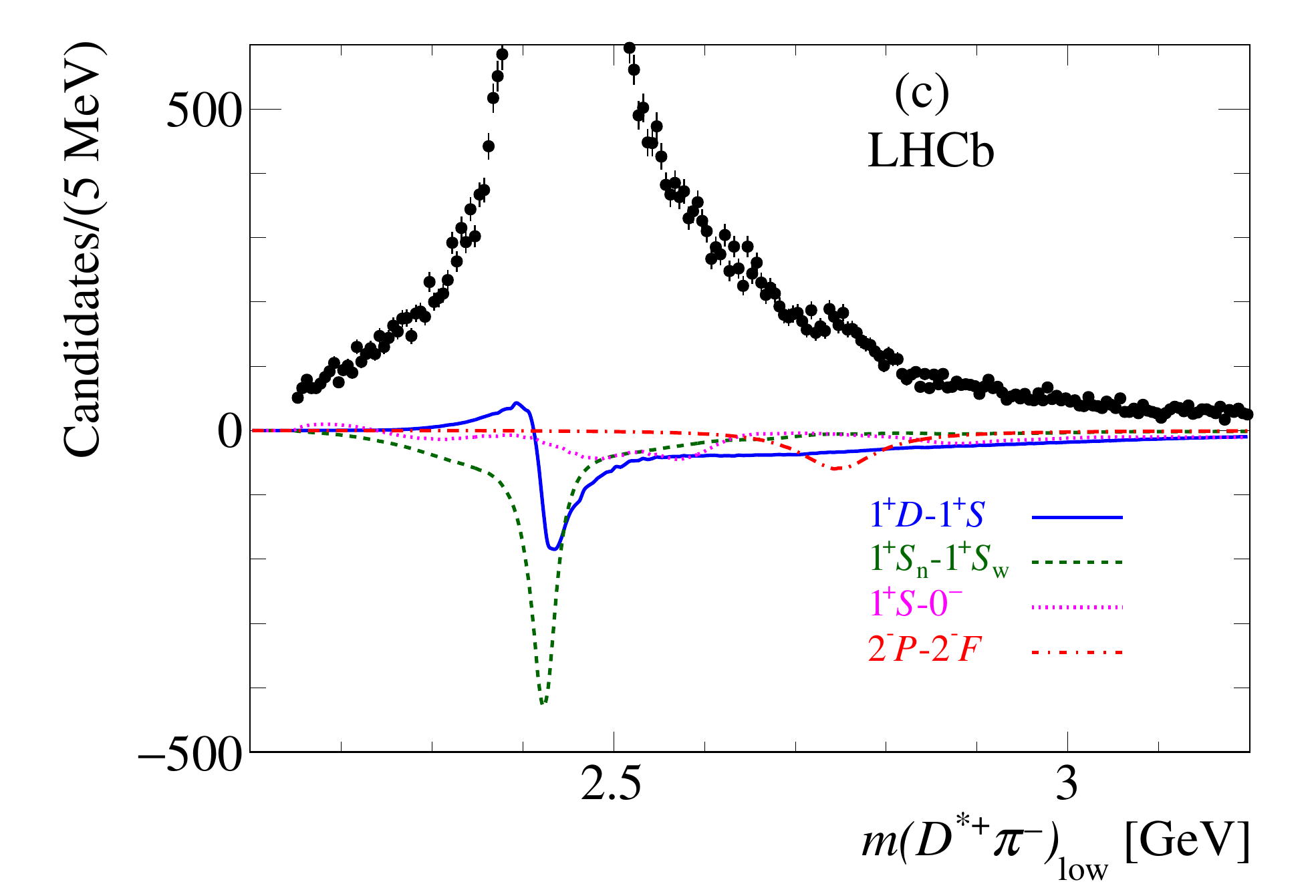}
\caption{\small\label{fig:fig15} Projections of the fit to the total dataset from the QMI fitting model with (a,b) all amplitude
contributions and (c) the significant interference terms.
  }
\end{figure}

The fit projections for Run 2 data (not biased by the $\pip \pim \pim$ mass cut) are shown in Fig.~\ref{fig:fig14}. Figure~\ref{fig:fig15} shows the fit projections on $m(\Dstarp \pim)_{\rm low}$ using the total dataset together with all the contributing amplitudes and the significant
interference contributions.
Using statistical uncertainties only, the separate fits for Run 1 and Run 2 give \mbox{$\chi^2/{\rm ndf}=2348/1748=1.34$} and \mbox{$\chi^2/{\rm ndf}=2111/1780=1.19$}, 
respectively. 
For a fit to the total dataset $\chi^2/{\rm ndf}=2551/1784 = 1.43$. However it has to be taken into account that in this fit 
the total sample size is double and therefore statistical uncertainties are smaller.
These $\chi^2/{\rm ndf}$ values indicate a good description of Run 2 data, but a worse description of the Run 1 data indicating some limitation in the handling of the efficiency for this data set.

\subsubsection{Systematic uncertainties}

Systematic uncertainties reported in Table~\ref{tab:tab2} and alternative fit models (described later), whose results are given in Tables~\ref{tab:tab6} and \ref{tab:tab7}, are evaluated as follows. When multiple contributions are needed to describe a given effect,
the average value of the absolute deviations from the reference fit is taken as a systematic uncertainty.

Table~\ref{tab:sys1} gives details on the contributions to the systematic uncertainties on the fractions and phases for the model where the $J^P=1^+S$ and $J^P=0^-$ amplitudes are described by QMI.
 \begin{table} [tb]
\centering
\caption{Absolute systematic uncertainties on the fractions (in \%) (top), and phases (bottom) for the model where the $J^P=1^+S$ and $J^P=0^-$ amplitudes are described by QMI.
}
\resizebox{\textwidth}{!}{
\begin{tabular}{llccccccccc}
\hline
Resonance  & $J^P$ & Purity & BW & Res.(a) & Res.(b) & Bkg size & Data/sim & Sim & Mod & Total\cr
\hline
$D_1(2420) $ & $1^+D$ & 0.36 & 2.88 & 0.05 & 0.01 & 0.30 & 0.19 & 0.33 & 0.12 & 2.9 \cr
$ 1^+S$ QMI & $1^+S$ & 0.54 & 1.37 & &  0.01 & 0.27 & 0.16 & 0.34 & 1.17 & 1.9 \cr
$D^*_2(2460)$ & $2^+$ & 0.14 & 0.04 & 0.05 & 0.01 & 0.05 & 0.04 & 0.26 & 0.00 & 0.3 \cr
$D_1(2420) $ & $1^+S$ & 0.03 & 0.46 & 0.02 & 0.01 & 0.01 & 0.02 & 0.21 & 0.00 & 0.5 \cr
$ 0^-$ QMI & $0^-$ & 0.07 & 0.08 & & 0.01 & 0.03 & 0.07 & 0.17 & 0.69 & 0.72 \cr
$D^*_1(2600)$ & $1^-$ & 0.05 & 0.53 & 0.01 & 0.01 & 0.02 & 0.06 & 0.14 & 0.04 & 0.6 \cr
$D_2(2740)$ & $2^-P$ &  0.11 & 0.36 & 0.01 & 0.00 & 0.06 & 0.06 & 0.10 & 0.07 & 0.4\cr
$D_2(2740)$ & $2^-F$ & 0.09 & 1.13 & 0.02 & 0.01 & 0.12 & 0.07 & 0.12 & 0.29 & 1.1 \cr
$D^*_3(2750)$ & $3^-$ & 0.02 & 0.02 &0.00 & 0.00 & 0.01 & 0.02 & 0.03 & 0.008 & 0.1 \cr
\hline
$ 1^+S$ QMI & $1^+S$ & 0.01 & 0.15 & & 0.00 & 0.00 & 0.00 & 0.00 & 0.010 & 0.15\cr
$D^*_2(2460)$ & $2^+$ & 0.01 & 0.48 &  0.01 & 0.00 & 0.00 & 0.00 & 0.01 & 0.01 & 0.48\cr
$D_1(2420) $ & $1^+S$ & 0.02 & 0.31 & 0.00 & 0.00 & 0.01 & 0.02 & 0.03 & 0.00 & 0.31\cr
$ 0^-$ QMI & $0^-$ & 0.01 & 0.08 &  0.00 & 0.01 & 0.17 & 0.00 & 0.01 & 0.04 & 0.19\cr
$D^*_1(2600)$ & $1^-$ & 0.03 & 0.03 & 0.01 & 0.00 & 0.01 & 0.00 & 0.02 & 0.02 & 0.05 \cr
$D_2(2740)$ & $2^-P$ &  0.04 & 0.14 & 0.02 &0.00 & 0.00 & 0.01 & 0.04 & 0.02 & 0.15 \cr
$D_2(2740)$ & $2^-F$ & 0.06 & 0.28 & 0.02& 0.00 & 0.01 & 0.00 & 0.03 & 0.02 & 0.29\cr
$D^*_3(2750)$ & $3^-$ & 0.13 & 0.28 & 0.04 & 0.00 & 0.01 & 0.03 & 0.04 & 0.04 & 0.31\cr
\hline
\end{tabular}
}
\label{tab:sys1}
 \end{table}
The effect of the background (labeled as Purity) is studied by changing the selection cut corresponding to lower (with $\calR>1.1$, $p=0.92$, 66\,064 candidates) or higher 
(with $\calR>0.2$, $p=0.87$, 85\,466 candidates) purity. 
The contribution due to the description of the resonance model (labeled as BW) is estimated by varying the Blatt--Weisskopf radius
between 1 and 5\,GeV$^{-1}$. 
The effect of the uncertainty on the resonance parameters is estimated by varying their values
within uncertainties. The label Res.(a) indicates a variation of the parameters of a given resonance, Res.(b) indicates a variation of the parameters of all the other resonances contributing to the \Bm decay.
The effect of the uncertainty of the background size (labeled as Bkg size) is estimated by modifying the value of the fixed purity value in the fit (90\%) by $\pm 0.5\%$.
The effect of the small discrepancy between the data and fit projections on the $\cos \theta$ and $\cos \theta_H$ distribution (labeled as Data/sim) is evaluated by weighting the efficiency distribution to match the data.
The effect of the limited simulation sample (labeled as Sim) is evaluated by fitting the data using 100 binned 2-dimensional efficiency tables obtained from the reference one through Poisson fluctuations of the entries in each bin.
Virtual contributions such as $\Bm \to B^{*0}_v \pim$~\cite{Aaij:2016fma} (labeled as Mod) are included and excluded in the fit.
The root-mean-square value of the deviations of the fraction from the reference fit are taken as systematic uncertainties. 
All the different contributions are added in quadrature. The dominant sources
of systematic uncertainties are found to be due to the Blatt--Weisskopf radius.

Table~\ref{tab:sys2} gives details on the contributions to the systematic uncertainties for the measured masses and widths of the resonances contributing to the \Bm decay.
In this case only the most relevant contributions are listed.
From the study of large control samples, a systematic uncertainty of $0.0015\,Q$ on the mass scale is added, where $Q$ is the $Q$-value involved in the resonance decay.
  \begin{table} [tb]
  \centering
 \caption{Systematic uncertainties contributions to the measured mass and width (in MeV) of the different resonances contributing to the \Bm decay.}
\begin{tabular}{llcccc}
\hline
Resonance  & Parameter & BW & Purity & Mass scale & Total\cr
\hline
$D_1(2420)$ & Mass     & \phz0.6 & \phz0.1 & \phz0.4 & \phz0.7 \cr 
            & Width    & \phz2.7 & \phz 0.4 & \phz    & \phz2.7 \cr
\hline
$D_0(2550)$ & Mass & \phz3.2 & \phz6.7 & \phz0.6 & \phz7.4 \cr
            & Width & 14.7 & \phz8.1 & \phz  & 16.8 \cr
$D^*_1(2600)$ & Mass & \phz2.9 & \phz2.9 & \phz0.7 & \phz4.5 \cr
             & Width & 14.9 & 12.7 & \phz  & 19.6 \cr
$D_2(2740)$ & \phz Mass & \phz4.3 & \phz5.6 & \phz0.9 & \phz7.1 \cr
            & Width & 25.1 & \phz8.0 & \phz   & 26.3 \cr
$D^*_3(2750)$ & Mass & \phz5.8 & \phz & \phz0.9 & \phz5.9 \cr
              & Width & 14.4 & \phz& \phz0.4 & 14.4 \cr
\hline
$D_1(2430)$ & Mass & \phz7.0 & \phz5.5 & \phz0.4 & \phz8.9 \cr
            & Width & 14.0 &24.0 & \phz & 27.8\cr
\hline
\end{tabular}
\label{tab:sys2}
\end{table}

The consistency between the Run 1 and Run 2 datasets is tested performing separate fits to the data and  good agreement is obtained, within the uncertainties, on fractions and relative phases.
Separate fits are performed to subsamples of the data where the $\Dstarp \pim \pim$ final state is directly (69\%) or undirectly (31\%) selected by the trigger conditions. The fitted fractions and phases are found  consistent within the statistical uncertainties. A test is performed by weighting the simulated \pt distribution to match the data and recomputing the efficiencies. The impact on the fitted fractions and phases is found to be negligible.

\subsubsection{Legendre polynomial moments projections}

\begin{figure}[tb]
\centering
\small
\includegraphics[width=0.49\textwidth]{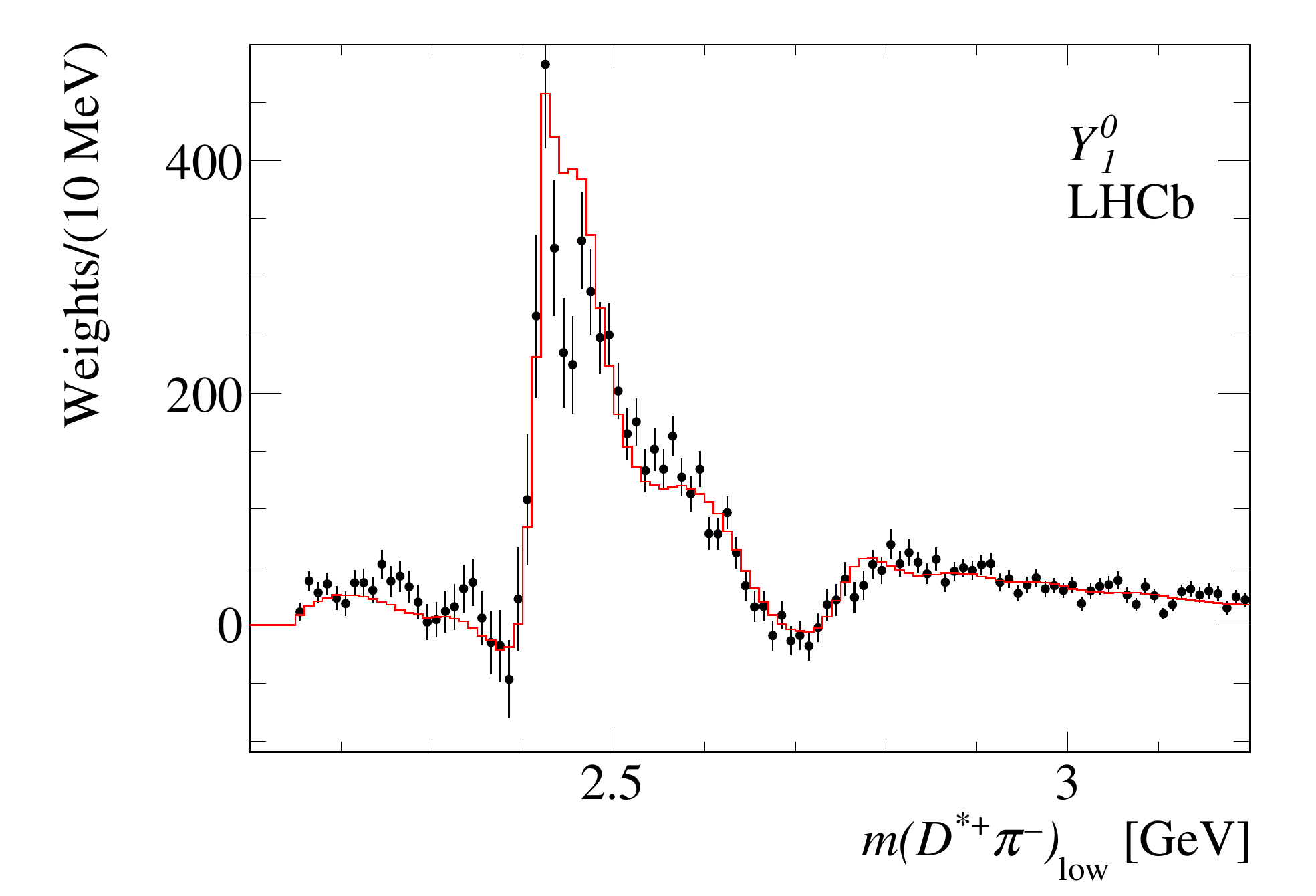}
\includegraphics[width=0.49\textwidth]{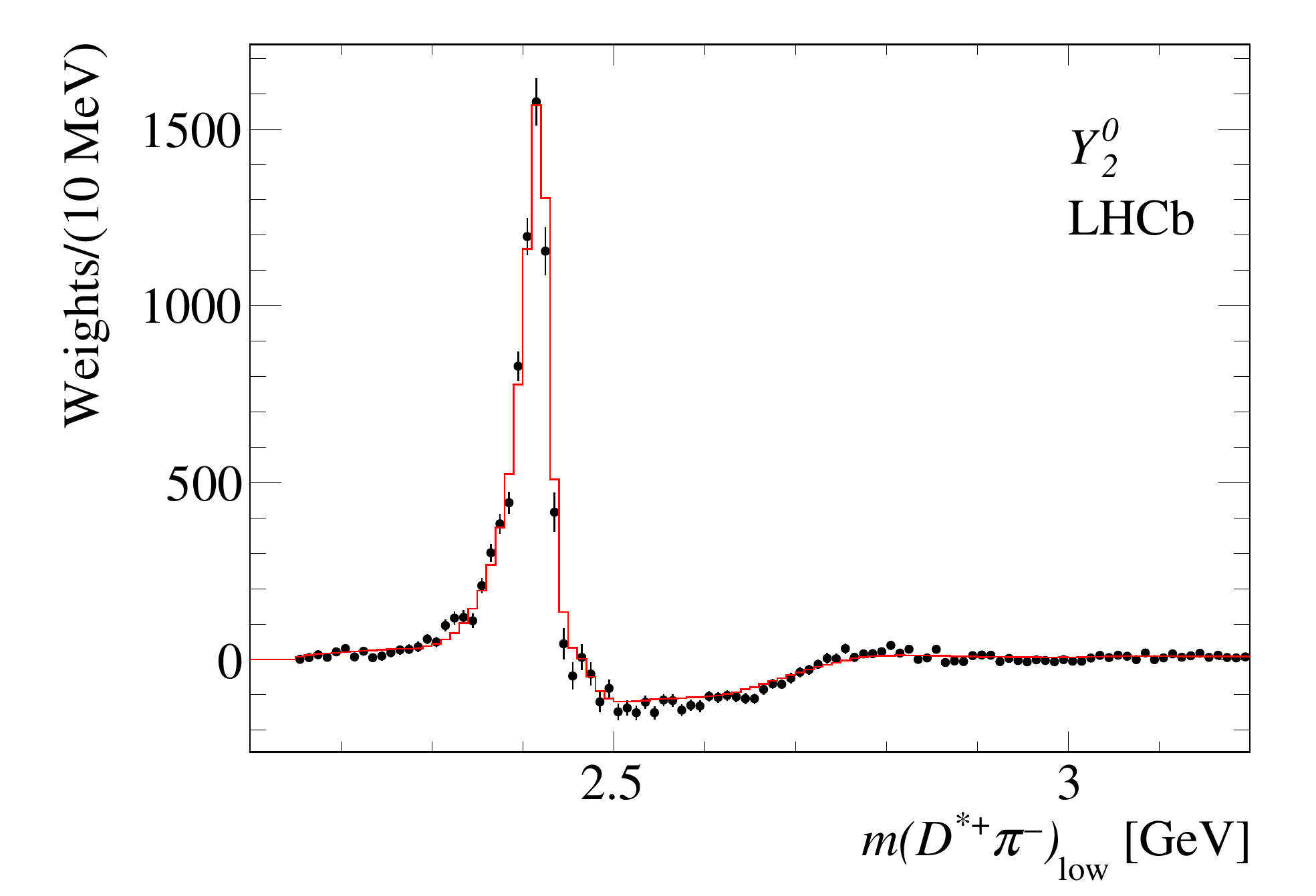}
\includegraphics[width=0.49\textwidth]{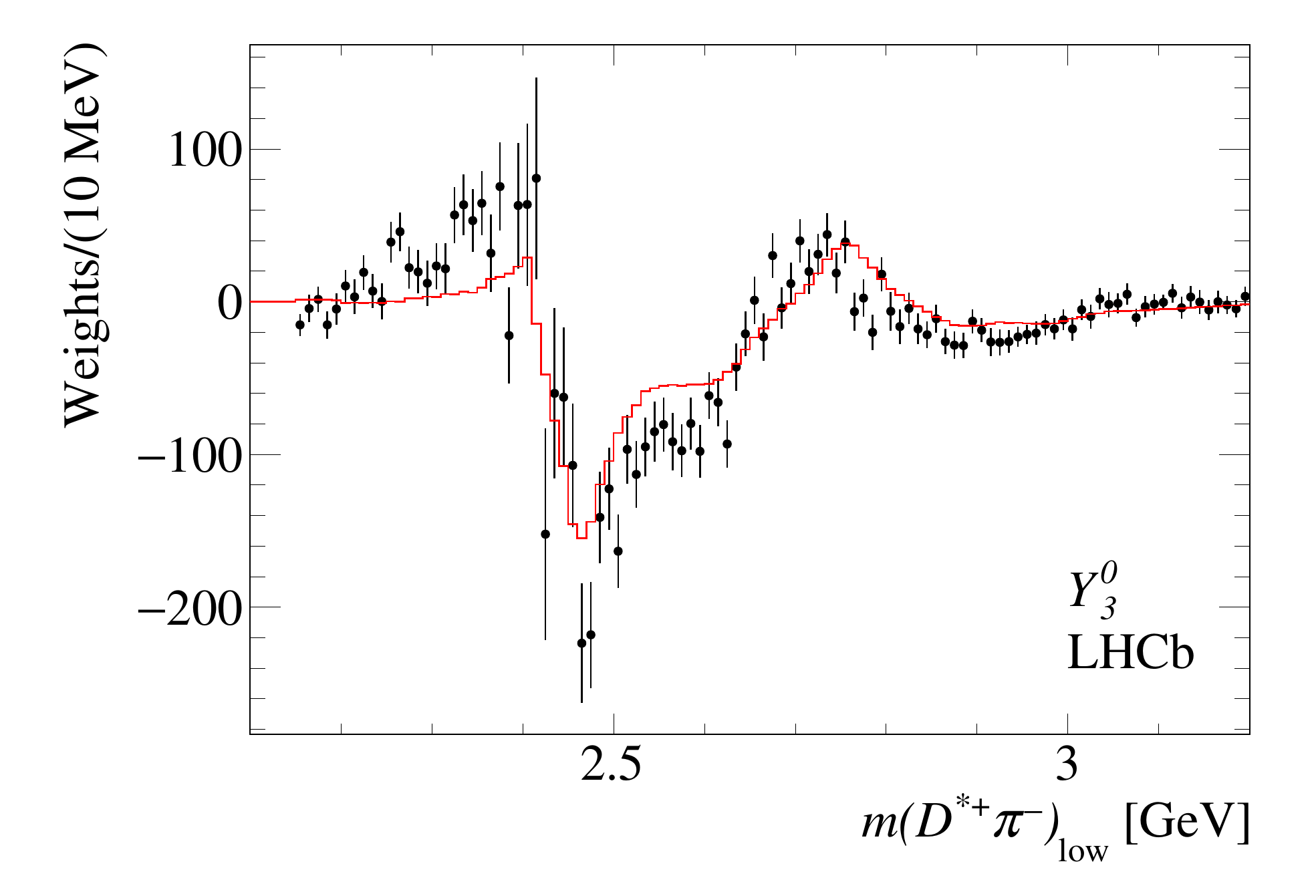}
\includegraphics[width=0.49\textwidth]{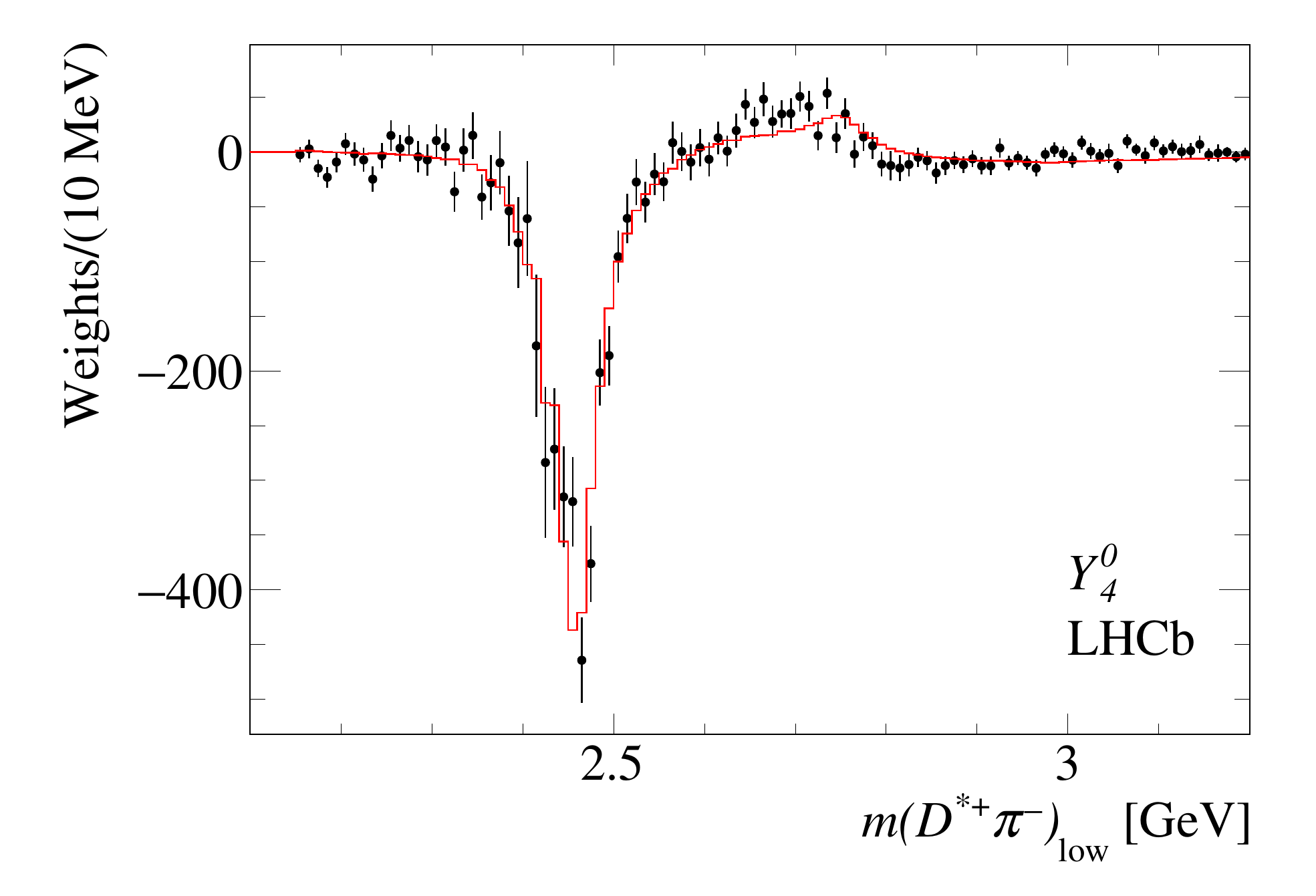}
\includegraphics[width=0.49\textwidth]{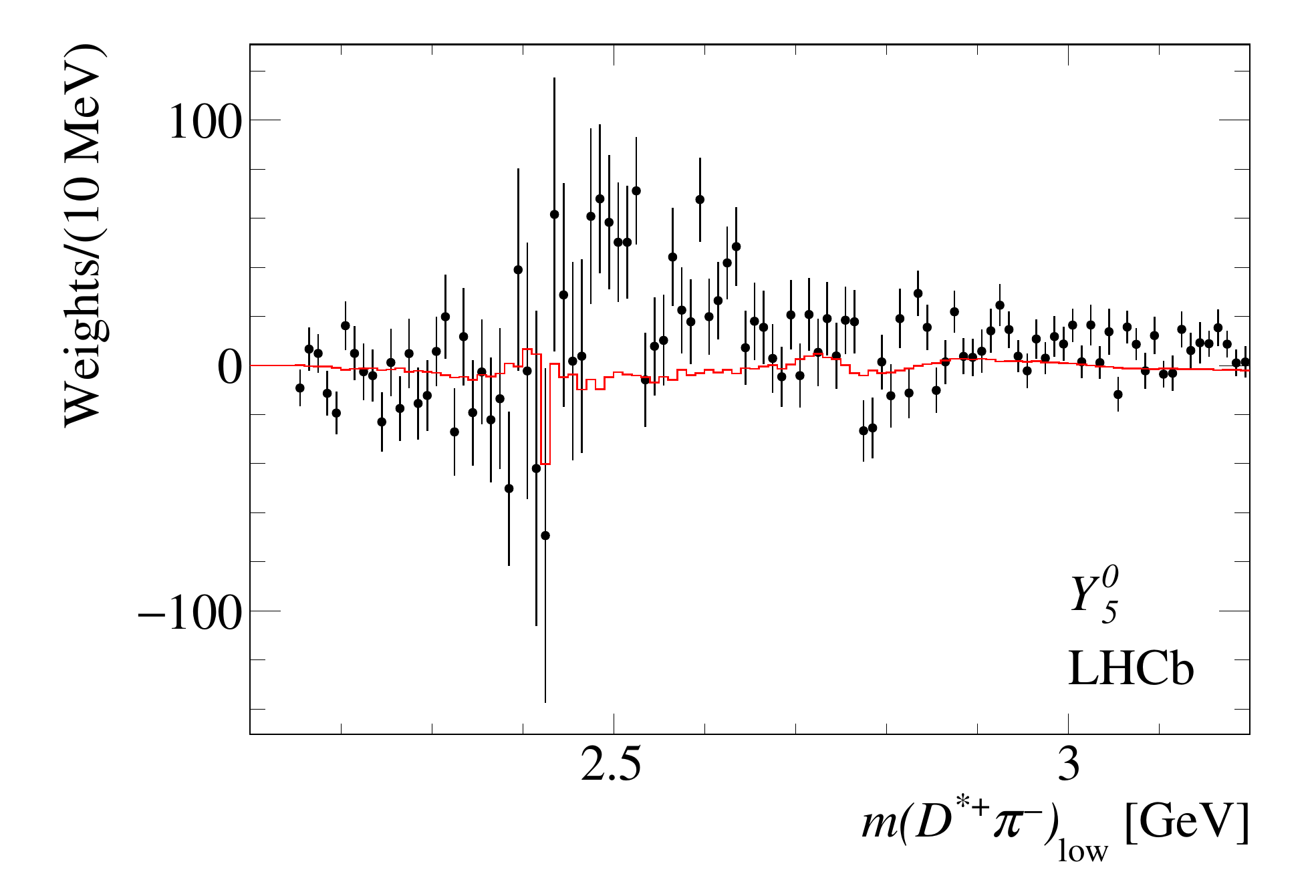}
\includegraphics[width=0.49\textwidth]{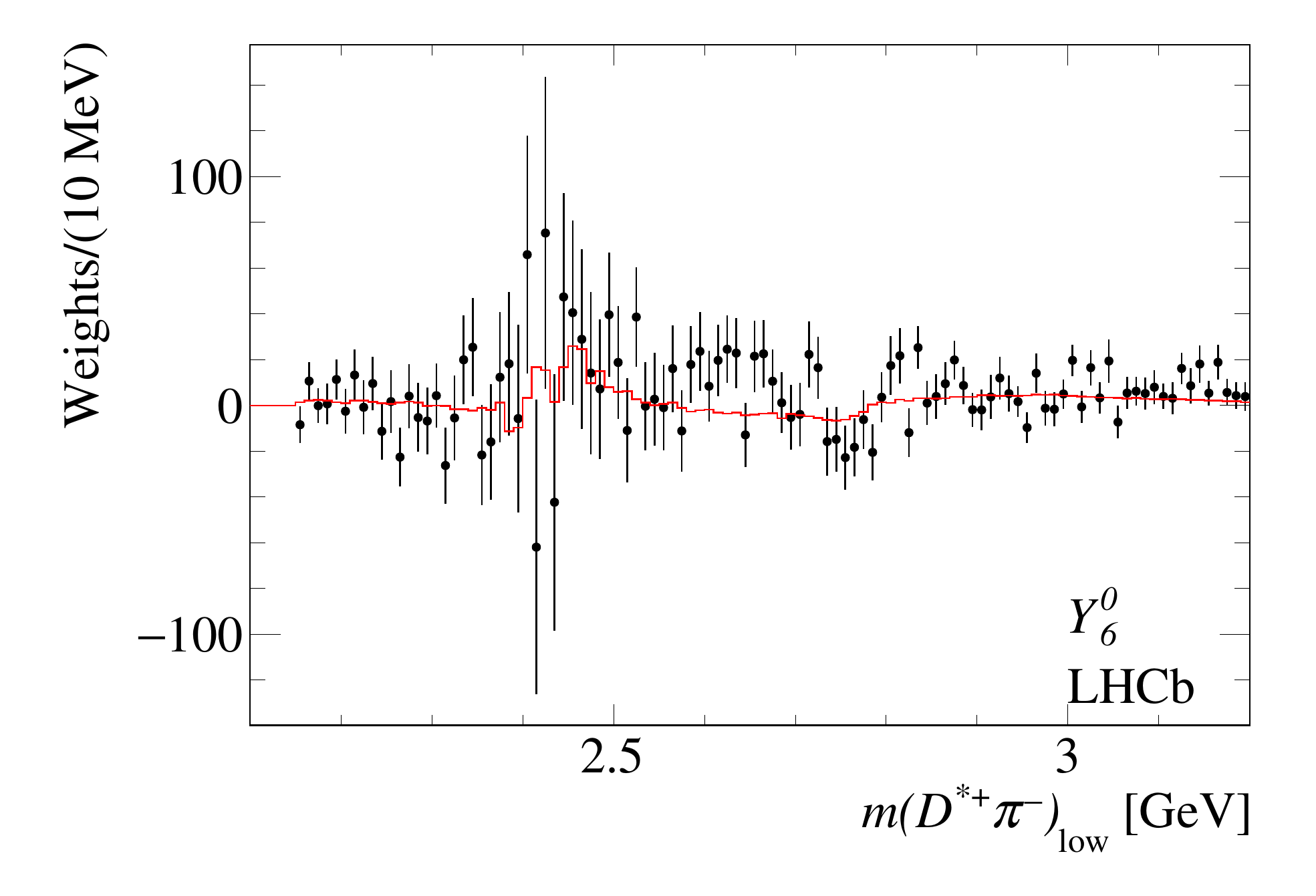}
\caption{\small\label{fig:fig16} Legendre polynomial moments $Y_L^0(\cos \theta)$ as functions of $m(\Dstarp \pim)$ for Run 2 data. The data are represented by filled dots and the superimposed histograms result from the amplitude analysis described in the text.
  }
\end{figure}

A more detailed understanding of the resonant structures present in the $\Dstarp \pim$ mass spectrum and of the agreement between data and fitting model is obtained by looking at the angular
distributions as functions of $\cos \theta$, $\cos \theta_H$ and $\cos \gamma$.
This is obtained by weighting the $\Dstarp \pim$ mass spectrum by the Legendre polynomial moments computed as functions of the above three angles. The $\Dstarp \pim$ mass spectrum weighted by Legendre polynomial moments expressed as functions of $\cos \theta$ is shown in Fig.~\ref{fig:fig16} for $L$ between 1 and 6 and reveals a rich structure. Higher moments are consistent with zero.
 Equations~(\ref{eq:leg}) relate the moments with orbital angular momentum between the \Dstarp and \pim mesons, assuming only partial waves between $L=0$ and $L=3$. Here $S$, $P$, $D$ and $F$ indicate the magnitudes of the amplitudes with angular momenta $L=0,\ 1,\ 2,\ 3$  and $\phi$ denotes their relative phases.
\begin{equation}
\begin{split}
  \sqrt{4 \pi}\langle Y_0^0\rangle& =S^2 + P^2 + D^2 + F^2\\
 \sqrt{4 \pi}\langle Y_1^0\rangle & = 2SP\cos\phi_{SP} + 1.789PD\cos\phi_{PD} + 1.757DF\cos\phi_{DF}\\
 \sqrt{4 \pi}\langle Y_2^0\rangle & = 2SD\cos\phi_{SD} + 0.894 P^2 + 1.757PF\cos\phi_{PF} + 0.639 D^2 + 0.596 F^2\\
 \sqrt{4 \pi}\langle Y_3^0\rangle & = 2SF\cos\phi_{SF} + 1.757PD\cos\phi_{PD} + 1.193DF\cos\phi_{DF}\\ 
 \sqrt{4 \pi}\langle Y_4^0\rangle & = 1.746PF\cos\phi_{PF} +  0.857D^2 + 0.545F^2\\
 \sqrt{4 \pi}\langle Y_5^0\rangle & = 1.699DF\cos\phi_{DF}\\
 \sqrt{4 \pi}\langle Y_6^0\rangle & = 0.840 F^2 \\
\end{split}
\label{eq:leg}
\end{equation}

A comparison with Table~\ref{tab:tab1} allows for the identification of the resonant contributions to each distribution, listed in Table~\ref{tab:tab_ylm}.

\begin{table} [tb]
 \caption{\small Relationship between the Legendre polynomial moments $Y_L^0$ and spin amplitudes. In the column describing the interfering amplitudes, the left side amplitude
 is intended to interfere with any of the amplitudes listed on the right side.}
  \label{tab:tab_ylm}
 \begin{center}
   \begin{tabular}{c|l|ll}
     \hline
  Moment & Squared amplitudes & Interfering amplitudes  \cr
\hline
$Y^0_1$ & & $1^+S$ & $0^-,1^-,2^-P$ \cr
&  & $0^-$ & $1^+D, 2^+$ \cr
&  & $1^-$ & $1^+D$ \cr
&  & $1^+D$& $2^-P,2^-F,3^-$ \cr
&  & $2^+$ & $3^-,2^-F$ \cr
\hline
$Y^0_2$ & $0^-,1^-,2^-P,1^+D,2^+,3^-$ & $1^+S$ & $1^+D,2^+$ \cr
&                           & $3^-$ & $0^-,1^-,2^-P$ \cr
&                           & $2^-F$ & $0^-,1^-,2^-P$ \cr
\hline
$Y_3^0$ &  & $1^+S$ & $3^-,2^-F$ \cr
&  & $1^+D$ & $0^-,1^-,2^-P,3^-,2^-F$ \cr
&  & $2^+$ & $0^+,1^-,3^-,2^-F$ \cr
\hline
$Y_4^0$ & $1^+D, 2^+$ &   $3^-$ & $0^-,2^-$ \cr
&             & $2^-F$ &  $0^-,1^-,2^-P$ \cr
\hline
$Y_5^0$ & & $3^-$ & $1^+D,2^+$ \cr
& & $2^-F$ & $1^+D,2^+$ \cr
\hline
$Y_6^0$ & $3^-,2^-F$ & & \cr
\hline
\end{tabular}
\end{center}
\end{table}

Significant interference effects between $1^+$ amplitudes can be observed in the $Y_2^0$ distribution and a clean $2^+$ signal due to \Dstartwo can be seen in the $Y_4^0$
distribution.
Other moments show rather complex structures. An overall good description of the data is obtained, although some small discrepancy can be seen in $Y_3^0$ and $Y_5^0$. This is expected, given the large number of physical contributions (see Table~\ref{tab:tab_ylm}) to the shape of the $Y_L^0$ moments which are also sensitive to efficiency effects.

\begin{figure}[tb]
\centering
\small
\includegraphics[width=0.49\textwidth]{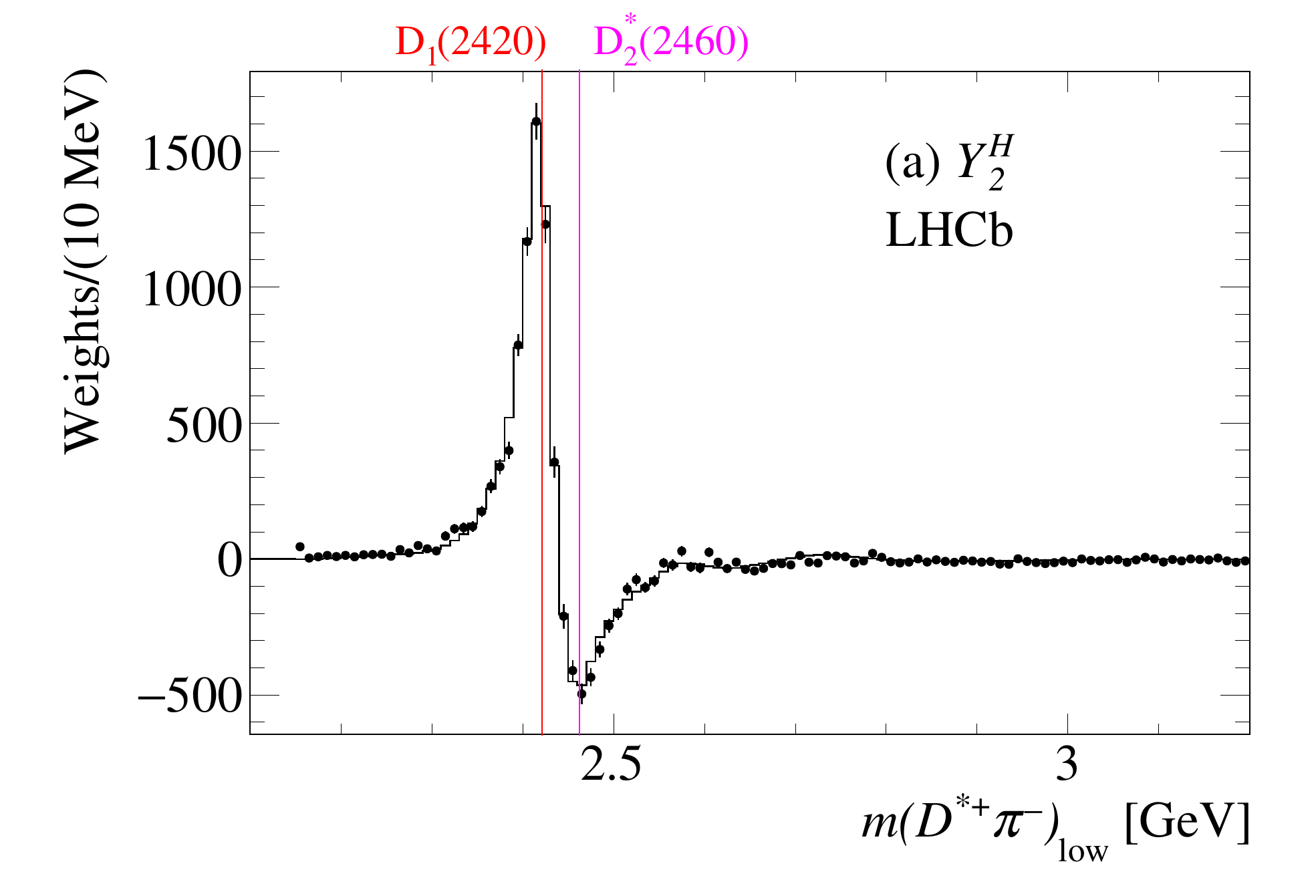}
\includegraphics[width=0.49\textwidth]{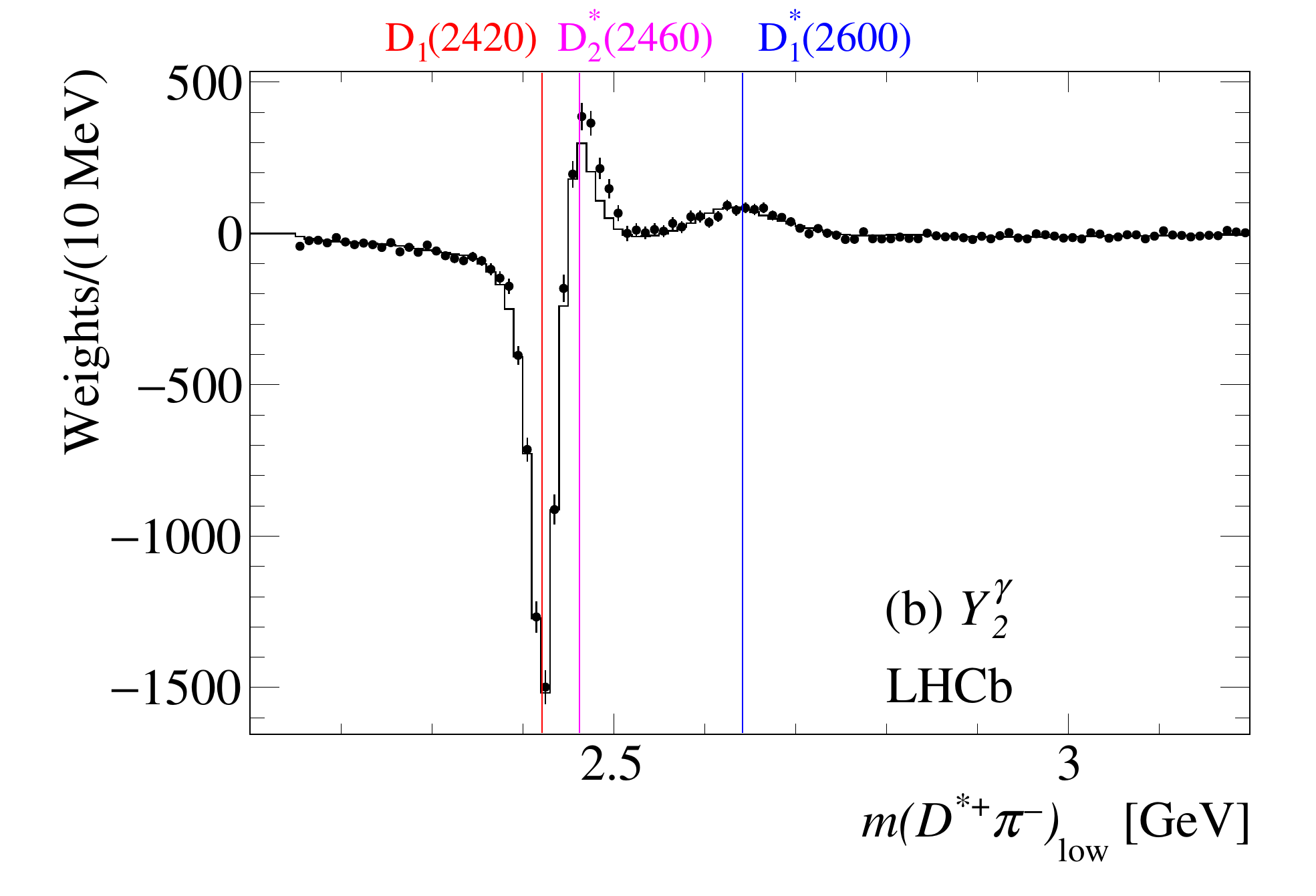}
\caption{\small\label{fig:fig17}
Distributions of the Legendre polynomial moments (a) $Y_2^H (\cos \theta_H)$ and (b) $Y_2^{\gamma}(\cos \gamma)$ as a function of $m(\Dstarp \pim)$ for Run 2 data.
The data are represented by filled dots while the superimposed histograms result from the amplitude analysis described in the text. The vertical lines indicate
the positions of the \Done, \Dstartwo and \Donem resonances.
  }
\end{figure}

Additional information can be obtained from the fit projections on the $\Dstarp \pim$ mass spectrum weighted by Legendre polynomial moments computed as functions
of $\cos \theta_H$ (labeled as $Y^H_L$) and $\cos \gamma$ (labeled as $Y^{\gamma}_L$), shown in Fig.~\ref{fig:fig17}(a) and Fig.~\ref{fig:fig17}(b) respectively.
The two $Y^H_2(\cos \theta_H)$ and $Y^{\gamma}_2(\cos \gamma)$ projections show interference effects between the \Done and \Dstartwo resonances.
The $Y^{\gamma}_2$ distribution also shows an enhancement at the position of the \Donem resonance. Other moments are consistent with zero.

\subsubsection{Search for additional contributions and spin-parity determination}

The presence of additional contributions is tested by adding them to the reference fit using the total dataset. The significance of each contribution is computed as its fitted fraction divided by its statistical uncertainty.
No evidence is found for the $D^*_1(2760)$ or $D^*_2(3000)$ contributions, previously observed in the $\Bm \to \Dp \pim \pim$ decay~\cite{Aaij:2016fma}. Their statistical significance is found to be 2.4 and 0.0, respectively. Virtual contributions, as described in Ref.~\cite{Aaij:2016fma}, are found to be small
with a statistical significance of $4.4\sigma$ but ignored because they have a small fraction ($0.12 \pm 0.03$)\%, an uncertain physical meaning~\cite{Aaij:2016fma} and do not significantly improve the fit $\chi^2$.

The presence of a $D^0 \pi^+ \pi^- \pi^-$ nonresonant contribution has been tested but excluded from the final fit. Its effect, due to the presence of broad $J^P=1^+$ resonances, is to produce large interference effects so that the total fraction increases to large and rather unphysical values without significantly improving the fit quality. 

It has been noted that the QMI $1^+S$ amplitude (Fig.~\ref{fig:fig9}) shows activity both in amplitude and phase in the mass region around 2.8\gev which could correspond to the presence
of an additional $D_1$ resonance. 
A test is performed including an additional $1^+S$ Breit--Wigner resonance in this mass region with free parameters. However, no significant contribution for this additional state is found.

The QMI approach is used for the most significant amplitudes and Breit--Wigner behavior is obtained for $J^P=1^+$, $J^P=2^+$, $J^P=0^-$ and $J^P=1^-$ resonances. For other contributions, such as the $J^P=2^-$ \Dtwom or $J^P=3^-$ \Dthree resonances, this is not possible due to the weakness of these contributions. For these two states, a spin analysis is performed.
For each state additional fits are performed where the masses and widths are fixed to the results given in Table~\ref{tab:tab2} but where the angular
distributions are replaced by those from other possible spin assignments. For the \Dtwom resonance, $J^P=0^-,\ 1^+D,\ 1^+S,\ 1^-,\ 2^+$ are tried but the likelihood and $\chi^2$ variations exclude all the alternative hypotheses with significances greater than $8 \sigma$.
The estimate of the significance is obtained using $\sqrt{\Delta \chi^2}$, where $\Delta \chi^2$ is the variation of the fit $\chi^2$ for the given spin hypothesis.
Similarly, for the \Dthree resonance, values of $J^P=0^-,1^+D,1^+S,1^-$ and $2^+$ are tried but excluded with significances greater than $8 \sigma$.
In conclusion, the present analysis measures the resonance parameters and establishes the quantum numbers of the \Dzero, \Donem, \Dtwom and \Dthree resonances.
The fitted parameters are compared with those measured by other analyses or other experiments in Table~\ref{tab:tab4}.
Note that different methods have been used to extract the resonances parameters. The results from the \babar~\cite{delAmoSanchez:2010vq} and \lhcb~\cite{Aaij:2013sza} collaborations come from inclusive studies of the $\Dp \pim$, $\Dz \pip$ and $\Dstarp \pim$ combinations where signals are fitted directly on the mass spectra. In the case of the $\Dstarp \pim$ mass spectrum, resonance production is enhanced by the use of selections on the helicity angle $\theta_H$. Cross feeds from the resonance production in the $\Dstarp \pim$ system are present in the $\Dp \pim$ and $\Dz \pip$ mass spectra. The \lhcb results from Ref.~\cite{Aaij:2015sqa} and \cite{Aaij:2016fma}, on the other hand, come from Dalitz plot analyses of $B$ decays.
\begin{table*}[t]
\caption{Comparison of the resonance parameters measured in the present work with previous measurements. The first uncertainty is statistical, the second systematic.
\label{tab:tab4}}
\begin{tabular}{l | c | l | r@{}c@{}l | r@{}c@{}l | l}
Resonance & $J^P$ & Decays & \multicolumn{3}{c|}{Mass [MeV]} & \multicolumn{3}{c|}{Width [MeV]} & References  \cr
\hline
$D_0(2550)^0$   & $0^-$  & $D^{*+}\pi^-$ & 2518   $\pm$ & \, 2   \, & $\pm$ 7  & 199  $\pm$ & \, 5   \, & $\pm$  17  & This work \cr
$D_J(2550)^0$   &        & $D^{*+}\pi^-$ & 2539.4 $\pm$ & \, 4.5 \, & $\pm$ 6.8 & 130  $\pm$ & \, 12  \, & $\pm$  13    & \babar \cite{delAmoSanchez:2010vq} \cr
$D_J(2580)^0$   &        & $D^{*+}\pi^-$ & 2579.5 $\pm$ & \, 3.4 \, & $\pm$ 3.5 & 177.5 $\pm$ & \, 17.8 \, & $\pm$ 46.0 &   \lhcb \cite{Aaij:2013sza} \cr
\hline
$D^*_1(2600)^0$ & $1^-$ & $D^{*+}\pi^-$  &  2641.9 $\pm$ & \, 1.8 \, & $\pm$ 4.5 & 149  $\pm$ & \, 4  \, & $\pm$ 20 & This work \cr
$D_J^*(2600)^0$ &       & $D^+\pi^-$    &  2608.7 $\pm$ & \, 2.4 \, & $\pm$ 2.5 &  93 $\pm$ & \, 6  \, & $\pm$ 13 & \babar \cite{delAmoSanchez:2010vq} \cr
$D_J^*(2650)^0$ &       & $D^{*+}\pi^-$  &  2649.2 $\pm$ & \, 3.5 \, & $\pm$ 3.5 & 140.2 $\pm$ & \, 17.1 \, & $\pm$ 18.6 &  \lhcb \cite{Aaij:2013sza} \cr
$D_1^*(2680)^0$ &       & $D^+\pi^-$  &  2681.1 $\pm$ & \, 5.6 \, & $\pm$ 4.9 & 186.7 $\pm$ & \, 8.5 \, & $\pm$ 8.6 &  \lhcb \cite{Aaij:2016fma}\cr
\hline
$D_2(2740)^0$   & $2^-$ & $D^{*+}\pi^-$ & 2751  $\pm$ & \, 3   \, & $\pm$ 7  & 102  $\pm$ & \, 6   \, & $\pm$  26  & This work \cr
$D_J(2750)^0$   &       & $D^{*+}\pi^-$ & 2752.4 $\pm$ & \, 1.7 \, & $\pm$ 2.7 & 71 $\pm$ & \, 6   \, & $\pm$ 11  & \babar \cite{delAmoSanchez:2010vq} \cr
$D_J(2740)^0$   &       & $D^{*+}\pi^-$ & 2737.0 $\pm$ & \, 3.5 \, & $\pm$ 11.2 & 73.2 $\pm$ & \, 13.4   \, & $\pm$ 25.0 & \lhcb \cite{Aaij:2013sza}  \cr
\hline
$D^*_3(2750)^0$ & $3^-$ & $D^{*+}\pi^-$ & 2753  $\pm$ & \, 4   \, & $\pm$ 6   & 66   $\pm$ & \, 10  \, & $\pm$  14  & This work \cr
$D_J^*(2760)^0$	&      & $D^{*+}\pi^-$ & 2761.1 $\pm$ & \, 5.1 \, & $\pm$ 6.5 & 74.4 $\pm$ & \, 3.4  \, & $\pm$ 37.0 & \lhcb \cite{Aaij:2013sza}\cr
                &      & $D^{+}\pi^-$ & 2760.1 $\pm$  & \, 1.1 \, & $\pm$ 3.7 & 74.4 $\pm$ & \, 3.4 \, & $\pm$ 19.1 & \lhcb \cite{Aaij:2013sza}\cr
                &      & $D^+\pi^-$  & 2763.3 $\pm$  & \, 2.3  \, & $\pm$ 2.3 & 60.9 $\pm$ & \, 5.1 \, & $\pm$ 3.6 & \babar \cite{delAmoSanchez:2010vq} \cr
$D_J^*(2760)^+$ &       & $D^0\pi^+$  & 2771.7 $\pm$ & \, 1.7   \, & $\pm$ 3.8 & 66.7 $\pm$ & \, 6.6 \, & $\pm$ 10.5 &  \lhcb \cite{Aaij:2013sza} \cr
$D_3^*(2760)^+$ & $3^-$ & $D^0\pi^-$ & 2798 $\pm$ & \, 7 \, & $\pm$ 1 & 105 $\pm$ & \, 18  \, & $\pm$ 6  & LHCb \cite{Aaij:2015sqa}\cr
$D_3^*(2760)^0$ & $3^-$ & $D^+\pi^-$ & 2775.5 $\pm$ & \, 4.5 \, & $\pm$ 4.5 & 95.3 $\pm$ & \, 9.6  \, & $\pm$ 7.9  & \lhcb \cite{Aaij:2016fma}\cr

\hline
\end{tabular}
\end{table*}

\subsection{Results from the Breit--Wigner model}

Table~\ref{tab:tab2} gives the resonance parameters for the \Donew and \Dzero states when they are described by relativistic Breit--Wigner functions. An amplitude analysis performed using this model gives the results shown in Table~\ref{tab:tab5}. In this case $\chi^2/{\rm ndf}=2348/1748=1.34$
and $\chi^2/{\rm ndf}=2205/1780=1.24$ for Run 1 and Run 2 data, respectively.
Somewhat reduced fractional contributions from the \Donew and \Dzero resonances with respect to the QMI approach can be seen. This effect can be
understood since in this model the $J^P= 1^+S$ and $J^P= 0^-$
contributions do not include possible additional contributions from higher mass resonances.
Systematic uncertainties are evaluated as described in Sec.~\ref{sec:qmi}.

\begin{table} [t]
\centering
\caption{\small Fit results from the amplitude analysis for the model where the \Donew and \Dzero resonances are described by relativistic Breit--Wigner functions. The first uncertainty is statistical, the second systematic.
}
\begin{tabular}{r | r | r@{$\:\pm\:$}l@{$\:\pm\:$}l | r@{$\:\pm\:$}c@{$\:\pm\:$}l}
Resonance & $J^P$ & \multicolumn{3}{c|}{fraction (\%)} & \multicolumn{3}{c}{phase (rad)} \cr
\hline
\Done & $1^+D$ & 56.5\phantom{0}    &  0.3   &  1.1  & \multicolumn{1}{r}{$0\phantom{.00\pm}$} \cr
\Donew & $ 1^+S$ & 26.0\phantom{0}   &  0.4   &  1.7  & $-$1.57   & 0.02   & 0.08 \cr
\Dstartwo & $2^+$ & 15.4\phantom{0}    &  0.2   &  0.1  & $-$0.77   & 0.01   & 0.01 \cr 
\Done & $1^+S$ & 5.9\phantom{0}     &  0.5   &  2.9  &  1.69   & 0.02   & 0.06  \cr 
\Dzero & $ 0^-$ &  5.3\phantom{0}     &  0.1   &  0.5  &  1.50   & 0.02   & 0.06  \cr
\Donem & $1^-$ &  5.0\phantom{0}    &  0.1   &  0.5  &  0.76   & 0.02   & 0.03  \cr
\Dtwom & $2^-P$ &  0.57    &  0.07  &  0.23  & $-$2.14   & 0.07   & 0.16  \cr
\Dtwom & $2^-F$ &  1.9\phantom{0}     &  0.1   &  1.0  &  0.49   & 0.04   & 0.40  \cr
\Dthree & $3^-$ & 0.78    &  0.06  &  0.13  &  $-$1.54  & 0.05   & 0.04  \cr
\hline
 Sum & &   117.3\phantom{0}   &  0.8  &  3.8 & \multicolumn{3}{c}{} \cr
\hline
\end{tabular}
\label{tab:tab5}
\end{table}

\subsection{Results from the {\boldmath\protect $J^P=1^+$} mixing model}
\label{sec:mix}

A consequence of the heavy-quark symmetry is that, in the infinite-mass heavy quark limit, heavy-light $Q \bar q$ mesons can be classified in
doublets labeled by the value of the total angular momentum $j_q$ of the light degrees of freedom with respect to the heavy quark $Q$~\cite{Isgur:1991wq}. In the quark model  $\boldsymbol{j}_q$ would be given by $\boldsymbol{j}_q=\boldsymbol{s}_q + \boldsymbol{L}$, where $\boldsymbol{L}$ is the light quark orbital angular momentum.
The Heavy Quark Effective Theory predicts that the two $J^P=1^+$ mesons, with $j_q=\frac{1}{2}$ and $j_q=\frac{3}{2}$, decay into the $D^*\pi$ final state via the $S$- and $D$-wave, respectively.
Due to the finite $c$-quark mass, the observed physical $1^+$ states can be a mixture of such pure states.
The mixing can occur for instance via the common $D^*\pi$ decay channel and the resulting $D'_1$ and $D_1$ amplitudes are 
a superposition of the $S$- and $D$-wave amplitudes
\begin{equation}
  A^{D_1'} = A^{1S} \cos \omega - A^{1D} \sin \omega e^{i\psi},
\end{equation}
\begin{equation}
  A^{D_1} = A^{1S} \sin \omega + A^{1D} \cos \omega e^{-i\psi},
\end{equation}
\noindent
where $\omega$ is the mixing angle and $\psi$ is a complex phase.

In this model the $J^P=1^+$ $D_1'$ amplitude is taken as reference. The $J^P=1^+S$ amplitudes are described by relativistic Breit--Wigner functions with free parameters, while the $J^P=0^-$ amplitude is described by the QMI model. All the other resonances are described by relativistic
Breit--Wigner functions with parameters fixed to the values reported in Table~\ref{tab:tab2}.
Table~\ref{tab:tab6} gives details on the fractions and relative phases.

  \begin{table} [tb]
\centering
\caption{\small Fit results from the mixing model. The first uncertainty is statistical, the second systematic.
}
\begin{tabular}{r | r | r@{$\:\pm\:$}l@{$\:\pm\:$}l | r@{$\:\pm\:$}c@{$\:\pm\:$}l}
Resonance & $J^P$ & \multicolumn{3}{c|}{fraction (\%)} & \multicolumn{3}{c}{phase (rad)} \cr
\hline
$D_1$           & $1^+$  & 58.9\phantom{0} &  0.7\phantom{0}  &  2.5 &  \multicolumn{1}{r}{$0\phantom{.00\pm}$} \cr
$D_1' $         & $1^+$  & 21.9\phantom{0}   &  2.2\phantom{0}  &  3.0 & $-$1.06  &  0.10  &  0.05 \cr
\Dstartwo    & $2^+$  & 14.0\phantom{0}   &  0.2\phantom{0}  &  0.3 & 2.66  &  0.09  &  0.15 \cr
$ 0^- \ QMI$    & $0^-$  &  6.5\phantom{0}   &  0.2\phantom{0}  &  1.5 &  2.03  &  0.09  &  0.28 \cr
\Donem   & $1^-$  &  4.9\phantom{0}   &  0.1\phantom{0}  &  0.5 & $-$2.24  &  0.09  &  0.11 \cr 
\Dtwom     & $2^-P$ &  0.72  &  0.08  &  0.30 & $-$2.59  &  0.10  &  0.53 \cr
\Dtwom     & $2^-F$ &  2.9\phantom{0}   &  0.2\phantom{0}  &  1.1 &  0.27  &  0.09  &  0.47\cr
\Dthree   & $3^-$  &  0.70  &  0.05  &  0.10 &   1.54  &  0.10  &  0.33 \cr
\hline
Sum & &   110.4\phantom{0}  &  2.3  &  4.4  & \multicolumn{3}{c}{}\cr
\hline
\end{tabular}
\label{tab:tab6}
 \end{table}
 The resulting mixing parameters are
 \begin{equation}
   \omega = -0.063 \pm 0.019 \pm 0.004, \qquad \psi=-0.29 \pm 0.09 \pm 0.07,
 \end{equation}
which are consistent with the results from the Belle collaboration~\cite{Abe:2003zm}
\begin{equation}
   \omega = -0.10 \pm 0.03 \pm 0.02 \pm 0.02, \qquad \psi=0.05 \pm 0.20 \pm 0.04 \pm 0.06.
\end{equation}
Combining statistical and systematic uncertainties the mixing angle deviates from zero by $2.3\sigma$.

The \chisqndf for the fit to the total dataset is $\chi^2/{\rm ndf}=2739/1780=1.54$.
Systematic uncertainties on the mixing parameters, fractional contributions and relative phases are computed as described in Sec.~\ref{sec:qmi}. The measured $D_1$ and $D_1'$ masses and widths are reported in Table~\ref{tab:tab2}.

\section{Measurement of the branching fractions}
\label{sec:br}

The known branching fraction of the $B^- \to \Dstarp \pi^- \pi^-$ decay mode is
\mbox{${\cal{B}} (B^- \to D^{*+} \pi^- \pi^-)=(1.35 \pm 0.22) \times 10^{-3}$}~\cite{Tanabashi:2018oca}.
Table~\ref{tab:tab7} reports the partial branching fractions for the resonances contributing to the total branching fraction.
They are obtained multiplying the \Bm branching fraction by the fractional contributions obtained from the amplitude analysis  performed using the Breit--Wigner model for all the resonances and reported in Table~\ref{tab:tab5}. For the $D_1$ and $D_1'$ branching fractions the fractional contributions
obtained from the mixing model and reported in Table~\ref{tab:tab6} are used.
Since the uncertainty on the absolute branching fraction is large, it has been separated from the other sources of systematic uncertainty.
The \Done resonance decays to $D$- and $S$-wave states and therefore the two contributions are
added; a similar procedure is followed for the \Dtwom resonance, which decays to $P$- and $F$-wave states.

\begin{table} [tb]
\centering
\caption{\small Summary of the measurements of the branching fractions. The first uncertainty is statistical, the second systematic, the third is due to the uncertainty on the measurement of the $B^- \to D^{*+} \pi^- \pi^-$ absolute branching fraction. A comparison with measurements obtained by the Belle collaboration~\cite{Abe:2003zm} is shown.}
\begin{tabular}{r | r | r@{}c@{}c@{}l | r@{}c@{}c@{}l}
Resonance & $J^P$ & 
          \multicolumn{8}{c}{${\cal{B}}(B^- \to R^0 \pi^-)\times{\cal{B}}(R^0 \to D^{*+} \pi^-) \times 10^{-4}$}  \cr
\hline
&       &   \multicolumn{4}{c|} {This analysis} & \multicolumn{4}{c}{Belle collaboration} \cr
\hline
\Done & $1^+$   &  8.42 $\pm$ & \,  0.08 \, & $\pm$  0.40 \, & $\pm$ 1.40 & \cr
\Donew & $ 1^+S$ &  3.51 $\pm$ & \,  0.06 \, & $\pm$ 0.23 \, & $\pm$ 0.57 &\cr
\Dstartwo & $2^+$  &  2.08 $\pm$ & \,  0.03 \, & $\pm$  0.14\, & $\pm$ 0.34 & 1.8 $\pm$ & \, 0.3 \, & $\pm$ 0.3  \, & $\pm$ 0.2\cr
\Dzero & $ 0^-$   &  0.72 $\pm$ & \,  0.01 \, & $\pm$ 0.07 \, & $\pm$ 0.12 &  \cr
\Donem & $1^-$  &  0.68 $\pm$ & \,  0.01 \, & $\pm$ 0.07 \, & $\pm$ 0.11 & \cr
\Dtwom & $2^-$    &  0.33 $\pm$ & \,  0.02 \, & $\pm$ 0.14 \, & $\pm$ 0.05 & \cr
\Dthree & $3^-$  &  0.11 $\pm$ & \,  0.01 \, & $\pm$ 0.02 \, & $\pm$ 0.02 & \cr 
\hline
$D_1$ & $1^+$          &  7.95 $\pm$ & \,  0.09 \, & $\pm$ 0.34 \, & $\pm$ 1.30  & 6.8 $\pm$ & \, 0.7 \, & $\pm$ 1.3  \, & $\pm$ 0.3 \cr
$D_1' $ & $1^+$         & 2.96 $\pm$ & \,  0.30 \, & $\pm$ 0.41 \, & $\pm$ 0.48 & 5.0 $\pm$ & \, 0.4 \, & $\pm$ 1.0  \,  & $\pm$ 0.4 \cr
\hline
\end{tabular}
\label{tab:tab7}
\end{table}

\section{Summary}
\label{sec:sum}

A four-body amplitude analysis of the $\Bm \to \Dstarp \pim \pim$ decay is performed using $pp$ collision data, corresponding to an integrated luminosity of 4.7\invfb, 
collected at center-of-mass energies of 7, 8 and 13\tev with the \lhcb detector.
Fractional contributions and relative phases
for the different resonances contributing in the decay are measured. The data allow for several quasi-model-independent searches
for the presence of new states. For the first time, the quantum numbers of previously observed charmed meson resonances are established.
In particular the resonance parameters, quantum numbers and partial branching fractions are measured for the \Done, \Donew, \Dzero, \Donem, \Dtwom and \Dthree resonances. The $J^P=1^+S$ and $J^P=0^-$ QMI amplitudes give indications for the presence of higher mass $D_1$ and $D_0'$ resonances
in the 2.80\gev mass region.
The data are fitted allowing for mixing between $D_1$ and $D_1'$ resonances and their mixing parameters are measured. In particular, the mixing angle deviates from zero by~$2.3\sigma$.

\input{acknowledgements}

\addcontentsline{toc}{section}{References}
\bibliographystyle{LHCb}
\bibliography{LHCb-PAPER-2019-027,LHCb-PAPER}
 
\newpage

\input{LHCb_Authorship_04-Jun-2019.tex}
\end{document}

%% file: preamble.tex
\usepackage[top=1in, bottom=1.25in, left=1in, right=1in]{geometry}

\usepackage{microtype}
\usepackage{lineno}  
\usepackage{xspace} 
\usepackage{caption} 

\usepackage{graphicx}  
\usepackage{color}
\usepackage{colortbl}
\graphicspath{{./figs/}} 
\DeclareGraphicsExtensions{.pdf,.PDF,png,.PNG}

\usepackage{amsmath} 
\usepackage{amssymb}
\usepackage{amsfonts}
\usepackage{upgreek} 

\newcommand*\patchAmsMathEnvironmentForLineno[1]{%
\expandafter\let\csname old#1\expandafter\endcsname\csname #1\endcsname
\expandafter\let\csname oldend#1\expandafter\endcsname\csname
end#1\endcsname
 \renewenvironment{#1}%
   {\linenomath\csname old#1\endcsname}%
   {\csname oldend#1\endcsname\endlinenomath}%
}
\newcommand*\patchBothAmsMathEnvironmentsForLineno[1]{%
  \patchAmsMathEnvironmentForLineno{#1}%
  \patchAmsMathEnvironmentForLineno{#1*}%
}
\AtBeginDocument{%
\patchBothAmsMathEnvironmentsForLineno{equation}%
\patchBothAmsMathEnvironmentsForLineno{align}%
\patchBothAmsMathEnvironmentsForLineno{flalign}%
\patchBothAmsMathEnvironmentsForLineno{alignat}%
\patchBothAmsMathEnvironmentsForLineno{gather}%
\patchBothAmsMathEnvironmentsForLineno{multline}%
\patchBothAmsMathEnvironmentsForLineno{eqnarray}%
}

\usepackage{hyperxmp}

\usepackage[pdftex,
            pdfauthor={\paperauthors},
            pdftitle={\paperasciititle},
            pdfkeywords={\paperkeywords},
            pdfcopyright={Copyright (C) \papercopyright},
            pdflicenseurl={\paperlicenceurl}]{hyperref}

\usepackage[colorinlistoftodos,textsize=scriptsize]{todonotes}

\usepackage[all]{hypcap} 

\input{lhcb-symbols-def} 

\usepackage{cite} 
\usepackage{mciteplus}

%% file: lhcb-symbols-def.tex
\usepackage{xspace} 
\usepackage{upgreek}

\def\lhcb   {\mbox{LHCb}\xspace}

\def\babar  {\mbox{BaBar}\xspace}
\def\belle  {\mbox{Belle}\xspace}

\def\MagUp {\mbox{\em Mag\kern -0.05em Up}\xspace}

\ifthenelse{\boolean{uprightparticles}}%
{

 \def\Pmu         {\ensuremath{\upmu}\xspace}

 \def\Ppi         {\ensuremath{\uppi}\xspace}

 \def\Ppsi        {\ensuremath{\uppsi}\xspace}

 \def\PDelta      {\ensuremath{\Delta}\xspace}                 
 \def\PXi         {\ensuremath{\Xi}\xspace}                 
 \def\PLambda     {\ensuremath{\Lambda}\xspace}                 
 \def\PSigma      {\ensuremath{\Sigma}\xspace}                 
 \def\POmega      {\ensuremath{\Omega}\xspace}                 
 \def\PUpsilon    {\ensuremath{\Upsilon}\xspace}

 \def\PB      {\ensuremath{\mathrm{B}}\xspace}                 
                  
 \def\PD      {\ensuremath{\mathrm{D}}\xspace}

 \def\PJ      {\ensuremath{\mathrm{J}}\xspace}                 
 \def\PK      {\ensuremath{\mathrm{K}}\xspace}

 \def\Pb      {\ensuremath{\mathrm{b}}\xspace}                 
 \def\Pc      {\ensuremath{\mathrm{c}}\xspace}                 
                  
 \def\Pe      {\ensuremath{\mathrm{e}}\xspace}

 \def\Pi      {\ensuremath{\mathrm{i}}\xspace}

 \def\Ps      {\ensuremath{\mathrm{s}}\xspace}

 \def\thebaroffset{0.0em}
}
{

 \def\Pmu         {\ensuremath{\mu}\xspace}

 \def\Ppi         {\ensuremath{\pi}\xspace}

 \def\Ppsi        {\ensuremath{\psi}\xspace}                 
                  
 \mathchardef\PDelta="7101
 \mathchardef\PXi="7104
 \mathchardef\PLambda="7103
 \mathchardef\PSigma="7106
 \mathchardef\POmega="710A
 \mathchardef\PUpsilon="7107
                  
 \def\PB      {\ensuremath{B}\xspace}                 
                  
 \def\PD      {\ensuremath{D}\xspace}

 \def\PJ      {\ensuremath{J}\xspace}                 
 \def\PK      {\ensuremath{K}\xspace}

 \def\Pb      {\ensuremath{b}\xspace}                 
 \def\Pc      {\ensuremath{c}\xspace}                 
                  
 \def\Pe      {\ensuremath{e}\xspace}

 \def\Pi      {\ensuremath{i}\xspace}

 \def\Ps      {\ensuremath{s}\xspace}

 \def\thebaroffset{0.18em}
}
\newcommand{\offsetoverline}[2][\thebaroffset]{\kern #1\overline{\kern -#1 #2}}%

\makeatletter
\ifcase \@ptsize \relax
  \newcommand{\miniscule}{\@setfontsize\miniscule{4}{5}}
\or
  \newcommand{\miniscule}{\@setfontsize\miniscule{5}{6}}
\or
  \newcommand{\miniscule}{\@setfontsize\miniscule{5}{6}}
\fi
\makeatother

\DeclareRobustCommand{\optbar}[1]{\shortstack{{\miniscule (\rule[.5ex]{1.25em}{.18mm})}
  \\ [-.7ex] $#1$}}

\def\epem       {{\ensuremath{\Pe^+\Pe^-}}\xspace}

\def\mumu       {{\ensuremath{\Pmu^+\Pmu^-}}\xspace}

\def\squark    {{\ensuremath{\Ps}}\xspace}

\def\cquark    {{\ensuremath{\Pc}}\xspace}

\def\bquark    {{\ensuremath{\Pb}}\xspace}

\def\pion   {{\ensuremath{\Ppi}}\xspace}
\def\piz    {{\ensuremath{\pion^0}}\xspace}
\def\pip    {{\ensuremath{\pion^+}}\xspace}
\def\pim    {{\ensuremath{\pion^-}}\xspace}

\def\kaon    {{\ensuremath{\PK}}\xspace}

\def\KorKbar {\kern \thebaroffset\optbar{\kern -\thebaroffset \PK}{}\xspace}

\def\Kp      {{\ensuremath{\kaon^+}}\xspace}
\def\Km      {{\ensuremath{\kaon^-}}\xspace}

\def\KS      {{\ensuremath{\kaon^0_{\mathrm{S}}}}\xspace}

\def\Dbar    {{\ensuremath{\offsetoverline{\PD}}}\xspace}
\def\D       {{\ensuremath{\PD}}\xspace}

\def\DorDbar {\kern \thebaroffset\optbar{\kern -\thebaroffset \PD}\xspace}
\def\Dz      {{\ensuremath{\D^0}}\xspace}
\def\Dzb     {{\ensuremath{\Dbar{}^0}}\xspace}
\def\Dp      {{\ensuremath{\D^+}}\xspace}
\def\Dm      {{\ensuremath{\D^-}}\xspace}

\def\DpDm    {\ensuremath{\Dp {\kern -0.16em \Dm}}\xspace}

\def\Dstarp  {{\ensuremath{\D^{*+}}}\xspace}

\def\B       {{\ensuremath{\PB}}\xspace}
\def\Bbar    {{\ensuremath{\offsetoverline{\PB}}}\xspace}

\def\BorBbar {\kern \thebaroffset\optbar{\kern -\thebaroffset \PB}\xspace}
\def\Bz      {{\ensuremath{\B^0}}\xspace}
\def\Bzb     {{\ensuremath{\Bbar{}^0}}\xspace}
\def\Bd      {{\ensuremath{\B^0}}\xspace}

\def\BdorBdbar {\kern \thebaroffset\optbar{\kern -\thebaroffset \Bd}\xspace}
\def\Bu      {{\ensuremath{\B^+}}\xspace}
\def\Bub     {{\ensuremath{\B^-}}\xspace}

\def\Bm      {{\ensuremath{\Bub}}\xspace}

\def\Bs      {{\ensuremath{\B^0_\squark}}\xspace}

\def\BsorBsbar {\kern \thebaroffset\optbar{\kern -\thebaroffset \Bs}\xspace}

\def\jpsi     {{\ensuremath{{\PJ\mskip -3mu/\mskip -2mu\Ppsi\mskip 2mu}}}\xspace}

\def\Y#1S{\ensuremath{\PUpsilon{(#1S)}}\xspace}

\def\LorLbar     {\kern \thebaroffset\optbar{\kern -\thebaroffset \PLambda}\xspace}

\newcommand{\decay}[2]{\ensuremath{#1\!\to #2}\xspace} 

\def\to                 {\ensuremath{\rightarrow}\xspace}

\def\AT#1     {\ensuremath{A_{\mathrm{T}}^{#1}}\xspace}           

\def\C#1      {\ensuremath{\mathcal{C}_{#1}}\xspace}                       
\def\Cp#1     {\ensuremath{\mathcal{C}_{#1}^{'}}\xspace}                    
\def\Ceff#1   {\ensuremath{\mathcal{C}_{#1}^{\mathrm{(eff)}}}\xspace}        
\def\Cpeff#1  {\ensuremath{\mathcal{C}_{#1}^{'\mathrm{(eff)}}}\xspace}       
\def\Ope#1    {\ensuremath{\mathcal{O}_{#1}}\xspace}                       
\def\Opep#1   {\ensuremath{\mathcal{O}_{#1}^{'}}\xspace}                    

\newcommand{\nospaceunit}[1]{\ensuremath{\text{#1}}}       
\newcommand{\aunit}[1]{\ensuremath{\text{\,#1}}}       

\newcommand{\tev}{\aunit{Te\kern -0.1em V}\xspace}
\newcommand{\gev}{\aunit{Ge\kern -0.1em V}\xspace}
\newcommand{\mev}{\aunit{Me\kern -0.1em V}\xspace}
\newcommand{\kev}{\aunit{ke\kern -0.1em V}\xspace}
\newcommand{\ev}{\aunit{e\kern -0.1em V}\xspace}
 
\newcommand{\mevc}{\ensuremath{\aunit{Me\kern -0.1em V\!/}c}\xspace}
\newcommand{\gevc}{\ensuremath{\aunit{Ge\kern -0.1em V\!/}c}\xspace}
\newcommand{\mevcc}{\ensuremath{\aunit{Me\kern -0.1em V\!/}c^2}\xspace}
\newcommand{\gevcc}{\ensuremath{\aunit{Ge\kern -0.1em V\!/}c^2}\xspace}

\def\mum  {\ensuremath{\,\upmu\nospaceunit{m}}\xspace}

\def\fb   {\ensuremath{\aunit{fb}}\xspace}
\def\invfb   {\ensuremath{\fb^{-1}}\xspace}

\newcommand{\chisq}{\ensuremath{\chi^2}\xspace}
\newcommand{\chisqndf}{\ensuremath{\chi^2/\mathrm{ndf}}\xspace}

\def\gsim{{~\raise.15em\hbox{$>$}\kern-.85em
          \lower.35em\hbox{$\sim$}~}\xspace}
\def\lsim{{~\raise.15em\hbox{$<$}\kern-.85em
          \lower.35em\hbox{$\sim$}~}\xspace}

\newcommand{\Real}{\ensuremath{\mathcal{R}e}\xspace}

\def\pt         {\ensuremath{p_{\mathrm{T}}}\xspace}

\def\evtgen     {\mbox{\textsc{EvtGen}}\xspace}

\def\geant      {\mbox{\textsc{Geant4}}\xspace}

\def\photos     {\mbox{\textsc{Photos}}\xspace}

\def\pythia     {\mbox{\textsc{Pythia}}\xspace}

\def\tell1  {TELL1\xspace}
\def\ukl1   {UKL1\xspace}

\newcommand{\ie}{\mbox{\itshape i.e.}\xspace}

\newcommand{\phz}{\phantom{0}}

%% file: title-LHCb-PAPER.tex

\begin{titlepage}
\pagenumbering{roman}

\vspace*{-1.5cm}
\centerline{\large EUROPEAN ORGANIZATION FOR NUCLEAR RESEARCH (CERN)}
\vspace*{1.5cm}
\noindent
\begin{tabular*}{\linewidth}{lc@{\extracolsep{\fill}}r@{\extracolsep{0pt}}}
\ifthenelse{\boolean{pdflatex}}
{\vspace*{-1.5cm}\mbox{\!\!\!\includegraphics[width=.14\textwidth]{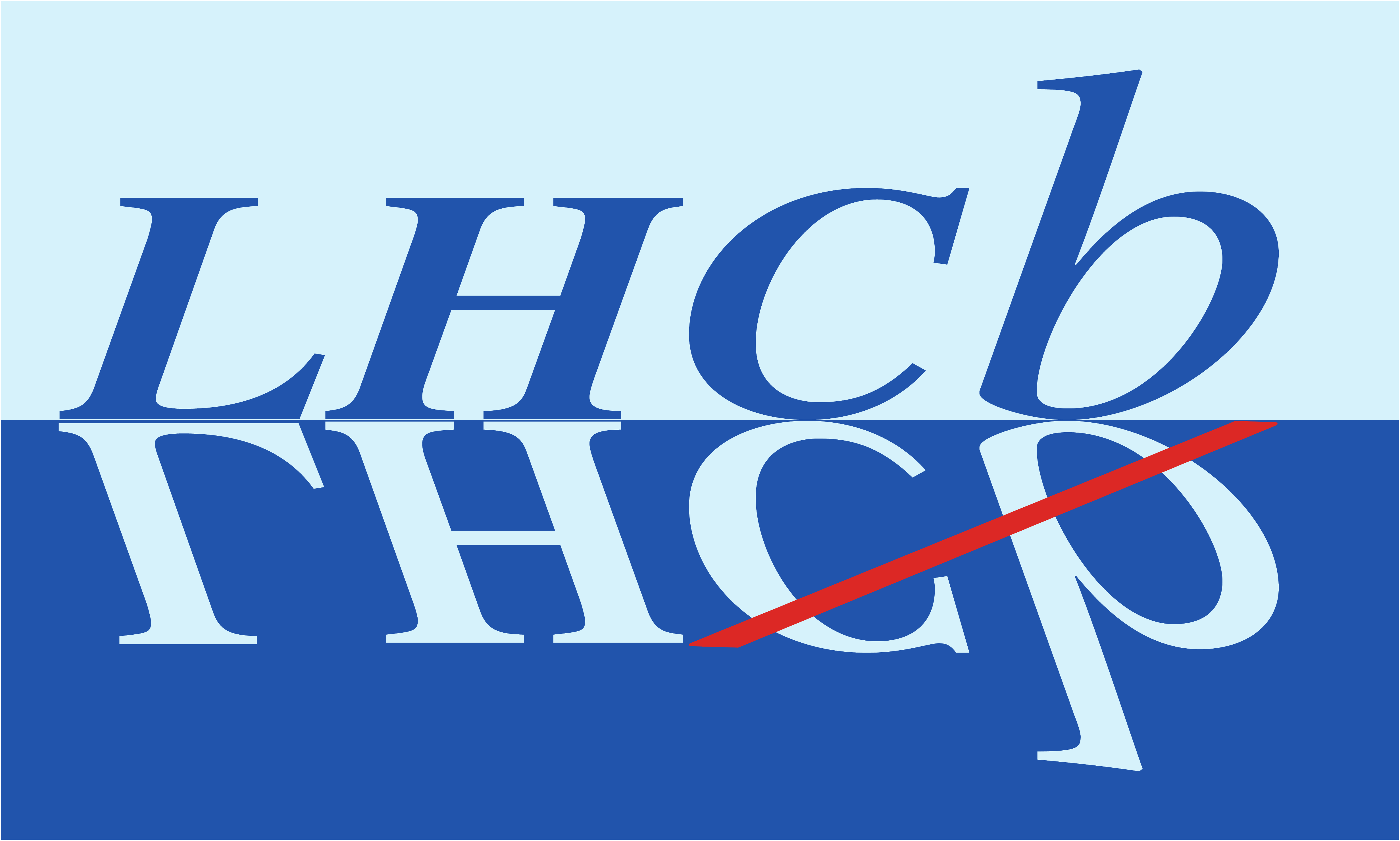}} & &}%
{\vspace*{-1.2cm}\mbox{\!\!\!\includegraphics[width=.12\textwidth]{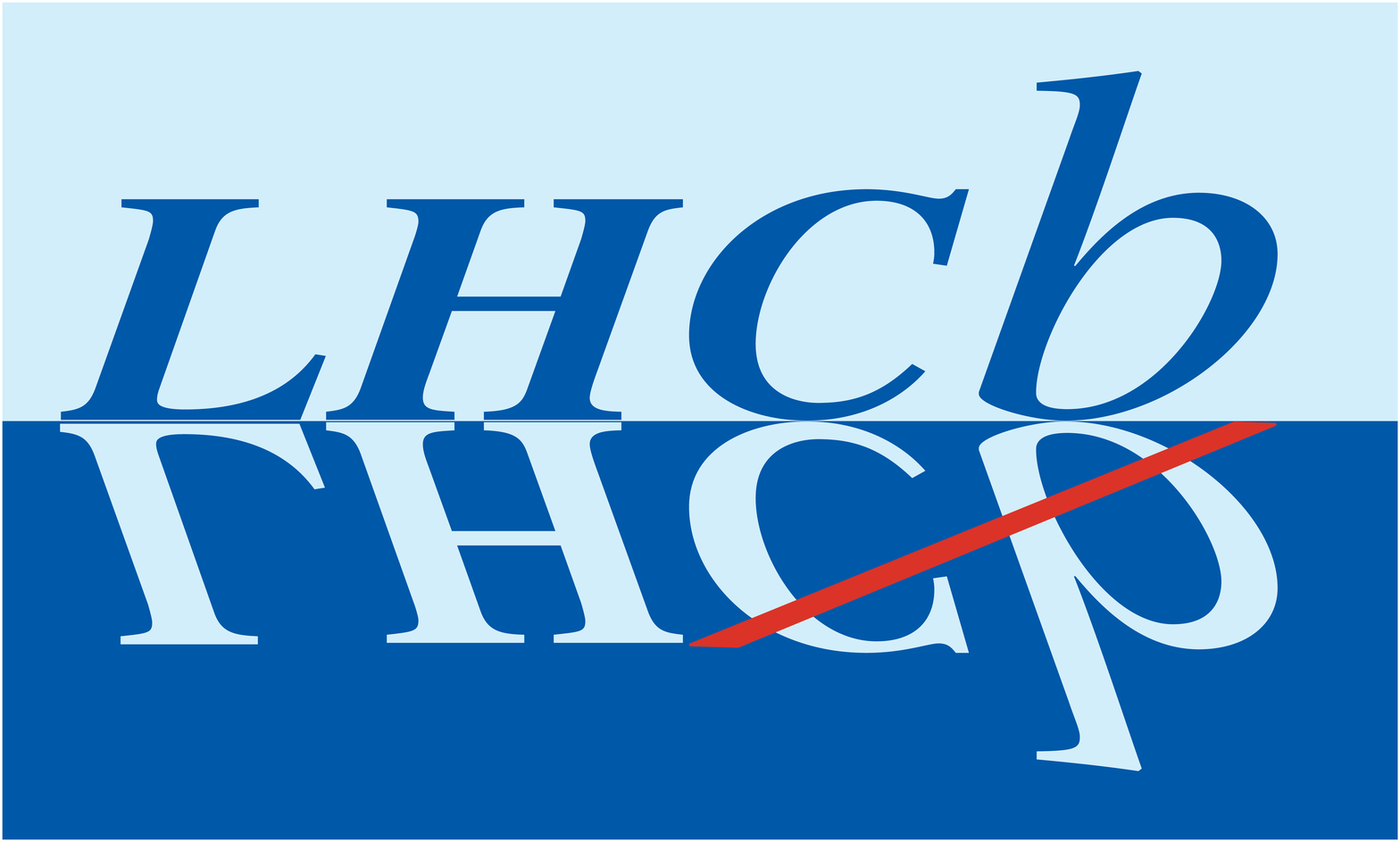}} & &}%
\\
 & & CERN-EP-2019-201 \\  
& & LHCb-PAPER-2019-027 \\  
 & & February 24, 2020 \\
 & & \\
\end{tabular*}

\vspace*{4.0cm}

{\normalfont\bfseries\boldmath\huge
\begin{center}
  \papertitle 
\end{center}
}

\vspace*{1.8cm}

\begin{center}
\paperauthors\footnote{Authors are listed at the end of this paper.}
\end{center}

\vspace{\fill}

\begin{abstract}
  \noindent
 A four-body amplitude analysis of the $\Bm \to \Dstarp \pim \pim$ decay is performed,
where fractions and relative phases
of the various resonances contributing to the decay are measured.
Several quasi-model-independent analyses are performed aimed at searching
for the presence of new states and establishing the quantum numbers of previously observed charmed meson resonances.
In particular the resonance parameters and quantum numbers are determined for the \Done, \Donew, \Dzero, \Donem, \Dtwom and \Dthree states. The mixing between the \Done and \Donew resonances is studied and the mixing parameters are measured.
The dataset corresponds to an integrated luminosity of 4.7\invfb, 
collected in proton-proton collisions at center-of-mass energies of 7, 8 and 13\tev with the \lhcb detector.  
\end{abstract}

\vspace*{2.0cm}

\begin{center}
Published in Phys. Rev. D101 (2019) 032005 
\end{center}

\vspace{\fill}

{\footnotesize 
\centerline{\copyright~\papercopyright. \href{\paperlicenceurl}{\paperlicence}.}}
\vspace*{2mm}

\end{titlepage}


\newpage
\setcounter{page}{2}
\mbox{~}

\cleardoublepage

%% file: acknowledgements.tex
\section*{Acknowledgements}
%
%
\noindent We express our gratitude to our colleagues in the CERN
accelerator departments for the excellent performance of the LHC. We
thank the technical and administrative staff at the LHCb
institutes.
We acknowledge support from CERN and from the national agencies:
CAPES, CNPq, FAPERJ and FINEP (Brazil); 
MOST and NSFC (China); 
CNRS/IN2P3 (France); 
BMBF, DFG and MPG (Germany); 
INFN (Italy); 
NWO (Netherlands); 
MNiSW and NCN (Poland); 
MEN/IFA (Romania); 
MSHE (Russia); 
MinECo (Spain); 
SNSF and SER (Switzerland); 
NASU (Ukraine); 
STFC (United Kingdom); 
DOE NP and NSF (USA).
We acknowledge the computing resources that are provided by CERN, IN2P3
(France), KIT and DESY (Germany), INFN (Italy), SURF (Netherlands),
PIC (Spain), GridPP (United Kingdom), RRCKI and Yandex
LLC (Russia), CSCS (Switzerland), IFIN-HH (Romania), CBPF (Brazil),
PL-GRID (Poland) and OSC (USA).
We are indebted to the communities behind the multiple open-source
software packages on which we depend.
Individual groups or members have received support from
AvH Foundation (Germany);
EPLANET, Marie Sk\l{}odowska-Curie Actions and ERC (European Union);
ANR, Labex P2IO and OCEVU, and R\'{e}gion Auvergne-Rh\^{o}ne-Alpes (France);
Key Research Program of Frontier Sciences of CAS, CAS PIFI, and the Thousand Talents Program (China);
RFBR, RSF and Yandex LLC (Russia);
GVA, XuntaGal and GENCAT (Spain);
the Royal Society
and the Leverhulme Trust (United Kingdom).

%% file: LHCb_Authorship_04-Jun-2019.tex
\centerline
{\large\bf LHCb collaboration}
\begin
{flushleft}
\small
R.~Aaij$^{30}$,
C.~Abell{\'a}n~Beteta$^{47}$,
T.~Ackernley$^{57}$,
B.~Adeva$^{44}$,
M.~Adinolfi$^{51}$,
H.~Afsharnia$^{8}$,
C.A.~Aidala$^{78}$,
S.~Aiola$^{24}$,
Z.~Ajaltouni$^{8}$,
S.~Akar$^{62}$
% ...

{\footnotesize \it

% ...
$ ^{8}$Universit{\'e} Clermont Auvergne, CNRS/IN2P3, LPC, Clermont-Ferrand, France\\
% ...
$ ^{24}$INFN Sezione di Milano, Milano, Italy\\
% ...
$ ^{30}$Nikhef National Institute for Subatomic Physics, Amsterdam, Netherlands\\
% ...
$ ^{44}$Instituto Galego de F{\'\i}sica de Altas Enerx{\'\i}as (IGFAE), Universidade de Santiago de Compostela, Santiago de Compostela, Spain\\
% ...
$ ^{47}$Physik-Institut, Universit{\"a}t Z{\"u}rich, Z{\"u}rich, Switzerland\\
% ...
$ ^{51}$H.H. Wills Physics Laboratory, University of Bristol, Bristol, United Kingdom\\
% ...
$ ^{57}$Oliver Lodge Laboratory, University of Liverpool, Liverpool, United Kingdom\\
% ...
$ ^{62}$University of Cincinnati, Cincinnati, OH, United States\\
% ...
$ ^{65}$Syracuse University, Syracuse, NY, United States\\
% ...
$ ^{78}$University of Michigan, Ann Arbor, United States, associated to $^{65}$\\
\bigskip
% ...
}
\end{flushleft}